\documentclass[prd,preprintnumbers,twocolumn,amsmath,nofootinbib,amssymb]{revtex4}
\usepackage{graphicx,color,dcolumn,booktabs,bm}
\usepackage{longtable,lscape}
\usepackage{txfonts}
\usepackage{overpic}
\usepackage{amssymb}
\usepackage{epstopdf}
\usepackage{indentfirst}
\usepackage{feynmf}   
\usepackage{slashed}  
\usepackage{cases}
\usepackage{color}
\usepackage{float}
\usepackage{multirow}
\usepackage{ulem}
\usepackage{graphicx,color,dcolumn,booktabs,bm}
\usepackage{epsfig,dsfont,amssymb,amsmath,amsfonts,amsbsy,mathrsfs}

\graphicspath{{Figures/}} %

\usepackage{hyperref}
\hypersetup{colorlinks,citecolor=blue,anchorcolor=red,menucolor=red, linkcolor=red,filecolor=red,runcolor=red,urlcolor=blue,frenchlinks=red}

\makeatletter
\@addtoreset{equation}{section}
\makeatother

\allowdisplaybreaks

\begin{document}

\title{Higher molecular $P_{\psi s}^{\Lambda/\Sigma}$ pentaquarks arising from the $\Xi_c^{(\prime,*)}\bar D_1/\Xi_c^{(\prime,*)}\bar D_2^*$ interactions}

\author{Fu-Lai Wang$^{1,2,3,5}$}
\email{wangfl2016@lzu.edu.cn}
\author{Xiang Liu$^{1,2,3,4,5}$}
\email{xiangliu@lzu.edu.cn}
\affiliation{$^1$School of Physical Science and Technology, Lanzhou University, Lanzhou 730000, China\\
$^2$Lanzhou Center for Theoretical Physics, Key Laboratory of Theoretical Physics of Gansu Province, Lanzhou University, Lanzhou 730000, China\\
$^3$Key Laboratory of Quantum Theory and Applications of MoE, Lanzhou University,
Lanzhou 730000, China\\
$^4$MoE Frontiers Science Center for Rare Isotopes, Lanzhou University, Lanzhou 730000, China\\
$^5$Research Center for Hadron and CSR Physics, Lanzhou University and Institute of Modern Physics of CAS, Lanzhou 730000, China}

\begin{abstract}
The discoveries of the $P_{\psi s}^\Lambda(4459)$ and $P_{\psi s}^\Lambda(4338)$ as the potential $\Xi_c\bar D^{(*)}$ molecules have sparked our curiosity in exploring a novel class of molecular $P_{\psi s}^{\Lambda/\Sigma}$ pentaquarks. In this study, we carry out an investigation into the higher molecular pentaquarks, specifically focusing on the $P_{\psi s}^{\Lambda/\Sigma}$ states arising from the $\Xi_c^{(\prime,*)}\bar D_1/\Xi_c^{(\prime,*)}\bar D_2^*$ interactions. Our approach employs the one-boson-exchange model, incorporating both the $S$-$D$ wave mixing effect and the coupled channel effect. Our numerical results suggest that the $\Xi_c\bar D_1$ states with $I(J^P)=0({1}/{2}^+,\,{3}/{2}^+)$, the $\Xi_c\bar D_2^*$ states with $I(J^P)=0({3}/{2}^+,\,{5}/{2}^+)$, the $\Xi_c^{\prime}\bar D_1$ states with $I(J^P)=0({1}/{2}^+,\,{3}/{2}^+)$, the $\Xi_c^{\prime}\bar D_2^*$ states with $I(J^P)=0({3}/{2}^+,\,{5}/{2}^+)$, the $\Xi_c^{*}\bar D_1$ states with $I(J^P)=0({1}/{2}^+,\,{3}/{2}^+,\,{5}/{2}^+)$, and the $\Xi_c^{*}\bar D_2^*$ states with $I(J^P)=0({1}/{2}^+,\,{3}/{2}^+,\,{5}/{2}^+,\,{7}/{2}^+)$ can be recommended as the most promising molecular $P_{\psi s}^\Lambda$ pentaquark candidates, and there may exist the potential molecular $P_{\psi s}^\Sigma$ pentaquark candidates for several isovector $\Xi_c^{(\prime,*)}\bar D_1/\Xi_c^{(\prime,*)}\bar D_2^*$ states. With the higher statistical data accumulation at the LHCb's Run II and Run III status, there is the possibility that our predicted $P_{\psi s}^{\Lambda/\Sigma}$ states can be detected through the weak decay of the $\Xi_b$ baryon, especially in hunting for the predicted $P_{\psi s}^\Lambda$ states.
\end{abstract}

\maketitle

\section{Introduction}\label{sec1}

If the hadronic states exhibit configurations or properties beyond the conventional $q\bar q$ meson and $qqq$ baryon schemes \cite{GellMann:1964nj,Zweig:1981pd}, they are commonly refereed to as the exotic hadron states, which include molecular states, compact multiquark states, hybrids, glueballs, and so on. In the past two decades, abundant candidates for exotic hadron states have been reported by different experiments, making the study of these states a central focus in the field of the hadron physics \cite{Liu:2013waa,Hosaka:2016pey,Chen:2016qju,Richard:2016eis,Lebed:2016hpi,Brambilla:2019esw,Liu:2019zoy,Chen:2022asf,Olsen:2017bmm,Guo:2017jvc,Meng:2022ozq}. This research has expanded our understanding of the hadron structures and provided valuable insights into the nonperturbative behavior of quantum chromodynamics (QCD). Since the masses of numerous observed exotic hadron states lie very close to the thresholds of two hadrons, the investigation of hadronic molecular states has gained popularity.

\begin{figure}[htbp]
\includegraphics[scale=0.38]{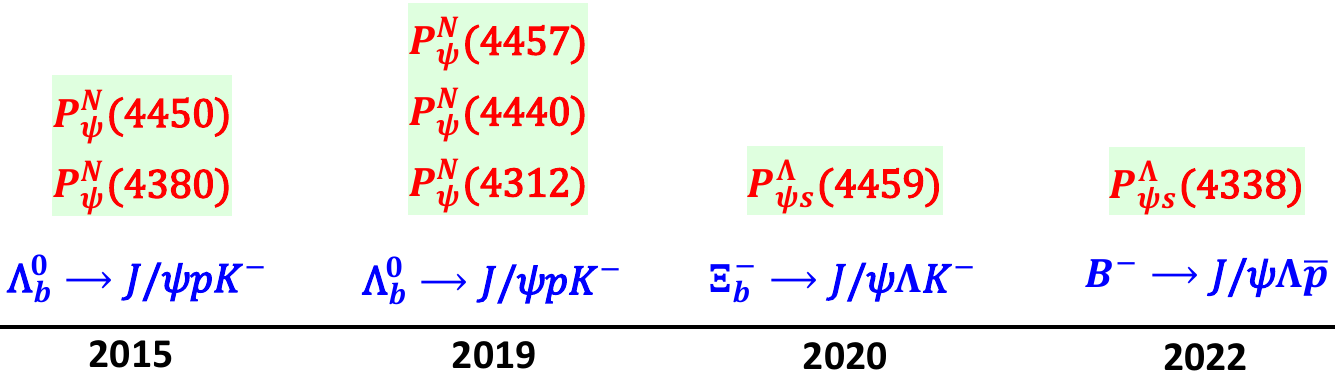}
\caption{The summary of the observed hidden-charm pentaquark states in the recent years \cite{Aaij:2015tga,Aaij:2019vzc,LHCb:2020jpq,LHCb:2022jad}. Here, we provide a list of these states along with their associated production processes and the years in which they were first observed.}\label{PcPcs}
\end{figure}

Significant progresses have been made in the study of the hidden-charm pentaquark states $P_{\psi}^{N}$ and $P_{\psi s}^{\Lambda}$ in recent years\footnote{In this work, we adopt the new naming scheme for the hidden-charm pentaquark states \cite{Gershon:2022xnn}.}, where we present the observed hidden-charm pentaquark states \cite{Aaij:2015tga,Aaij:2019vzc,LHCb:2020jpq,LHCb:2022jad} in Fig. \ref{PcPcs}. In 2015, the LHCb Collaboration reported the first discovery of the hidden-charm pentaquarks, namely $P_{\psi}^{N}(4380)$ and $P_{\psi}^{N}(4450)$, through an analysis of the $J/\psi p$ invariant mass spectrum of the $\Lambda_b^0 \to J/\psi p K^-$ process \cite{Aaij:2015tga}. However, in 2015, the experimental information alone did not allow for the distinction between various theoretical explanations for these observed hidden-charm pentaquark structures \cite{Chen:2016qju}. Especially, the LHCb claimed that the observed $P_{\psi}^{N}$ structures possess the opposite parities \cite{Aaij:2015tga}, which posed challenge within a unified framework of the hadronic molecular state \cite{Chen:2016qju}.

In 2019, LHCb conducted a more precise measurement of the $\Lambda_b^0 \to J/\psi p K^-$ process, utilizing  experimental data from both Run I and Run II \cite{Aaij:2019vzc}. This analysis revealed that the previously observed $P_{\psi}^{N}(4450)$ consists of two distinct substructures, $P_{\psi}^{N}(4440)$ and $P_{\psi}^{N}(4457)$. Furthermore, a new enhancement structure named $P_{\psi}^{N}(4312)$ was discovered, which can naturally be attributed to the $\Sigma_c \bar D^{(*)}$ molecular states \cite{Li:2014gra,Karliner:2015ina,Wu:2010jy,Wang:2011rga,Yang:2011wz,Wu:2012md,Chen:2015loa}. This updated experimental analysis from LHCb offers substantial support for the existence of the hidden-charm baryon-meson molecular pentaquark states in the realm of hadron spectroscopy \cite{Li:2014gra,Karliner:2015ina,Wu:2010jy,Wang:2011rga,Yang:2011wz,Wu:2012md,Chen:2015loa}. However, the explorations of the hidden-charm pentaquark states, both experimentally and theoretically, remain an ongoing process.

In 2020, LHCb presented the evidence for a potential hidden-charm pentaquark structure with strangeness in the $J/\psi \Lambda$ invariant mass spectrum of the $\Xi_b \to J/\psi \Lambda K^-$ process \cite{LHCb:2020jpq}, which was named $P_{\psi s}^\Lambda(4459)$. To date, the experimental measurement has not determined its spin-parity quantum number. Subsequently, in 2022, LHCb reported a new hidden-charm pentaquark structure with strangeness, $P_{\psi s}^\Lambda(4338)$, observed in the $B^- \to J/\psi \Lambda \bar p$ process by analyzing the $J/\psi \Lambda$ invariant mass spectrum \cite{LHCb:2022jad}. The preferred spin-parity quantum number for this state is $J^P=1/2^-$.

For revealing the properties of these observed hidden-charm pentaquark structures with strangeness like $P_{\psi s}^\Lambda(4459)$ \cite{LHCb:2020jpq} and $P_{\psi s}^\Lambda(4338)$ \cite{LHCb:2022jad},  theoretical studies, employing  the hadronic molecule scenario, have been proposed in Refs. \cite{Chen:2016ryt,Wu:2010vk,Hofmann:2005sw,Anisovich:2015zqa,Wang:2015wsa,Feijoo:2015kts,Lu:2016roh,Xiao:2019gjd,Shen:2020gpw,Chen:2015sxa,Zhang:2020cdi,Wang:2019nvm,Weng:2019ynv,Paryev:2023icm,Azizi:2023foj,Wang:2022neq,Wang:2022mxy,Karliner:2022erb,Yan:2022wuz,Meng:2022wgl,Ozdem:2023htj,Feijoo:2022rxf,Garcilazo:2022edi,Yang:2022ezl,Zhu:2022wpi,Chen:2022wkh,Ortega:2022uyu,Giachino:2022pws,Nakamura:2022jpd,Wang:2022tib,Ozdem:2022kei,Xiao:2022csb,Wang:2022gfb,Clymton:2022qlr,Chen:2022onm,Chen:2021spf,Gao:2021hmv,Ferretti:2021zis,Giron:2021fnl,Cheng:2021gca,Du:2021bgb,Chen:2021cfl,Hu:2021nvs,Yang:2021pio,Li:2021ryu,Lu:2021irg,Zou:2021sha,Wang:2021itn,Wu:2021caw,Clymton:2021thh,Xiao:2021rgp,Ozdem:2021ugy,Zhu:2021lhd,Chen:2021tip,Azizi:2021utt,Dong:2021juy,Liu:2020hcv,Wang:2020eep,Peng:2020hql,Chen:2020uif,Peng:2019wys,Chen:2020opr,Chen:2020kco,Burns:2022uha}. However, several puzzling phenomena arise when attempting to explain the observed $P_{\psi s}^\Lambda(4459)$ and $P_{\psi s}^\Lambda(4338)$ as the $\Xi_c\bar D^{(*)}$ molecular states \cite{Wang:2022mxy}.  For the $P_{\psi s}^\Lambda(4459)$ \cite{LHCb:2020jpq}, the presence of the double peak structures slightly below the threshold of the $\Xi_c\bar D^*$ channel poses a challenge. Regarding the $P_{\psi s}^\Lambda(4338)$ \cite{LHCb:2022jad}, its measured mass is close to and above the threshold of the $\Xi_c\bar D$ channel, making it difficult to assign it as the $\Xi_c\bar D$ molecular state. This difficulty arises from the requirement of the hadronic molecule explanation that the observed hadron state's mass should be close to and below the sum of the thresholds of its constituent hadrons \cite{Chen:2016qju, Liu:2019zoy}.

To address these puzzling phenomena, further investigation into the properties of the $P_{\psi s}^\Lambda(4459)$ and $P_{\psi s}^\Lambda(4338)$ is needed. Extensive discussions on these topics have taken place in recent years \cite{Paryev:2023icm,Azizi:2023foj,Wang:2022neq,Wang:2022mxy,Karliner:2022erb,Yan:2022wuz,Meng:2022wgl,Ozdem:2023htj,Feijoo:2022rxf,Garcilazo:2022edi,Yang:2022ezl,Zhu:2022wpi,Chen:2022wkh,Ortega:2022uyu,Giachino:2022pws,Nakamura:2022jpd,Wang:2022tib,Ozdem:2022kei,Xiao:2022csb,Wang:2022gfb,Clymton:2022qlr,Chen:2022onm,Chen:2021spf,Gao:2021hmv,Ferretti:2021zis,Giron:2021fnl,Cheng:2021gca,Du:2021bgb,Chen:2021cfl,Hu:2021nvs,Yang:2021pio,Li:2021ryu,Lu:2021irg,Zou:2021sha,Wang:2021itn,Wu:2021caw,Clymton:2021thh,Xiao:2021rgp,Ozdem:2021ugy,Zhu:2021lhd,Chen:2021tip,Azizi:2021utt,Dong:2021juy,Liu:2020hcv,Wang:2020eep,Peng:2020hql,Chen:2020uif,Peng:2019wys,Chen:2020opr,Chen:2020kco,Burns:2022uha} and should be checked in future experiments. Additionally, it is worth studying whether similar behaviors exist in other molecular $P_{\psi s}^{\Lambda}$ pentaquarks. This approach holds potential for unraveling the aforementioned puzzling phenomena. Moreover, the exploration of a novel class of molecular $P_{\psi s}^{\Lambda/\Sigma}$ pentaquarks could further enrich our knowledge in this field, providing valuable insights for future experimental searches and contributing to a more comprehensive understanding of these molecular pentaquarks.

The main focus of our study is on the $\Xi_c^{(\prime,*)}\bar D_1/\Xi_c^{(\prime,*)}\bar D_2^*$ systems that can be regarded as a new class of molecular pentaquark candidates, namely molecular $P_{\psi s}^{\Lambda/\Sigma}$ pentaquarks, with masses ranging approximately from 4.87 to 5.10 GeV. Here, the $D_1$ and $D_2^*$ states stand for the $D_1(2420)$ and $D_2^*(2460)$ charmed mesons listed in the Particle Data Group \cite{ParticleDataGroup:2022pth}. In our calculations, we deduce the effective potentials of the $\Xi_c^{(\prime,*)}\bar D_1/\Xi_c^{(\prime,*)}\bar D_2^*$ systems using the one-boson-exchange (OBE) model. These potentials incorporate contributions from the exchange of the $\sigma$, $\pi$, $\eta$, $\rho$, and $\omega$ particles \cite{Chen:2016qju,Liu:2019zoy}. To ensure comprehensive and systematic results, we account for both the $S$-$D$ wave mixing effect and the coupled channel effect. This consideration enables a more extensive mass spectrum of the $\Xi_c^{(\prime,*)}\bar D_1/\Xi_c^{(\prime,*)}\bar D_2^*$-type hidden-charm molecular pentaquark candidates with strangeness. By employing the obtained OBE effective potentials, we can then solve the coupled channel Schrödinger equation to search for the bound state solutions. This process allows us to predict a novel class of molecular $P_{\psi s}^{\Lambda/\Sigma}$ pentaquark candidates comprising the charmed baryons $\Xi_c^{(\prime,*)}$ and the anticharmed mesons $\bar D_1/\bar D_2^*$.

This paper is organized as follows. After presenting the Introduction in Sec. \ref{sec1}, we deduce the OBE effective potentials for the $\Xi_c^{(\prime,*)}\bar D_1/\Xi_c^{(\prime,*)}\bar D_2^*$ systems in Sec. \ref{sec2}. With these obtained OBE effective potentials, we discuss the bound state properties for the $\Xi_c^{(\prime,*)}\bar D_1/\Xi_c^{(\prime,*)}\bar D_2^*$ systems, and predict a novel class of molecular $P_{\psi s}^{\Lambda/\Sigma}$ pentaquark candidates composed of the charmed baryons $\Xi_c^{(\prime,*)}$ and the anticharmed mesons $\bar D_1/\bar D_2^*$ in Sec. \ref{sec3}. Finally, this work ends with the discussions and conclusions in Sec. \ref{sec4}.

\section{The deduction of the OBE effective potentials of the $\Xi_c^{(\prime,*)}\bar D_1/\Xi_c^{(\prime,*)}\bar D_2^*$ systems}\label{sec2}

The main task of the present work is to investigate a novel class of molecular $P_{\psi s}^{\Lambda/\Sigma}$ pentaquark candidates comprising the charmed baryons $\Xi_c^{(\prime,*)}$ and the anticharmed mesons $\bar D_1/\bar D_2^*$, so the $\Xi_c^{(\prime,*)}\bar D_1/\Xi_c^{(\prime,*)}\bar D_2^*$ interactions are the important inputs in the whole calculations. In this work, we deduce the effective potentials of the $\Xi_c^{(\prime,*)}\bar D_1/\Xi_c^{(\prime,*)}\bar D_2^*$ systems by adopting the OBE model  \cite{Chen:2016qju,Liu:2019zoy}, which is one of the powerful tool to discuss the interactions between hadrons by exchanging the allowed light pseudoscalar, scalar, and vector mesons, such as $\pi$, $\eta$, $\sigma$, $\rho$, $\omega$, and so on. During the past few decades,  the OBE model is widely adopted to study the hadron-hadron interactions. Especially, this model was applied to reproduce the masses of the observed $P_{\psi}^{N}$ \cite{Aaij:2019vzc} and $P_{\psi s}^{\Lambda}$ \cite{LHCb:2020jpq,LHCb:2022jad} under the baryon-meson molecule picture \cite{Li:2014gra,Karliner:2015ina,Wu:2010jy,Wang:2011rga,Yang:2011wz,Wu:2012md,Chen:2015loa,Paryev:2023icm,Azizi:2023foj,Wang:2022neq,Wang:2022mxy,Karliner:2022erb,Yan:2022wuz,Meng:2022wgl,Ozdem:2023htj,Feijoo:2022rxf,Garcilazo:2022edi,Yang:2022ezl,Zhu:2022wpi,Chen:2022wkh,Ortega:2022uyu,Giachino:2022pws,Nakamura:2022jpd,Wang:2022tib,Ozdem:2022kei,Xiao:2022csb,Wang:2022gfb,Clymton:2022qlr,Chen:2022onm,Chen:2021spf,Gao:2021hmv,Ferretti:2021zis,Giron:2021fnl,Cheng:2021gca,Du:2021bgb,Chen:2021cfl,Hu:2021nvs,Yang:2021pio,Li:2021ryu,Lu:2021irg,Zou:2021sha,Wang:2021itn,Wu:2021caw,Clymton:2021thh,Xiao:2021rgp,Ozdem:2021ugy,Zhu:2021lhd,Chen:2021tip,Azizi:2021utt,Dong:2021juy,Liu:2020hcv,Wang:2020eep,Peng:2020hql,Chen:2020uif,Peng:2019wys,Chen:2020opr,Chen:2020kco,Burns:2022uha}, which encourages us to predict the $\Xi_c^{(\prime,*)}\bar D_1/\Xi_c^{(\prime,*)}\bar D_2^*$-type hidden-charm molecular pentaquark candidates with strangeness within the OBE model.

\subsection{Effective Lagrangians}\label{subsec1}

When taking the OBE model to estimate the interactions between hadrons quantitatively, the previous theoretical studies usually adopt the effective Lagrangian approach to calculate the scattering amplitudes \cite{Chen:2016qju,Liu:2019zoy}, and it is necessary to construct the relevant effective Lagrangians. For describing the $\Xi_c^{(\prime,*)}\bar D_1/\Xi_c^{(\prime,*)}\bar D_2^*$ interactions, the contributions from the exchange of the scalar meson $\sigma$, the pseudoscalar mesons $\pi$ and $\eta$, and the vector mesons $\rho$ and $\omega$ are considered \cite{Chen:2016qju,Liu:2019zoy}, and we need to calculate them out one by one and sum them in the concrete calculations.

For the sake of completeness, we briefly recall the properties of the baryons $\Xi_c^{(\prime,*)}$ and the mesons $\bar D_1/\bar D_2^*$, which can provide the crucial information to construct the effective Lagrangians. By taking the heavy quark spin symmetry \cite{Wise:1992hn}, the total angular momentum of the light degrees of freedom $j_{\ell}$ including both the light quark spin $s_q$ and the orbital angular momentum $\ell$ is a good quantum number for the hadron containing the single heavy quark, and the hadronic states with the total angular momentum $J=j_{\ell} \pm 1/2$ can form the doublet, except for the case for $j_{\ell}=0$. Thus, the properties of the single heavy hadrons in the same doublet are degenerate approximatively, which can be written as the superfield to construct the effective Lagrangians. For these discussed singly charmed baryons, $\Xi_c$ with $J^P=1/2^+$ is the $S$-wave charmed baryon in the $\bar{3}_F$ flavor representation with $j_{\ell}=0$, while $\Xi_c^\prime$ with $J^P=1/2^+$ and $\Xi_c^*$ with $J^P=3/2^+$ are the $S$-wave charmed baryons in the $6_F$ flavor representation with $j_{\ell}=1$ \cite{ParticleDataGroup:2022pth}. In general, the singly charmed baryon matrices $\mathcal{B}_{\bar{3}}$ and $\mathcal{B}_6^{(*)}$ are defined as \cite{Wise:1992hn,Casalbuoni:1992gi,Casalbuoni:1996pg,Yan:1992gz,Bando:1987br,Harada:2003jx,Chen:2017xat}
\begin{eqnarray}
\mathcal{B}_{\bar{3}} = \left(\begin{array}{ccc}
        0    &\Lambda_c^+      &\Xi_c^+\\
        -\Lambda_c^+       &0      &\Xi_c^0\\
        -\Xi_c^+      &-\Xi_c^0     &0
\end{array}\right),
\mathcal{B}_6^{(*)} = \left(\begin{array}{ccc}
         \Sigma_c^{{(*)}++}                  &\frac{\Sigma_c^{{(*)}+}}{\sqrt{2}}     &\frac{\Xi_c^{(\prime,*)+}}{\sqrt{2}}\\
         \frac{\Sigma_c^{{(*)}+}}{\sqrt{2}}      &\Sigma_c^{{(*)}0}    &\frac{\Xi_c^{(\prime,*)0}}{\sqrt{2}}\\
         \frac{\Xi_c^{(\prime,*)+}}{\sqrt{2}}    &\frac{\Xi_c^{(\prime,*)0}}{\sqrt{2}}      &\Omega_c^{(*)0}
\end{array}\right),
\end{eqnarray}
respectively. Under the heavy quark spin symmetry, the $S$-wave singly charmed baryons in $6_F$ flavor representation $\mathcal{B}_6$ and $\mathcal{B}^*_6$ can be expressed as the superfield $\mathcal{S}_{\mu}$, which is given by \cite{Wise:1992hn,Casalbuoni:1992gi,Casalbuoni:1996pg,Yan:1992gz,Bando:1987br,Harada:2003jx,Chen:2017xat}
\begin{eqnarray}
\mathcal{S}_{\mu}=-\sqrt{\frac{1}{3}}(\gamma_{\mu}+v_{\mu})\gamma^5\mathcal{B}_6+\mathcal{B}_{6\mu}^*.
\end{eqnarray}
Here, the four velocity has the form $v_{\mu}=(1, 0, 0, 0)$ within the nonrelativistic approximation, and its conjugate field is $\bar{\mathcal{S}}_{\mu}=\mathcal{S}_{\mu}^\dag\gamma^0$. For these focused mesons, $\bar D_1$ with $J^P=1^+$ and $\bar D_2^*$ with $J^P=2^+$ are the $P$-wave anticharmed mesons in the $T$ doublet with $j_{\ell}=3/2$ \cite{ParticleDataGroup:2022pth}, which can be constructed as the superfield $T^{(\bar Q)\mu}_a$ as follows \cite{Ding:2008gr}
\begin{eqnarray}
T^{(\bar Q)\mu}_a=\left[\bar D^{*\mu\nu}_{2a}\gamma_{\nu}-\sqrt{\frac{3}{2}}\bar D_{1a\nu}\gamma_5\left(g^{\mu\nu}-\frac{1}{3}(\gamma^{\mu}-v^{\mu})\gamma^{\nu}
\right)\right]\frac{1-\slash \!\!\!v}{2},\nonumber\\
\end{eqnarray}
where the corresponding conjugate field is $\overline{T}^{(\bar Q)\mu}_a=\gamma^0T_a^{(\bar Q)\mu\dag}\gamma^0$. For convenience, we take the column matrices to describe the anticharmed meson fields in the $T$ doublet, i.e., $\bar D_1=(\bar D_1^0,\,\bar D_1^-,\, D_{s1}^-)^T$ and $\bar D_{2}^{*}=(\bar D_2^{*0},\,\bar D_2^{*-},\, D_{s2}^{*-})^T$.

Now we move on to construct the effective Lagrangians adopted in the present work by taking into account the symmetry requirements. With the help of the constraints of the heavy quark symmetry, the chiral symmetry, and the hidden local symmetry \cite{Casalbuoni:1992gi,Casalbuoni:1996pg,Yan:1992gz,Harada:2003jx,Bando:1987br}, the effective Lagrangians related to the interactions between the singly charmed baryons $\Xi_c^{(\prime,*)}$ and the light scalar, pseudoscalar, and vector mesons are constructed as \cite{Wise:1992hn,Casalbuoni:1992gi,Casalbuoni:1996pg,Yan:1992gz,Bando:1987br,Harada:2003jx,Chen:2017xat}
\begin{eqnarray}
\mathcal{L}_{\mathcal{B}_{\bar{3}}\mathcal{B}_{\bar{3}}\mathcal{E}} &=& l_B\langle\bar{\mathcal{B}}_{\bar{3}}\sigma\mathcal{B}_{\bar{3}}\rangle
          +i\beta_B\langle\bar{\mathcal{B}}_{\bar{3}}v^{\mu}(\mathcal{V}_{\mu}
          -\rho_{\mu})\mathcal{B}_{\bar{3}}\rangle,\\
\mathcal{L}_{\mathcal{S}\mathcal{S}\mathcal{E}} &=& l_S\langle\bar{\mathcal{S}}_{\mu}\sigma\mathcal{S}^{\mu}\rangle -\frac{3}{2}g_1\varepsilon^{\mu\nu\lambda\kappa}v_{\kappa}\langle\bar{\mathcal{S}}_{\mu}\mathcal{A}_{\nu}
\mathcal{S}_{\lambda}\rangle\nonumber\\
    &&+i\beta_{S}\langle\bar{\mathcal{S}}_{\mu}v_{\alpha}\left(\mathcal{V}^{\alpha}
    -\rho^{\alpha}\right) \mathcal{S}^{\mu}\rangle +\lambda_S\langle\bar{\mathcal{S}}_{\mu}F^{\mu\nu}(\rho)\mathcal{S}_{\nu}\rangle,\nonumber\\\\
\mathcal{L}_{\mathcal{B}_{\bar{3}}\mathcal{S}\mathcal{E}}&=&ig_4\langle\bar{\mathcal{S}^{\mu}}
\mathcal{A}_{\mu}\mathcal{B}_{\bar{3}}\rangle+i\lambda_I\varepsilon^{\mu\nu\lambda
\kappa}v_{\mu}\langle \bar{\mathcal{S}}_{\nu}F_{\lambda\kappa}\mathcal{B}_{\bar{3}}\rangle+h.c.,\nonumber\\
\end{eqnarray}
where the notation $\mathcal{E}$ in the subscript stands for the exchanged light mesons. Furthermore, the effective Lagrangians depicting the interactions of the anticharmed mesons in the $T$ doublet $\bar{D}_1/\bar{D}_2^*$ and the light scalar, pseudoscalar, and vector mesons can be constructed as \cite{Ding:2008gr}
\begin{eqnarray}
\mathcal{L}_{\bar {T}\bar {T}\mathcal{E}}&=&g''_{\sigma}\langle \overline{T}^{(\bar Q)\mu}_a\sigma T^{\,({\bar Q})}_{a\mu}\rangle+ik\langle \overline{T}^{(\bar Q)\mu}_{b}{\cal A}\!\!\!\slash_{ba}\gamma_5{T}^{\,(\bar Q)}_{a\mu}\rangle\nonumber\\
    &&-i\beta^{\prime\prime}\langle \overline{T}^{(\bar Q)}_{b\lambda}v^{\mu}({\cal V}_{\mu}-\rho_{\mu})_{ba}T^{\,(\bar Q)\lambda}_{a}\rangle\nonumber\\
    &&+i\lambda^{\prime\prime}\langle \overline{T}^{(\bar Q)}_{b\lambda}\sigma^{\mu\nu}F_{\mu\nu}(\rho)_{ba}T^{\,(\bar Q)\lambda}_{a}\rangle.
\end{eqnarray}
In the constructed effective Lagrangians, the axial current $\mathcal{A}_\mu$ and the vector current ${\mathcal V}_{\mu}$ are given by
\begin{eqnarray}
{\mathcal A}_{\mu}=\frac{1}{2}\left(\xi^{\dagger}\partial_{\mu}\xi-\xi\partial_{\mu}\xi^{\dagger}\right),~~~~
{\mathcal V}_{\mu}=\frac{1}{2}\left(\xi^{\dagger}\partial_{\mu}\xi+\xi\partial_{\mu}\xi^{\dagger}\right),
\end{eqnarray}
respectively. Here, the pseudo-Goldstone meson field is $\xi=e^{i\mathbb{P}/f_\pi}$, and $f_{\pi}$ stands for the pion decay constant with $f_{\pi}=0.132~{\rm GeV}$. Furthermore, $\mathcal{A}_\mu$ and ${\mathcal V}_{\mu}$ at the leading order of $\xi$ can be simplified to be
\begin{eqnarray}
\mathcal{A}_\mu=\frac{i}{f_\pi}\partial_\mu{\mathbb P},~~~~~\mathcal{V}_{\mu}=0.
\end{eqnarray}
In addition, we define the vector meson field $\rho_{\mu}$ and the vector meson field strength tensor $F_{\mu\nu}$ as
\begin{eqnarray}
\rho_{\mu}=\frac{i{g_V}}{\sqrt{2}}\mathbb{V}_{\mu},~~~~~
F_{\mu\nu}&=&\partial_{\mu}\rho_{\nu}-\partial_{\nu}\rho_{\mu}+\left[\rho_{\mu},\rho_{\nu}\right],
\end{eqnarray}
respectively. Explicitly, the light pseudoscalar meson matrix $\mathbb{P}$ and the light vector meson matrix $\mathbb{V}_{\mu}$ are defined as
\begin{eqnarray}
{\mathbb{P}} &=& {\left(\begin{array}{ccc}
       \frac{\pi^0}{\sqrt{2}}+\frac{\eta}{\sqrt{6}} &\pi^+ &K^+\\
       \pi^-       &-\frac{\pi^0}{\sqrt{2}}+\frac{\eta}{\sqrt{6}} &K^0\\
       K^-         &\bar K^0   &-\sqrt{\frac{2}{3}} \eta     \end{array}\right)},\\
{\mathbb{V}}_{\mu} &=& {\left(\begin{array}{ccc}
       \frac{\rho^0}{\sqrt{2}}+\frac{\omega}{\sqrt{2}} &\rho^+ &K^{*+}\\
       \rho^-       &-\frac{\rho^0}{\sqrt{2}}+\frac{\omega}{\sqrt{2}} &K^{*0}\\
       K^{*-}         &\bar K^{*0}   & \phi     \end{array}\right)}_{\mu},
\end{eqnarray}
respectively.

Combined with the constructed effective Lagrangians and the defined physical quantities, the concrete effective Lagrangians can be further obtained after expanding the above constructed effective Lagrangians to the leading order of $\xi$, which are needed in the realistic calculations. Specifically, the effective Lagrangians for the heavy hadrons $\Xi_c^{(\prime,*)}/\bar{D}_1/\bar{D}_2^*$ coupling with the light scalar meson $\sigma$ are expressed as
\begin{eqnarray}
\mathcal{L}_{\mathcal{B}_{\bar{3}}\mathcal{B}_{\bar{3}}\sigma}&=& l_B\langle \bar{\mathcal{B}}_{\bar{3}}\sigma\mathcal{B}_{\bar{3}}\rangle,\\
\mathcal{L}_{\mathcal{B}_{6}^{(*)}\mathcal{B}_{6}^{(*)}\sigma}&=&-l_S\langle\bar{\mathcal{B}}_6\sigma\mathcal{B}_6\rangle+l_S\langle
\bar{\mathcal{B}}_{6\mu}^{*}
\sigma\mathcal{B}_6^{*\mu}\rangle\nonumber\\
    &&-\frac{l_S}{\sqrt{3}}\langle\bar{\mathcal{B}}_{6\mu}^{*}\sigma
    \left(\gamma^{\mu}+v^{\mu}\right)\gamma^5\mathcal{B}_6\rangle+h.c.,\\
\mathcal{L}_{\bar{T}\bar{T}\sigma}&=&-2g_\sigma^{\prime\prime}\bar{D}_{1a\mu}\bar{D}^{\mu\dagger}_{1a}\sigma
+2g_\sigma^{\prime\prime}\bar{D}^{*\dagger}_{2a\mu\nu}\bar{D}^{*\mu\nu}_{2a}\sigma,
\end{eqnarray}
and the effective Lagrangians depicting the interactions of the heavy hadrons $\Xi_c^{(\prime,*)}/\bar{D}_1/\bar{D}_2^*$ and the light pseudoscalar mesons $\mathbb{P}$ are
\begin{eqnarray}
\mathcal{L}_{\mathcal{B}_6^{(*)}\mathcal{B}_6^{(*)}\mathbb{P}}&=&i\frac{g_1}{2f_{\pi}}\varepsilon^{\mu\nu\lambda\kappa}v_{\kappa}\langle\bar{\mathcal{B}}_6
\gamma_{\mu}\gamma_{\lambda}\partial_{\nu}\mathbb{P}\mathcal{B}_6\rangle\nonumber\\
    &&-i\frac{3g_1}{2f_{\pi}}\varepsilon^{\mu\nu\lambda\kappa}v_{\kappa}
    \langle\bar{\mathcal{B}}_{6\mu}^{*}\partial_{\nu}\mathbb{P}\mathcal{B}_{6\lambda}^*\rangle\nonumber\\
    &&+i\frac{\sqrt{3}g_1}{2f_{\pi}}v_{\kappa}\varepsilon^{\mu\nu\lambda\kappa}
    \langle\bar{\mathcal{B}}_{6\mu}^*\partial_{\nu}\mathbb{P}{\gamma_{\lambda}\gamma^5}
      \mathcal{B}_6\rangle+h.c.,\nonumber\\\\
\mathcal{L}_{\mathcal{B}_{\bar{3}}\mathcal{B}_6^{(*)}\mathbb{P}} &=& -\sqrt{\frac{1}{3}}\frac{g_4}{f_{\pi}}\langle\bar{\mathcal{B}}_6\gamma^5\left(\gamma^{\mu}
+v^{\mu}\right)\partial_{\mu}\mathbb{P}\mathcal{B}_{\bar{3}}\rangle\nonumber\\
    &&-\frac{g_4}{f_{\pi}}\langle\bar{\mathcal{B}}_{6\mu}^*\partial^{\mu} \mathbb{P}\mathcal{B}_{\bar{3}}\rangle+h.c.,\\
\mathcal {L}_{\bar{T} \bar{T}\mathbb{P}} &=&-\frac{5ik}{3f_\pi}\varepsilon^{\mu\nu\rho\tau}v_\nu\bar{D}^{\dag}_{1a\rho}
\bar{D}_{1b\tau}\partial_\mu\mathbb{P}_{ba}\nonumber\\
    &&+\frac{2ik}{f_\pi}\varepsilon^{\mu\nu\rho\tau}v_\nu\bar{D}^{*\alpha\dag}_{2a\rho}
    \bar{D}^{*}_{2b\alpha\tau}\partial_\mu\mathbb{P}_{ba}\nonumber\\
    &&+\sqrt{\frac{2}{3}}\frac{k}{f_\pi}\left(\bar{D}^{\dagger}_{1a\mu}
    \bar{D}^{*\mu\lambda}_{2b}+\bar{D}_{1b\mu}
    \bar{D}^{*\mu\lambda\dagger}_{2a}\right)\partial_\lambda\mathbb{P}_{ba},\nonumber\\
\end{eqnarray}
and the effective Lagrangians describing the interactions between the heavy hadrons $\Xi_c^{(\prime,*)}/\bar{D}_1/\bar{D}_2^*$ and the light vector mesons $\mathbb{V}$ are
\begin{eqnarray}
\mathcal{L}_{\mathcal{B}_{\bar{3}}\mathcal{B}_{\bar{3}}\mathbb{V}}&=&
\frac{1}{\sqrt{2}}\beta_Bg_V\langle\bar{\mathcal{B}}_{\bar{3}}v\cdot\mathbb{V}
\mathcal{B}_{\bar{3}}\rangle,\\
\mathcal{L}_{\mathcal{B}_6^{(*)}\mathcal{B}_6^{(*)}\mathbb{V}}&=&
-\frac{\beta_Sg_V}{\sqrt{2}}\langle\bar{\mathcal{B}}_6v\cdot\mathbb{V}
\mathcal{B}_6\rangle+\frac{\beta_Sg_V}{\sqrt{2}}\langle\bar{\mathcal{B}}_{6\mu}^*v\cdot {V}\mathcal{B}_6^{*\mu}\rangle\nonumber\\
    &&-i\frac{\lambda_S g_V}{3\sqrt{2}}\langle\bar{\mathcal{B}}_6\gamma_{\mu}\gamma_{\nu}
    \left(\partial^{\mu}\mathbb{V}^{\nu}-\partial^{\nu}\mathbb{V}^{\mu}\right)
    \mathcal{B}_6\rangle\nonumber\\
    &&+i\frac{\lambda_Sg_V}{\sqrt{2}}\langle\bar{\mathcal{B}}_{6\mu}^*
    \left(\partial^{\mu}\mathbb{V}^{\nu}-\partial^{\nu}\mathbb{V}^{\mu}\right)
    \mathcal{B}_{6\nu}^*\rangle\nonumber\\
    &&-i\frac{\lambda_Sg_V}{\sqrt{6}}\langle\bar{\mathcal{B}}_{6\mu}^*
    \left(\partial^{\mu}\mathbb{V}^{\nu}-\partial^{\nu}\mathbb{V}^{\mu}\right)
    \left(\gamma_{\nu}+v_{\nu}\right)\gamma^5\mathcal{B}_6\rangle\nonumber\\
    &&-\frac{\beta_Sg_V}{\sqrt{6}}\langle\bar{\mathcal{B}}_{6\mu}^*v\cdot \mathbb{V}\left(\gamma^{\mu}+v^{\mu}\right)\gamma^5\mathcal{B}_6\rangle+h.c.,\nonumber\\\\
\mathcal{L}_{\mathcal{B}_{\bar{3}}\mathcal{B}_6^{(*)}\mathbb{V}} &=&
-\frac{\lambda_Ig_V}{\sqrt{2}}\varepsilon^{\mu\nu\lambda\kappa}v_{\mu}\langle \bar{\mathcal{B}}_{6\nu}^*\left(\partial_{\lambda}\mathbb{V}_{\kappa}
-\partial_{\kappa}\mathbb{V}_{\lambda}\right)
          \mathcal{B}_{\bar{3}}\rangle\nonumber\\
    &&-\frac{\lambda_Ig_V}{\sqrt{6}}\varepsilon^{\mu\nu\lambda\kappa}v_{\mu}\langle \bar{\mathcal{B}}_6\gamma^5\gamma_{\nu}
        \left(\partial_{\lambda}\mathbb{V}_{\kappa}-\partial_{\kappa} \mathbb{V}_{\lambda}\right)\mathcal{B}_{\bar{3}}\rangle+h.c.,\nonumber\\\\
\mathcal {L}_{\bar{T}\bar{T}\mathbb{V}} &=&\sqrt{2}\beta^{\prime\prime}g_{V}\left(v\cdot\mathbb{V}_{ba}\right)\bar{D}_{1b\mu}
\bar{D}^{\mu\dagger}_{1a}\nonumber\\
    &&+\frac{5\sqrt{2}i\lambda^{\prime\prime}g_{V}}{3}\left(\bar{D}^{\nu}_{1b}
    \bar{D}^{\mu\dagger}_{1a}
    -\bar{D}^{\nu\dagger}_{1a}\bar{D}^{\mu}_{1b}\right)\partial_\mu\mathbb{V}_{ba\nu}\nonumber\\
    &&-\sqrt{2}\beta^{\prime \prime}g_{V}\left(v\cdot\mathbb{V}_{ba}\right) \bar{D}_{2b}^{*\lambda\nu} \bar{D}^{*\dagger}_{2a{\lambda\nu}}\nonumber\\
    &&+2\sqrt{2}i\lambda^{\prime\prime} g_{V}\left(\bar{D}^{*\lambda\nu\dagger}_{2a}
    \bar{D}^{*\mu}_{2b\lambda}-\bar{D}^{*\lambda\nu}_{2b} \bar{D}^{*\mu\dagger}_{2a\lambda}\right)\partial_\mu \mathbb{V}_{ba\nu}\nonumber\\
    &&+\frac{i\beta^{\prime\prime}g_{V}}{\sqrt{3}}\varepsilon^{\lambda\alpha\rho\tau}v_{\rho}
    \left(v\cdot\mathbb{V}_{ba}\right)\left(\bar{D}^{\dagger}_{1a\alpha} \bar{D}^{*}_{2b\lambda\tau}-\bar{D}_{1b\alpha}
    \bar{D}^{\dagger*}_{2a\lambda\tau}\right)\nonumber\\
    &&+\frac{2\lambda^{\prime\prime}g_{V}}{\sqrt{3}}
    \left[3\varepsilon^{\mu\lambda\nu\tau}v_\lambda\left(\bar{D}^{\alpha\dagger}_{1a}
    \bar{D}^{*}_{2b\alpha\tau}+\bar{D}^{\alpha}_{1b}
    \bar{D}^{*\dagger}_{2a\alpha\tau}\right)\partial_\mu\mathbb{V}_{ba\nu}\right.\nonumber\\
    &&\left.+2\varepsilon^{\lambda\alpha\rho\nu}v_\rho
    \left(\bar{D}^{\dagger}_{1a\alpha}\bar{D}^{*\mu}_{2b\lambda}
    +\bar{D}_{1b\alpha}\bar{D}^{\dagger\mu*}_{2a\lambda}\right)\right.\nonumber\\
    &&\left.\times\left(\partial_\mu \mathbb{V}_{ba\nu}-\partial_\nu \mathbb{V}_{ba\mu}\right)\right].
\end{eqnarray}

For these obtained effective Lagrangians, the coupling constants are the important input parameters to describe the strengths of the interaction vertices quantitatively. In general, we can extract the coupling constants through reproducing the experimental data when there exists the relevant experimental information, and the coupling constants also can be deduced by taking the theoretical models and approaches. Furthermore, the phase factors of the related coupling constants can be fixed with the help of the quark model \cite{Riska:2000gd}. In the following numerical analysis, we take $l_B=-3.65$, $l_S=6.20$, $g_{\sigma}^{\prime\prime}=-0.76$, $g_1=0.94$, $g_4=1.06$, $k=-0.59$, $\beta_B g_{V}=-6.00$, $\beta_S g_{V}=12.00$, $\lambda_Ig_V=-6.80~\rm {GeV^{-1}}$, $\lambda_Sg_V=19.20~\rm {GeV^{-1}}$, $\beta^{\prime\prime} g_{V}=5.25$, and $\lambda^{\prime\prime} g_{V}=3.27 ~\rm {GeV^{-1}}$, which were given in Refs. \cite{Wang:2022mxy,Chen:2017xat,Chen:2019asm,Chen:2020kco,Wang:2020bjt,Wang:2019nwt,Chen:2018pzd,Wang:2021hql,Wang:2019aoc,Wang:2021ajy,Wang:2021yld,Wang:2021aql,Yang:2021sue,Wang:2020dya}. In the past decades, these coupling constants are widely applied to discuss the hadron-hadron interactions, especially after the observed $P_{\psi}^N$ and $P_{\psi s}^{\Lambda}$  \cite{Aaij:2015tga,Aaij:2019vzc,LHCb:2020jpq,LHCb:2022jad}.

\subsection{The OBE potentials}\label{subsec2}

Now we illustrate how to deduce the OBE effective potentials for the $\Xi_c^{(\prime,*)}\bar D_1/\Xi_c^{(\prime,*)}\bar D_2^*$ systems based on the constructed effective Lagrangians \cite{Wang:2022mxy,Wang:2020dya,Wang:2019nwt,Wang:2019aoc,Wang:2020bjt,Wang:2021hql,Chen:2018pzd,Yang:2021sue,Wang:2021ajy,Wang:2021yld,Wang:2021aql}. In the context of the effective Lagrangian approach, we can calculate the scattering amplitude $\mathcal{M}^{h_1h_2\to h_3h_4}(\bm{q})$ of the $h_1h_2\to h_3h_4$ scattering process by exchanging the allowed light mesons $\mathcal{E}$ with the help of the Feynman rule, which can be obtained by the following relation \cite{Wang:2021ajy}
\begin{eqnarray}
i\mathcal{M}^{h_1h_2\to h_3h_4}(\bm{q})=\sum_{\mathcal{E}=\sigma,\,\mathbb{P},\,\mathbb{V}}i\Gamma^{h_1h_3\mathcal{E}}_{(\mu)} P_{\mathcal{E}}^{(\mu\nu)} i\Gamma^{h_2h_4\mathcal{E}}_{(\nu)}.
\end{eqnarray}
Here, $P_{\mathcal{E}}^{(\mu\nu)}$ is the propagator of the exchanged light meson, which can be defined as
\begin{eqnarray}
P_{\sigma}&=&\frac{i}{q^2-m_\sigma^2},~~~~~~~~~~~~~~~~~~~~P_{{\mathbb P}}=\frac{i}{q^2-m_{{\mathbb P}}^2},\nonumber\\
P_{\mathbb{V}}^{\mu\nu}&=&-i\frac{g^{\mu\nu}-{q^{\mu} q^{\nu}}/{m_{\mathbb{V}}^2}}{q^2-m_{\mathbb{V}}^2}
\end{eqnarray}
for the scalar, pseudoscalar, and vector mesons,  respectively. Here, $q$ and $m_{\mathcal{E}}$ are the four momentum and the mass of the exchanged light meson, respectively. $\Gamma^{h_1h_3\mathcal{E}}_{(\mu)}$ and $\Gamma^{h_2h_4\mathcal{E}}_{(\nu)}$ are the corresponding interaction vertices for the $h_1h_2\to h_3h_4$ scattering process, which can be extracted from the constructed effective Lagrangians $\mathcal{L}_{h_1h_3\mathcal{E}}$ and $\mathcal{L}_{h_2h_4\mathcal{E}}$, respectively. In the above subsection, we have constructed the effective Lagrangians adopted in the present work. In Appendix \ref{app00}, we present the related interaction vertices. In addition, we also need to define the normalization relations for the heavy hadrons $\Xi_c^{(\prime,*)}/\bar{D}_1/\bar{D}_2^*$ to write down the scattering amplitude $\mathcal{M}^{h_1h_2\to h_3h_4}(\bm{q})$. In our calculations, we take the normalization relations for the heavy hadrons $\Xi_c^{(\prime,*)}/\bar{D}_1/\bar{D}_2^*$ as \cite{Ding:2008gr}
\begin{eqnarray}
\langle 0|\Xi_{c}^{(\prime)}|cqs\left({1}/{2}^+\right)\rangle &=& \sqrt{2m_{\Xi_{c}^{(\prime)}}}{\left(\chi_{\frac{1}{2}m},\frac{\bm{\sigma}\cdot\bm{p}}{2m_{\Xi_{c}^{(\prime)}}}\chi_{\frac{1}{2}m}\right)^T},\nonumber\\
\langle 0|\Xi_{c}^{*\mu}|cqs\left({3}/{2}^+\right)\rangle &=&\sqrt{2m_{\Xi_{c}^*}}\left(\Phi_{\frac{3}{2}m}^{\mu},\frac{\bm{\sigma}\cdot\bm{p}}{2m_{\Xi_{c}^*}}\Phi_{\frac{3}{2}m}^{\mu}\right)^T,\nonumber\\
\langle 0|\bar D_{1}^{\mu}|\bar{c}q(1^+)\rangle&=&\sqrt{m_{\bar D_{1}}}\epsilon^\mu,\nonumber\\
\langle 0|\bar D_{2}^{*\mu\nu}|\bar{c}q(2^+)\rangle&=&\sqrt{m_{\bar D_{2}^*}}\zeta^{\mu\nu},
\end{eqnarray}
respectively. Here, $m_{i}$ ($i=\Xi_{c},\,\Xi_{c}^{\prime},\,\Xi_{c}^{*},\,\bar D_{1},\,\bar D_{2}^{*}$) is the mass of the heavy hadron $i$, while $\bm{\sigma}$ and $\bm{p}$ are the Pauli matrix and the momentum of the charmed baryon, respectively. $\epsilon^\mu$ and $\zeta^{\mu\nu}$ are the polarization vector and the polarization tensor, respectively. In the static limit, the polarization vector $\epsilon_{m}^{\mu}\,(m=0,\,\pm1)$ can be explicitly written as $\epsilon_{-1}^{\mu}= \left(0,\,-1,\,i,\,0\right)/\sqrt{2}$, $\epsilon_{0}^{\mu}= \left(0,0,0,-1\right)$, and $\epsilon_{+1}^{\mu}= \left(0,\,1,\,i,\,0\right)/\sqrt{2}$, while the polarization tensor $\zeta^{\mu\nu}_{m}$ can be constructed by the coupling of both polarization vectors $\epsilon^{\mu}_{m_1}$ and $\epsilon^{\nu}_{m_2}$ \cite{Cheng:2010yd}, which can be represented as
\begin{eqnarray}
\zeta^{\mu\nu}_{m}=\sum_{m_1,m_2}C^{2, m}_{1m_1,1m_2}\epsilon^{\mu}_{m_1}\epsilon^{\nu}_{m_2},
\end{eqnarray}
where the Clebsch-Gordan coefficient $C^{2, m}_{1m_1,1m_2}$ is used to describe the related coupling. In addition, the spin wave function of the charmed baryon $\Xi_{c}^{(\prime)}$ is defined as $\chi_{\frac{1}{2}m}$, and the polarization tensor $\Phi_{\frac{3}{2}m}^{\mu}$ of the charmed baryon $\Xi_{c}^{*}$ can be constructed by the coupling of the spin wave function $\chi_{\frac{1}{2}m_1}$ and the polarization vector $\epsilon_{m_2}^{\mu}$, which can be given by
\begin{eqnarray}
\Phi_{\frac{3}{2}m}^{\mu}=\sum_{m_1,m_2}C^{\frac{3}{2},m}_{\frac{1}{2}m_1,1m_2}\chi_{\frac{1}{2}m_1}\epsilon_{m_2}^{\mu}.
\end{eqnarray}

Up to now, we have obtained the scattering amplitude $\mathcal{M}^{h_1h_2\to h_3h_4}(\bm{q})$ of the $h_1h_2\to h_3h_4$ process. Taking into account both the Breit approximation and the nonrelativistic normalization \cite{Breitapproximation}, the effective potential in the momentum space $\mathcal{V}_E^{h_1h_2\to h_3h_4}(\bm{q})$ can be extracted based on the obtained scattering amplitude $\mathcal{M}^{h_1h_2\to h_3h_4}(\bm{q})$. To be more specific, the relation between the effective potential in the momentum space $\mathcal{V}^{h_1h_2\to h_3h_4}(\bm{q})$ and the scattering amplitude $\mathcal{M}^{h_1h_2\to h_3h_4}(\bm{q})$ can be written in a general form of \cite{Breitapproximation}
\begin{eqnarray}
\mathcal{V}_E^{h_1h_2\to h_3h_4}(\bm{q}) &=&-\frac{\mathcal{M}^{h_1h_2\to h_3h_4}(\bm{q})} {4\sqrt{m_{h_1}m_{h_2}m_{h_3}m_{h_4}}},
\end{eqnarray}
where $m_{h_i}\,(i=1, \,2,\,3, \,4)$ is the mass of the hadron $h_i$. The obtained effective potential in the momentum space $\mathcal{V}_E^{h_1h_2\to h_3h_4}(\bm{q})$ is the function of the momentum of the exchanged light mesons $\bm{q}$, but we discuss the bound state properties of the $\Xi_c^{(\prime,*)}\bar D_1/\Xi_c^{(\prime,*)}\bar D_2^*$ systems by solving the Schr$\ddot{\rm o}$dinger equation in the coordinate space in the present work. Thus, we need to obtain the effective potentials in the coordinate space $\mathcal{V}_E^{h_1h_2\to h_3h_4}(\bm{r})$ for these discussed systems. After taking the Fourier transformation for the effective potential in the momentum space $\mathcal{V}_E^{h_1h_2\to h_3h_4}(\bm{q})$ together with the form factor, we can deduce the effective potential in the coordinate space $\mathcal{V}_E^{h_1h_2\to h_3h_4}(\bm{r})$ by the following relation
\begin{eqnarray}
\mathcal{V}_E^{h_1h_2\to h_3h_4}(\bm{r}) =\int\frac{d^3\bm{q}}{(2\pi)^3}e^{i\bm{q}\cdot\bm{r}}\mathcal{V}_E^{h_1h_2\to h_3h_4}(\bm{q})\mathcal{F}^2(q^2,m_{\mathcal{E}}^2).\nonumber\\
\end{eqnarray}
Given that the conventional baryons and mesons are not point particles, the form factor $\mathcal{F}(q^2,m_{\mathcal{E}}^2)$ was introduced in each interaction vertex for the Feynman diagram, which can be taken to compensate the roles of the inner structure of the discussed hadrons and the off shell of the exchanged light mesons. Generally speaking, there exist many different kinds of form factors \cite{Chen:2016qju}, and we choose the monopole-type form factor in the present work, i.e.,
\begin{eqnarray}
\mathcal{F}(q^2,m_{\mathcal{E}}^2) = \frac{\Lambda^2-m_{\mathcal{E}}^2}{\Lambda^2-q^2},
\end{eqnarray}
which is similar to the case for studying the bound state properties of the deuteron \cite{Tornqvist:1993ng,Tornqvist:1993vu}. In the above monopole-type form factor, $\Lambda$ is the cutoff parameter, and we define the mass and the four momentum of the exchanged light meson as $m_{\mathcal{E}}$ and $q$, respectively.

In the following, we further discuss the related wave functions for the $\Xi_c^{(\prime,*)}\bar D_1/\Xi_c^{(\prime,*)}\bar D_2^*$ systems, which contain the color, the spin-orbital, the flavor, and the spatial wave functions. For the hadronic molecular states composed of two color-singlet hadrons, the color wave function is simply taken as unity. The spin-orbital and flavor wave functions can be constructed by taking into account the coupling of the constituent hadrons, which can be used to calculate the operator matrix elements and the isospin factors for the OBE effective potentials, respectively. In addition, the spatial wave function can be obtained by solving the Schr$\ddot{\rm o}$dinger equation, which can be regarded as the important inputs to study their properties in future, such as the strong decay properties, the electromagnetic properties, and so on. The spin-orbital wave functions $|{}^{2S+1}L_{J}\rangle$ of the $\Xi_c^{(\prime,*)}\bar D_1/\Xi_c^{(\prime,*)}\bar D_2^*$ systems can be constructed as
\begin{eqnarray}
\Xi_{c}^{(\prime)}\bar{D}_{1}:|{}^{2S+1}L_{J}\rangle&=&\sum_{m,m',m_Sm_L}C^{S,m_S}_{\frac{1}{2}m,1m'}
C^{J,M}_{Sm_S,Lm_L}\chi_{\frac{1}{2}m}\epsilon^{m'}|Y_{L,m_L}\rangle,\nonumber\\
\Xi_{c}^{(\prime)}\bar{D}_{2}^{*}:|{}^{2S+1}L_{J}\rangle&=&\sum_{m,m'',m_Sm_L}
C^{S,m_S}_{\frac{1}{2}m,2m'}C^{J,M}_{Sm_S,Lm_L}\chi_{\frac{1}{2}m}\zeta^{m'}
|Y_{L,m_L}\rangle,\nonumber\\
\Xi_{c}^{*}\bar{D}_{1}:|{}^{2S+1}L_{J}\rangle&=&\sum_{m,m',m_Sm_L}
C^{S,m_S}_{\frac{3}{2}m,1m'}C^{J,M}_{Sm_S,Lm_L}\Phi_{\frac{3}{2}m}\epsilon^{m'}
|Y_{L,m_L}\rangle,\nonumber\\
\Xi_{c}^{*}\bar{D}_{2}^{*}:|{}^{2S+1}L_{J}\rangle&=&\sum_{m,m',m_Sm_L}
C^{S,m_S}_{\frac{3}{2}m,2m'}C^{J,M}_{Sm_S,Lm_L}\Phi_{\frac{3}{2}m}\zeta^{m'}
|Y_{L,m_L}\rangle,\nonumber\\\label{spinorbitalwavefunctions}
\end{eqnarray}
where $|Y_{L,m_L}\rangle$ is the spherical harmonics function. Since the isospin quantum numbers for the charmed baryons $\Xi_c^{(\prime,*)}$ and the charmed mesons $D_1/D_2^*$ are $1/2$, the $\Xi_c^{(\prime,*)}\bar D_1/\Xi_c^{(\prime,*)}\bar D_2^*$ systems have the isospin quantum numbers either 0 or 1, where we summarize the flavor wave functions of the isoscalar and isovector $\Xi_c^{(\prime,*)}\bar D_1/\Xi_c^{(\prime,*)}\bar D_2^*$ systems in Table \ref{flavorwavefunctions}. Finally, we can derive the OBE effective potentials in the coordinate space for the $\Xi_c^{(\prime,*)}\bar D_1/\Xi_c^{(\prime,*)}\bar D_2^*$ systems by the standard strategy listed above, which is collected in Appendix \ref{app01}.
\renewcommand\tabcolsep{0.46cm}
\renewcommand{\arraystretch}{1.50}
\begin{table}[!htbp]
\caption{The flavor wave functions of the isoscalar and isovector  $\Xi_c^{(\prime,*)}\bar D_1/\Xi_c^{(\prime,*)}\bar D_2^*$ systems. Here, the notation $\mathcal{D}$ stands for either $D_1$ or $D_2^*$ meson, while $I$ and $I_3$ are used to denote the isospin and its third component of the discussed system, respectively. }
\label{flavorwavefunctions}
\begin{tabular}{l|l|l}
\toprule[1.0pt]
\toprule[1.0pt]
Isospins&$\left|I,I_3\right\rangle$ & Flavor wave functions \\
\hline
Isoscalar&$\left|0,0\right\rangle$ & $\frac{1}{\sqrt{2}}\left|\Xi_c^{(\prime,*)+}\mathcal{D}^{-}\right\rangle-\frac{1}{\sqrt{2}}\left|\Xi_c^{(\prime,*)0}\bar{\mathcal{D}}^{0}\right\rangle$ \\\hline
\multirow{3}{*}{Isovector}&$\left|1,1\right\rangle$ & $\left|\Xi_c^{(\prime,*)+}\bar{\mathcal{D}}^{0}\right\rangle$ \\
&$\left|1,0\right\rangle$ & $\frac{1}{\sqrt{2}}\left|\Xi_c^{(\prime,*)+}\mathcal{D}^{-}\right\rangle+\frac{1}{\sqrt{2}}\left|\Xi_c^{(\prime,*)0}\bar{\mathcal{D}}^{0}\right\rangle$ \\
&$\left|1,-1\right\rangle$ & $\left|\Xi_c^{(\prime,*)0}\mathcal{D}^{-}\right\rangle$ \\
\bottomrule[1.0pt]
\bottomrule[1.0pt]
\end{tabular}
\end{table}

In the past decades, the OBE model has made a lot of progress when studying the interactions between hadrons \cite{Chen:2016qju,Liu:2019zoy}, and the previous theoretical works have introduced a series of important effects to discuss the fine structures of the hadron-hadron interactions, such as the $S$-$D$ wave mixing effect, the coupled channel effect, and so on. Specifically, the contributions of the $S$-$D$ wave mixing effect and the coupled channel effect may result in the interesting and important phenomena, such as the influence of the coupled channel effect can reproduce the double peak structures of the $P_{\psi s}^{\Lambda}(4459)$ \cite{LHCb:2020jpq} existing in the $J/\psi \Lambda$ invariant mass spectrum \cite{Wang:2022mxy}, which inspires our interest to consider the $S$-$D$ wave mixing effect and the coupled channel effect when discussing the $\Xi_c^{(\prime,*)}\bar D_1/\Xi_c^{(\prime,*)}\bar D_2^*$ interactions. After considering the roles of the $S$-$D$ wave mixing effect and the coupled channel effect, the mass spectrum of the $\Xi_c^{(\prime,*)}\bar D_1/\Xi_c^{(\prime,*)}\bar D_2^*$-type hidden-charm molecular pentaquark candidates with strangeness may become more abundant. When studying the $\Xi_c^{(\prime,*)}\bar D_1/\Xi_c^{(\prime,*)}\bar D_2^*$ interactions, the $S$-wave and $D$-wave channels $|{}^{2S+1}L_{J}\rangle$ are given by
\begin{eqnarray}
\begin{array}{l}
\Xi_{c}^{(\prime)}\bar{D}_1\left\{
  \begin{array}{l}
    J^P=\frac{1}{2}^+:\left|{}^2\mathbb{S}_{\frac{1}{2}}\right\rangle,\,\left|{}^4\mathbb{D}_{\frac{1}{2}}\right\rangle\nonumber\\
    J^P=\frac{3}{2}^+:\left|{}^4\mathbb{S}_{\frac{3}{2}}\right\rangle,\,\left|{}^2\mathbb{D}_{\frac{3}{2}}\right\rangle,\,\left|{}^4\mathbb{D}_{\frac{3}{2}}\right\rangle
  \end{array}
\right.,\\
\Xi_{c}^{(\prime)}\bar{D}_2^*\left\{
  \begin{array}{l}
    J^P=\frac{3}{2}^+:\left|{}^4\mathbb{S}_{\frac{3}{2}}\right\rangle,\,\left|{}^4\mathbb{D}_{\frac{3}{2}}\right\rangle,\,\left|{}^6\mathbb{D}_{\frac{3}{2}}\right\rangle\nonumber\\
    J^P=\frac{5}{2}^+:\left|{}^6\mathbb{S}_{\frac{5}{2}}\right\rangle,\,\left|{}^4\mathbb{D}_{\frac{5}{2}}\right\rangle,\,\left|{}^6\mathbb{D}_{\frac{5}{2}}\right\rangle
  \end{array}
\right.,\\
\Xi_{c}^{*}\bar{D}_1~\left\{
  \begin{array}{l}
    J^P=\frac{1}{2}^+:\left|{}^2\mathbb{S}_{\frac{1}{2}}\right\rangle,\,\left|{}^4\mathbb{D}_{\frac{1}{2}}\right\rangle,\,\left|{}^6\mathbb{D}_{\frac{1}{2}}\right\rangle\nonumber\\
    J^P=\frac{3}{2}^+:\left|{}^4\mathbb{S}_{\frac{3}{2}}\right\rangle,\,\left|{}^2\mathbb{D}_{\frac{3}{2}}\right\rangle,\,\left|{}^4\mathbb{D}_{\frac{3}{2}}\right\rangle,\,\left|{}^6\mathbb{D}_{\frac{3}{2}}\right\rangle\nonumber\\
    J^P=\frac{5}{2}^+:\left|{}^6\mathbb{S}_{\frac{5}{2}}\right\rangle,\,\left|{}^2\mathbb{D}_{\frac{5}{2}}\right\rangle,\,\left|{}^4\mathbb{D}_{\frac{5}{2}}\right\rangle,\,\left|{}^6\mathbb{D}_{\frac{5}{2}}\right\rangle
  \end{array}
\right.,\\
\Xi_{c}^{*}\bar{D}_2^*~\left\{
  \begin{array}{l}
    J^P=\frac{1}{2}^+:\left|{}^2\mathbb{S}_{\frac{1}{2}}\right\rangle,\,\left|{}^4\mathbb{D}_{\frac{1}{2}}\right\rangle,\,\left|{}^6\mathbb{D}_{\frac{1}{2}}\right\rangle\nonumber\\
    J^P=\frac{3}{2}^+:\left|{}^4\mathbb{S}_{\frac{3}{2}}\right\rangle,\,\left|{}^2\mathbb{D}_{\frac{3}{2}}\right\rangle,\,\left|{}^4\mathbb{D}_{\frac{3}{2}}\right\rangle,\,\left|{}^6\mathbb{D}_{\frac{3}{2}}\right\rangle,\,\left|{}^8\mathbb{D}_{\frac{3}{2}}\right\rangle\nonumber\\
    J^P=\frac{5}{2}^+:\left|{}^6\mathbb{S}_{\frac{5}{2}}\right\rangle,\,\left|{}^2\mathbb{D}_{\frac{5}{2}}\right\rangle,\,\left|{}^4\mathbb{D}_{\frac{5}{2}}\right\rangle,\,\left|{}^6\mathbb{D}_{\frac{5}{2}}\right\rangle,\,\left|{}^8\mathbb{D}_{\frac{5}{2}}\right\rangle\nonumber\\
    J^P=\frac{7}{2}^+:\left|{}^8\mathbb{S}_{\frac{7}{2}}\right\rangle,\,\left|{}^2\mathbb{D}_{\frac{7}{2}}\right\rangle,\,\left|{}^4\mathbb{D}_{\frac{7}{2}}\right\rangle,\,\left|{}^6\mathbb{D}_{\frac{7}{2}}\right\rangle,\,\left|{}^8\mathbb{D}_{\frac{7}{2}}\right\rangle
  \end{array}
\right.,
\end{array}
\end{eqnarray}
where the notation $|^{2S+1}L_J\rangle$ is applied to illustrate the information of the spin $S$, the orbital angular momentum $L$, and the total angular momentum $J$ for the corresponding channels, while $L=\mathbb{S}$ and $\mathbb{D}$ are introduced to distinguish the $S$-wave and $D$-wave interactions for the corresponding mixing channels in the present work.

\section{Mass spectrum of the predicted hidden-charm molecular pentaquarks with strangeness}\label{sec3}

By employing the obtained OBE effective potentials in the coordinate space for the $\Xi_c^{(\prime,*)}\bar D_1/\Xi_c^{(\prime,*)}\bar D_2^*$ systems, we can further discuss their bound state properties, by which a novel class of molecular $P_{\psi s}^{\Lambda/\Sigma}$ pentaquark candidates composed of the charmed baryons $\Xi_c^{(\prime,*)}$ and the anticharmed mesons $\bar D_1/\bar D_2^*$ can be predicted. As is well known, the Schr$\ddot{\rm o}$dinger equation is a powerful tool to discuss the two-body bound state problems\footnote{We should illustrate the limitation inherent in the approach we have adopted. For these observed $P_\psi^N$ states \cite{Aaij:2015tga,Aaij:2019vzc}, which can decay into $J/\psi p$, they also embody to some extent the nature of the  resonance states, which potentially manifests as the hadronic molecular states. Our approach of solving the Schr\"odinger equation can only reflect the bound state property and cannot describe the resonance behavior. There are plausible ways to elucidate the mechanism governing the production of these $P_\psi^N$ states \cite{Aaij:2015tga,Aaij:2019vzc}, involving the application of the Bethe-Salpeter or Lippmann-Schwinger equation within the framework of the coupled-channel formalism, as discussed in Refs. \cite{Uchino:2015uha,He:2015cea,He:2016pfa,Xiao:2019aya,He:2019ify,He:2019rva,Xiao:2020frg,Zhu:2021lhd,Zhu:2022wpi,Lin:2023iww,Feijoo:2022rxf}. } \cite{Chen:2016qju}. After solving the coupled channel Schr$\ddot{\rm o}$dinger equation, the bound state solutions including the binding energy $E$ and the spatial wave functions of the individual channel $\psi_i(r)$ can be obtained for the $\Xi_c^{(\prime,*)}\bar D_1/\Xi_c^{(\prime,*)}\bar D_2^*$ systems. Based on the obtained spatial wave functions of the individual channel $\psi_i(r)$, we can further estimate the root-mean-square radius $r_{\rm RMS}$ and the probabilities of the individual channel $P_i$ by the following relations
\begin{eqnarray}
r_{\rm RMS}&=&\sqrt {\int \sum_i \psi_i^{\dag}(r) \psi_i(r) r^4 dr},\\
P_i&=&\int \psi_i^{\dag}(r) \psi_i(r) r^2 dr,
\end{eqnarray}
where the spatial wave functions of the discussed system $\psi_i(r)$ satisfy the normalization condition, i.e., $\int \sum_i \psi_i^{\dag}(r) \psi_i(r) r^2 dr=1$. In short, the bound state solutions containing the binding energy $E$, the root-mean-square radius $r_{\rm RMS}$, and the probabilities of the individual channel $P_i$ can offer the important information to discuss the possibilities of the $\Xi_c^{(\prime,*)}\bar D_1/\Xi_c^{(\prime,*)}\bar D_2^*$ systems as the molecular $P_{\psi s}^{\Lambda/\Sigma}$ pentaquark candidates.

When solving the Schr$\ddot{\rm o}$dinger equation, the repulsive centrifugal potential $\ell(\ell+1)/2\mu r^2$ arises for the higher partial wave states $\ell \geqslant 1$, which shows that the $S$-wave state is more easily to form the hadronic molecular state compared with the higher partial wave states for the certain hadronic system \cite{Chen:2016qju,Guo:2017jvc}. Consequently, the $S$-wave $\Xi_c^{(\prime,*)}\bar D_1/\Xi_c^{(\prime,*)}\bar D_2^*$ systems will be the main research objects of the present work, which is also inspired by the explanations of the observed $P_{\psi}^{N}$ \cite{Aaij:2019vzc}, $P_{\psi s}^{\Lambda}$ \cite{LHCb:2020jpq,LHCb:2022jad}, and $T_{cc}$ \cite{LHCb:2021vvq} as the $S$-wave hadronic molecular states \cite{Li:2014gra,Karliner:2015ina,Wu:2010jy,Wang:2011rga,Yang:2011wz,Wu:2012md,Chen:2015loa,Paryev:2023icm,Azizi:2023foj,Wang:2022neq,Wang:2022mxy,Karliner:2022erb,Yan:2022wuz,Meng:2022wgl,Ozdem:2023htj,Feijoo:2022rxf,Garcilazo:2022edi,Yang:2022ezl,Zhu:2022wpi,Chen:2022wkh,Ortega:2022uyu,Giachino:2022pws,Nakamura:2022jpd,Wang:2022tib,Ozdem:2022kei,Xiao:2022csb,Wang:2022gfb,Clymton:2022qlr,Chen:2022onm,Chen:2021spf,Gao:2021hmv,Ferretti:2021zis,Giron:2021fnl,Cheng:2021gca,Du:2021bgb,Chen:2021cfl,Hu:2021nvs,Yang:2021pio,Li:2021ryu,Lu:2021irg,Zou:2021sha,Wang:2021itn,Wu:2021caw,Clymton:2021thh,Xiao:2021rgp,Ozdem:2021ugy,Zhu:2021lhd,Chen:2021tip,Azizi:2021utt,Dong:2021juy,Liu:2020hcv,Wang:2020eep,Peng:2020hql,Chen:2020uif,Peng:2019wys,Chen:2020opr,Chen:2020kco,Burns:2022uha,Manohar:1992nd,Ericson:1993wy,Tornqvist:1993ng,Janc:2004qn,Ding:2009vj,Molina:2010tx,Ding:2020dio,Li:2012ss,Xu:2017tsr,Liu:2019stu,Ohkoda:2012hv,Tang:2019nwv,Li:2021zbw,Chen:2021vhg,Ren:2021dsi,Xin:2021wcr,Chen:2021tnn,Albaladejo:2021vln,Dong:2021bvy,Baru:2021ldu,Du:2021zzh,Kamiya:2022thy,Padmanath:2022cvl,Agaev:2022ast,Ke:2021rxd,Zhao:2021cvg,Deng:2021gnb,Santowsky:2021bhy,Dai:2021vgf,Feijoo:2021ppq,Wang:2023ovj,Peng:2023lfw,Dai:2023cyo,Du:2023hlu,Kinugawa:2023fbf,Lyu:2023xro,Li:2023hpk,Dai:2023mxm,Wang:2022jop,Wu:2022gie,Ortega:2022efc,Jia:2022qwr,Praszalowicz:2022sqx,Chen:2022vpo,Lin:2022wmj,Cheng:2022qcm,Mikhasenko:2022rrl,He:2022rta}. Furthermore, the hadronic molecular state is the loosely bound state composed of the color-singlet hadrons \cite{Chen:2016qju}. Thus, the most promising hadronic molecular candidate should exist the key features of the small binding energy and the large size \cite{Chen:2016qju}, which is the important lesson learned from the deuteron studies and can guide us to discuss the $\Xi_c^{(\prime,*)}\bar D_1/\Xi_c^{(\prime,*)}\bar D_2^*$-type hidden-charm molecular pentaquark candidates with strangeness. Of course, the observed $P_{\psi}^{N}$ \cite{Aaij:2019vzc}, $P_{\psi s}^{\Lambda}$ \cite{LHCb:2020jpq,LHCb:2022jad}, and $T_{cc}$ \cite{LHCb:2021vvq} have the characteristics of the small binding energy and the large size under the hadronic molecule picture \cite{Li:2014gra,Karliner:2015ina,Wu:2010jy,Wang:2011rga,Yang:2011wz,Wu:2012md,Chen:2015loa,Paryev:2023icm,Azizi:2023foj,Wang:2022neq,Wang:2022mxy,Karliner:2022erb,Yan:2022wuz,Meng:2022wgl,Ozdem:2023htj,Feijoo:2022rxf,Garcilazo:2022edi,Yang:2022ezl,Zhu:2022wpi,Chen:2022wkh,Ortega:2022uyu,Giachino:2022pws,Nakamura:2022jpd,Wang:2022tib,Ozdem:2022kei,Xiao:2022csb,Wang:2022gfb,Clymton:2022qlr,Chen:2022onm,Chen:2021spf,Gao:2021hmv,Ferretti:2021zis,Giron:2021fnl,Cheng:2021gca,Du:2021bgb,Chen:2021cfl,Hu:2021nvs,Yang:2021pio,Li:2021ryu,Lu:2021irg,Zou:2021sha,Wang:2021itn,Wu:2021caw,Clymton:2021thh,Xiao:2021rgp,Ozdem:2021ugy,Zhu:2021lhd,Chen:2021tip,Azizi:2021utt,Dong:2021juy,Liu:2020hcv,Wang:2020eep,Peng:2020hql,Chen:2020uif,Peng:2019wys,Chen:2020opr,Chen:2020kco,Burns:2022uha,Manohar:1992nd,Ericson:1993wy,Tornqvist:1993ng,Janc:2004qn,Ding:2009vj,Molina:2010tx,Ding:2020dio,Li:2012ss,Xu:2017tsr,Liu:2019stu,Ohkoda:2012hv,Tang:2019nwv,Li:2021zbw,Chen:2021vhg,Ren:2021dsi,Xin:2021wcr,Chen:2021tnn,Albaladejo:2021vln,Dong:2021bvy,Baru:2021ldu,Du:2021zzh,Kamiya:2022thy,Padmanath:2022cvl,Agaev:2022ast,Ke:2021rxd,Zhao:2021cvg,Deng:2021gnb,Santowsky:2021bhy,Dai:2021vgf,Feijoo:2021ppq,Wang:2023ovj,Peng:2023lfw,Dai:2023cyo,Du:2023hlu,Kinugawa:2023fbf,Lyu:2023xro,Li:2023hpk,Dai:2023mxm,Wang:2022jop,Wu:2022gie,Ortega:2022efc,Jia:2022qwr,Praszalowicz:2022sqx,Chen:2022vpo,Lin:2022wmj,Cheng:2022qcm,Mikhasenko:2022rrl,He:2022rta}. With the above considerations, we expect that the reasonable binding energy should be at most tens of MeV, and the two constituent hadrons should not overlap too much in the spatial distributions when discussing the $\Xi_c^{(\prime,*)}\bar D_1/\Xi_c^{(\prime,*)}\bar D_2^*$-type hidden-charm molecular pentaquark candidates with strangeness \cite{Chen:2016qju,Chen:2017xat}.

Besides the coupling constants listed in the above section, the masses of the involved hadrons are the important inputs to obtain the numerical results, and these conventional mesons and baryons have been observed experimentally. In the following numerical analysis, we take the masses of the relevant mesons and baryons from the Particle Data Group \cite{ParticleDataGroup:2022pth} and adopt the averaged masses for the multiple hadrons, i.e., $m_\sigma=600.00$ MeV, $m_\pi=137.27$ MeV, $m_\eta=547.86$ MeV, $m_\rho=775.26$ MeV, $m_\omega=782.66$ MeV, $m_{\Xi_{c}}=2469.08$ MeV, $m_{\Xi_{c}^{\prime}}=2578.45$ MeV, $m_{\Xi_{c}^{*}}=2645.10$ MeV, $m_{D_1}=2422.10$ MeV, and $m_{D_2^*}=2461.10$ MeV.

In our numerical analysis, the cutoff $\Lambda$ arises from the form factor is the crucial parameter to study the $\Xi_c^{(\prime,*)}\bar D_1/\Xi_c^{(\prime,*)}\bar D_2^*$-type hidden-charm molecular pentaquark candidates with strangeness. At present, the cutoff value $\Lambda$ cannot be exactly extracted for the $\Xi_c^{(\prime,*)}\bar D_1/\Xi_c^{(\prime,*)}\bar D_2^*$ systems, which is due to the absence of the relevant experimental data. Fortunately, the experience of studying the bound state properties of the deuteron within the OBE model can offer the valuable hints, and the cutoff parameter in the monopole-type form factor about 1.0 GeV is the reasonable input to discuss the hadronic molecular states \cite{Machleidt:1987hj,Epelbaum:2008ga,Esposito:2014rxa,Chen:2016qju,Tornqvist:1993ng,Tornqvist:1993vu,Wang:2019nwt,Chen:2017jjn}. Furthermore, the masses of the observed $P_{\psi}^{N}$ \cite{Aaij:2019vzc}, $P_{\psi s}^{\Lambda}$ \cite{LHCb:2020jpq,LHCb:2022jad}, and $T_{cc}$ \cite{LHCb:2021vvq} can be reproduced within the hadronic molecule picture \cite{Li:2014gra,Karliner:2015ina,Wu:2010jy,Wang:2011rga,Yang:2011wz,Wu:2012md,Chen:2015loa,Paryev:2023icm,Azizi:2023foj,Wang:2022neq,Wang:2022mxy,Karliner:2022erb,Yan:2022wuz,Meng:2022wgl,Ozdem:2023htj,Feijoo:2022rxf,Garcilazo:2022edi,Yang:2022ezl,Zhu:2022wpi,Chen:2022wkh,Ortega:2022uyu,Giachino:2022pws,Nakamura:2022jpd,Wang:2022tib,Ozdem:2022kei,Xiao:2022csb,Wang:2022gfb,Clymton:2022qlr,Chen:2022onm,Chen:2021spf,Gao:2021hmv,Ferretti:2021zis,Giron:2021fnl,Cheng:2021gca,Du:2021bgb,Chen:2021cfl,Hu:2021nvs,Yang:2021pio,Li:2021ryu,Lu:2021irg,Zou:2021sha,Wang:2021itn,Wu:2021caw,Clymton:2021thh,Xiao:2021rgp,Ozdem:2021ugy,Zhu:2021lhd,Chen:2021tip,Azizi:2021utt,Dong:2021juy,Liu:2020hcv,Wang:2020eep,Peng:2020hql,Chen:2020uif,Peng:2019wys,Chen:2020opr,Chen:2020kco,Burns:2022uha,Manohar:1992nd,Ericson:1993wy,Tornqvist:1993ng,Janc:2004qn,Ding:2009vj,Molina:2010tx,Ding:2020dio,Li:2012ss,Xu:2017tsr,Liu:2019stu,Ohkoda:2012hv,Tang:2019nwv,Li:2021zbw,Chen:2021vhg,Ren:2021dsi,Xin:2021wcr,Chen:2021tnn,Albaladejo:2021vln,Dong:2021bvy,Baru:2021ldu,Du:2021zzh,Kamiya:2022thy,Padmanath:2022cvl,Agaev:2022ast,Ke:2021rxd,Zhao:2021cvg,Deng:2021gnb,Santowsky:2021bhy,Dai:2021vgf,Feijoo:2021ppq,Wang:2023ovj,Peng:2023lfw,Dai:2023cyo,Du:2023hlu,Kinugawa:2023fbf,Lyu:2023xro,Li:2023hpk,Dai:2023mxm,Wang:2022jop,Wu:2022gie,Ortega:2022efc,Jia:2022qwr,Praszalowicz:2022sqx,Chen:2022vpo,Lin:2022wmj,Cheng:2022qcm,Mikhasenko:2022rrl,He:2022rta} when the cutoff values in the monopole-type form factor are around 1.0 GeV. Thus, we try to search for the loosely bound state solutions of the $\Xi_c^{(\prime,*)}\bar D_1/\Xi_c^{(\prime,*)}\bar D_2^*$ systems by scanning the cutoff parameters $\Lambda$ from 0.8 to 2.5 GeV, and select three typical values to present their bound state properties. Generally speaking, a loosely bound state can be recommended as the most promising hadronic molecular candidate with the cutoff parameter closed to 1.0 GeV.

In the following, we discuss the bound state properties for the $\Xi_c^{(\prime,*)}\bar D_1/\Xi_c^{(\prime,*)}\bar D_2^*$ systems, and predict a novel class of molecular $P_{\psi s}^{\Lambda/\Sigma}$ pentaquark candidates comprising the charmed baryons $\Xi_c^{(\prime,*)}$ and the anticharmed mesons $\bar D_1/\bar D_2^*$. To ensure comprehensive and systematic results, both the $S$-$D$ wave mixing effect and the coupled channel effect are explicitly taken into account in our calculations.

\subsection{$\Xi_c\bar D_1$ system}

\begin{figure}[htbp]
\includegraphics[scale=0.36]{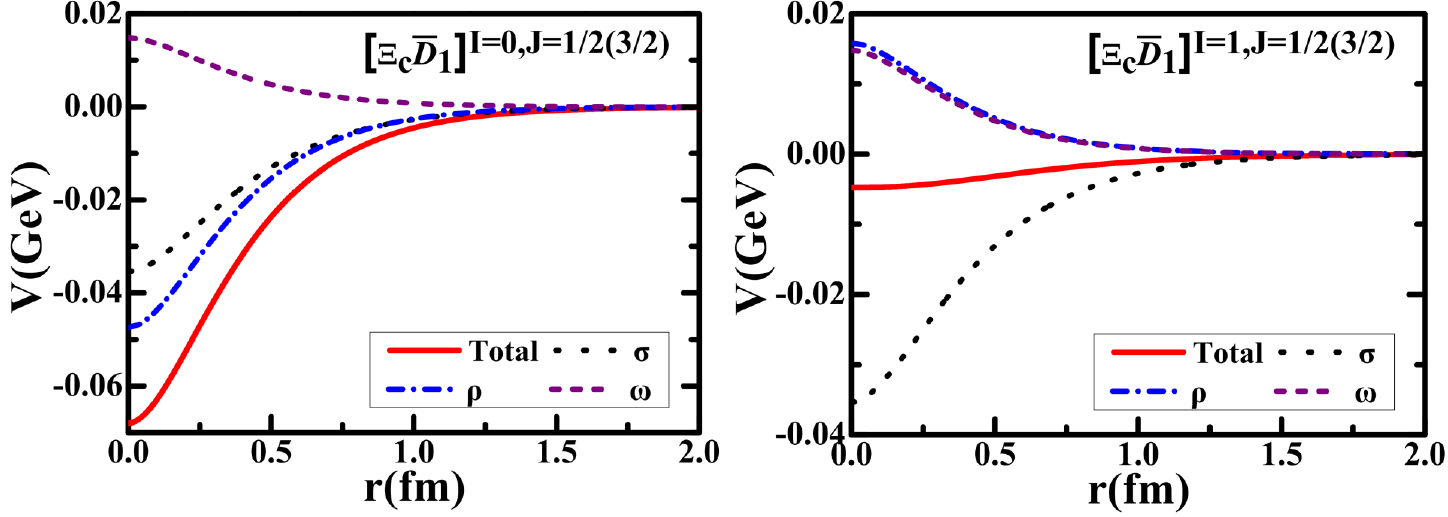}
\caption{The OBE effective potentials for the $\Xi_c\bar D_1$ states with $I(J^P)=0,1(1/2^+,\,3/2^+)$, where the cutoff $\Lambda$ is fixed as the typical value $1.0$ GeV.}\label{potentialshape1}
\end{figure}

The interaction of the $\Xi_c\bar D_1$ system is quite simple, and there only exists the $\sigma$, $\rho$, and $\omega$ exchange interactions due to the symmetry constraints \cite{Wise:1992hn}.  In Fig. \ref{potentialshape1}, we present the OBE effective potentials for the $\Xi_c\bar D_1$ states with $I(J^P)=0,1(1/2^+,\,3/2^+)$, where the cutoff $\Lambda$ is fixed as the typical value $1.0$ GeV. For the isoscalar $\Xi_c\bar D_1$ states with $J^P=1/2^+$ and $3/2^+$, the $\omega$ exchange is the repulsive  potential, while both the $\sigma$ and $\rho$ exchanges provide the attractive potential, which lead to the strong attractive interaction. For the isovector $\Xi_c\bar D_1$ states with $J^P=1/2^+$ and $3/2^+$, the attractive part of the effective potential comes from the $\sigma$ exchange, while the $\rho$ and $\omega$ exchanges give the repulsion potential, which make the total effective potential is weakly attractive. As given in Ref. \cite{Chen:2017vai}, the effective potential from the $\sigma$ exchange is attractive, and the $\omega$ exchange potential is repulsive by analyzing the quark configuration of the $\Xi_c\bar D_1$ system, which is consistent with our obtained results. Furthermore, the tensor force from the $S$-$D$ wave mixing effect is scarce for the $\Xi_c\bar D_1$ system. Thus, the single channel case and the $S$-$D$ wave mixing case give the same bound state solutions, and the probabilities for the $D$-wave channels are zero, which can be reflected in our obtained numerical results.

In the following, we study the bound state solutions for the $\Xi_c\bar D_1$ system by solving the Schr$\ddot{\rm o}$dinger equation. First, we discuss the bound state properties of the $\Xi_c\bar D_1$ system by considering the $S$-$D$ wave mixing analysis, and the relevant numerical results are presented in Table \ref{XicD1}. For the isovector $\Xi_c\bar D_1$ states with $J^P=1/2^+$ and $3/2^+$, we cannot find the bound state solutions by scanning the cutoff values $\Lambda=0.8\sim2.5~{\rm GeV}$ in the $S$-$D$ wave mixing case, since the OBE effective potentials are weakly attractive for both states as illustrated in Fig. \ref{potentialshape1}. For the isoscalar $\Xi_c\bar D_1$ states with $J^P=1/2^+$ and $3/2^+$, the bound state solutions can be found by choosing the cutoff values $\Lambda$ around 1.32 GeV, which is close to the reasonable range around 1.0 GeV. Thus, the isoscalar $\Xi_c\bar D_1$ states with $J^P=1/2^+$ and $3/2^+$ can be regarded as the most promising hidden-charm molecular pentaquark candidates with strangeness. Nevertheless, if we use the same cutoff value as input, there exists the same bound state properties for the isoscalar $\Xi_c\bar D_1$ states with $J^P=1/2^+$ and $3/2^+$ in the context of the $S$-$D$ wave mixing analysis, since the $\Xi_c\bar D_1$ system does not exist the spin-spin interaction to split into the isoscalar $\Xi_c\bar D_1$ bound states with $J^P=1/2^+$ and $3/2^+$. Thus, there exists the phenomenon of the mass degeneration for the isoscalar $\Xi_c\bar D_1$ bound states with $J^P=1/2^+$ and $3/2^+$ when adopting the same cutoff value in the $S$-$D$ wave mixing case, and such phenomenon is also found for the isoscalar $\Xi_c\bar D^*$ system \cite{Wang:2022mxy}.

\renewcommand\tabcolsep{0.07cm}
\renewcommand{\arraystretch}{1.50}
\begin{table}[!htbp]
\centering
\caption{The cutoff values dependence of the bound state solutions for the $\Xi_c\bar D_1$ system by considering the $S$-$D$ wave mixing case and the coupled channel case. Here, the dominant contribution channel is shown in bold font, while the units of the cutoff $\Lambda$, binding energy $E$, and root-mean-square radius $r_{{\rm RMS}}$ are $\rm{GeV}$, $\rm {MeV}$, and $\rm {fm}$, respectively.}\label{XicD1}
\begin{tabular}{c|cccc}\toprule[1.0pt]\toprule[1.0pt]
\multicolumn{5}{c}{$S$-$D$ wave mixing case}\\\midrule[1.0pt]
$I(J^P)$&$\Lambda$  &$E$ &$r_{\rm RMS}$ &P(${}^2\mathbb{S}_{\frac{1}{2}}/{}^4\mathbb{D}_{\frac{1}{2}}$)\\\hline
\multirow{3}{*}{$0(\frac{1}{2}^+)$}&1.32&$-0.27$ &4.87&\textbf{100.00}/$o(0)$     \\
&1.49&$-4.66$ &1.58&\textbf{100.00}/$o(0)$      \\
&1.65&$-12.46$ &1.05&\textbf{100.00}/$o(0)$     \\\midrule[1.0pt]
$I(J^P)$&$\Lambda$ &$E$ &$r_{\rm RMS}$ &P(${}^4\mathbb{S}_{\frac{3}{2}}/{}^2\mathbb{D}_{\frac{3}{2}}/{}^4\mathbb{D}_{\frac{3}{2}}$)\\\hline
\multirow{3}{*}{$0(\frac{3}{2}^+)$}&1.32&$-0.27$ &4.87&\textbf{100.00}/$o(0)$/$o(0)$\\
&1.49&$-4.66$ &1.58&\textbf{100.00}/$o(0)$/$o(0)$\\
&1.65&$-12.46$ &1.05&\textbf{100.00}/$o(0)$/$o(0)$\\\midrule[1.0pt]
\multicolumn{5}{c}{Coupled channel case}\\\midrule[1.0pt]
$I(J^P)$&$\Lambda$  &$E$ &$r_{\rm RMS}$ &P($\Xi_c\bar D_1/\Xi_c^{\prime}\bar D_1/\Xi_c^{*}\bar D_1/\Xi_c^{*}\bar D_2^*$)\\\hline
\multirow{3}{*}{$0(\frac{1}{2}^+)$}&1.04 &$-0.56$ & 3.76& \textbf{95.90}/3.84/0.07/0.18     \\
                           &1.07&$-4.57$&1.44& \textbf{85.32}/14.14/0.03/0.51      \\
                           &1.09&$-10.06$&0.98&\textbf{75.85}/23.41/0.06/0.68 \\\midrule[1.0pt]
\multirow{3}{*}{$1(\frac{1}{2}^+)$}&1.90& $-0.97$ &2.75&\textbf{93.47}/5.24/0.08/1.21 \\
                           &1.92&$-4.38$& 1.27& \textbf{88.00}/9.64/0.13/2.24\\
                           &1.94&$-9.62$&0.84& \textbf{84.19}/12.69/0.16/2.97\\\midrule[1.0pt]
$I(J^P)$&$\Lambda$ &$E$ &$r_{\rm RMS}$ &P($\Xi_c\bar D_1/\Xi_c\bar D_2^*/\Xi_c^{\prime}\bar D_1/\Xi_c^{\prime}\bar D_2^*/\Xi_c^{*}\bar D_1/\Xi_c^{*}\bar D_2^*$)\\\hline
\multirow{3}{*}{$0(\frac{3}{2}^+)$}& 1.09&$-0.32$ &4.59& \textbf{97.98}/0.13/0.43/0.39/0.58/0.49\\
&1.12&$-2.70$& 1.88& \textbf{91.21}/1.24/1.43/1.87/1.26/3.00\\
&1.15&$-10.55$&0.93&\textbf{69.84}/6.78/2.66/6.45/0.76/13.51\\\midrule[1.0pt]
\multirow{3}{*}{$1(\frac{3}{2}^+)$}& 1.71&$-0.68$&2.79&\textbf{69.91}/10.87/16.33/0.02/1.67/1.21\\
& 1.72&$-4.65$&0.96& \textbf{49.00}/18.10/27.89/0.027/2.95/2.03\\
& 1.73&$-9.97$&0.64&\textbf{41.06}/20.45/32.57/0.03/3.56/2.34\\
\bottomrule[1.0pt]\bottomrule[1.0pt]
\end{tabular}
\end{table}

For the $\Xi_c\bar D_1$ system, we can further take into account the contribution of the coupled channel effect, and the obtained numerical results are given in Table \ref{XicD1}. After including the role of the coupled channel effect, the isoscalar $\Xi_c\bar D_1$ states with $J^P=1/2^+$ and $3/2^+$ still exist the bound state solutions, while the isovector $\Xi_c\bar D_1$ states with $J^P=1/2^+$ and $3/2^+$ can form the bound states with the cutoff values restricted to be below 2.0 GeV. For the isoscalar $\Xi_c\bar D_1$ states with $J^P=1/2^+$ and $3/2^+$, the bound state solutions can be obtained by choosing the cutoff values $\Lambda$ around 1.04 GeV and 1.09 GeV, respectively, where the dominant component is the $\Xi_c\bar D_1$ channel. Furthermore, when taking the same cutoff value, the isoscalar $\Xi_c\bar D_1$ states with $J^P=1/2^+$ and $3/2^+$ have different bound state solutions after considering the influence of the coupled channel effect, which is similar to the case for the isoscalar $\Xi_c\bar D^*$ system \cite{Wang:2022mxy}. We hope that the future experiments can focus on the phenomenon of the mass difference for the isoscalar $\Xi_c\bar D_1$ states with $J^P=1/2^+$ and $3/2^+$, which can test the importance of the coupled channel effect for studying the hadron-hadron interactions and the double peak hypothesis of the $P_{\psi s}^{\Lambda}(4459)$ \cite{LHCb:2020jpq} existing in the $J/\psi \Lambda$ invariant mass spectrum. For the isovector $\Xi_c\bar D_1$ states with $J^P=1/2^+$ and $3/2^+$, there exist the bound state solutions when we tune the cutoff values $\Lambda$ to be around 1.90 GeV and 1.72 GeV, respectively, where both bound states have a main part of the $\Xi_c\bar D_1$ channel. Based on the analysis mentioned above, it is clear that the contribution of the coupled channel effect cannot be neglected when discussing the bound state properties of the $\Xi_c\bar D_1$ system.

By comparing the  obtained bound state solutions of the isoscalar $\Xi_c\bar D_1$ states with $J^P=1/2^+$ and $3/2^+$, there is no priority for the isovector $\Xi_c\bar D_1$ states with $J^P=1/2^+$ and $3/2^+$ as the hidden-charm molecular pentaquark candidates with strangeness. Thus, we strongly suggest that the experiments should first search for the isoscalar $\Xi_c\bar D_1$ molecular  states with $J^P=1/2^+$ and $3/2^+$ in future. Of course, the isovector $\Xi_c\bar D_1$ states with $J^P=1/2^+$ and $3/2^+$ as the possible candidates of the hidden-charm molecular pentaquarks with strangeness can be acceptable, since the obtained cutoff values are not especially away from the reasonable range around 1.0 GeV when appearing the isovector $\Xi_c\bar D_1$ bound states with $J^P=1/2^+$ and $3/2^+$.

Within the OBE model, the coupling constants serve as the crucial inputs to describe the interaction strengths. As a rule, we prefer to derive the coupling constants by reproducing the experimental widths with the available experimental data. In addition, we can only estimate several coupling constants by utilising various theoretical models if the pertinent experimental data are unavailable. At present, there is no experimental data available regarding the coupling constant $g_{\sigma}^{\prime\prime}$, that can be estimated using the phenomenological model in this study. However, there are various values for the coupling constant $g_{\sigma}^{\prime\prime}$ from different approaches, such as $-0.76$, $-3.40$, and $-5.21$, which are determined by the spontaneously broken chiral symmetry \cite{Wang:2019nwt}, the quark model \cite{Liu:2019stu}, and the correlated two-pion exchange with the pole approximation \cite{Kim:2019rud}. Here, it should be noted that the $\sigma \bar D_1 \bar D_1$ coupling constant is identical to the $\sigma \bar D^* \bar D^*$ coupling constant in the quark model. In the following, we discuss the bound state solutions for the isoscalar $\Xi_c\bar D_1$ state with $J^P=1/2^+$ by considering the uncertainties of the coupling constant $g_{\sigma}^{\prime\prime}$. In Table~\ref{sr}, we display the obtained bound state solutions for the isoscalar $\Xi_c\bar D_1$ state with $J^P=1/2^+$ by taking $g_{\sigma}^{\prime\prime}= -0.76, -3.40$, and $-5.21$. From Table~\ref{sr}, it can be observed that the bound state solutions for the isoscalar $\Xi_c\bar D_1$ state with $J^P=1/2^+$ will change, but the isoscalar $\Xi_c\bar D_1$ state with $J^P=1/2^+$ still can be recommended as the most promising hidden-charm molecular pentaquark candidate with strangeness when considering the uncertainties of the coupling constant $g_{\sigma}^{\prime\prime}$.

\renewcommand\tabcolsep{0.13cm}
\renewcommand{\arraystretch}{1.50}
\begin{table}[!htbp]
\centering
\caption{Bound state solutions for the isoscalar $\Xi_c\bar D_1$ state with $J^P=1/2^+$ by taking $g_{\sigma}^{\prime\prime}= -0.76, -3.40$, and $-5.21$. The units of the cutoff $\Lambda$, binding energy $E$, and root-mean-square radius $r_{{\rm RMS}}$ are $\rm{GeV}$, $\rm {MeV}$, and $\rm {fm}$, respectively.}\label{sr}
\begin{tabular}{ccc|ccc|ccc}\toprule[1pt]\toprule[1pt]
\multicolumn{3}{c|}{$g_{\sigma}^{\prime\prime}= -0.76$}&\multicolumn{3}{c|}{$g_{\sigma}^{\prime\prime}= -3.40$}&\multicolumn{3}{c}{$g_{\sigma}^{\prime\prime}= -5.21$}\\\midrule[1.0pt]
$\Lambda$ &$E$&$r_{\rm RMS}$       &$\Lambda$ &$E$&$r_{\rm RMS}$          &$\Lambda$ &$E$&$r_{\rm RMS}$\\
1.32&$-0.27$ &4.87                  &0.94&$-0.28$&4.91                    &0.86&$-0.29$&4.87\\
1.49&$-4.66$ &1.58                  &1.01&$-4.02$&1.76                     &0.92&$-4.74$&1.67\\
1.65&$-12.46$ &1.05                 &1.08&$-11.71$&1.10                    &0.97&$-12.28$&1.16\\
\bottomrule[1pt]\bottomrule[1pt]
\end{tabular}
\end{table}

\subsection{$\Xi_c\bar D_2^*$ system}

Similar to the $\Xi_c\bar D_1$ system, the $\sigma$, $\rho$, and $\omega$ exchanges provide the total effective potential for the $\Xi_c\bar D_2^*$ system within the OBE model. For the isoscalar $\Xi_c\bar D_2^*$ states with $J^P=3/2^+$ and $5/2^+$, the $\sigma$ and $\rho$ exchange potentials are the attractive, and the $\omega$ exchange provides the repulsive potential. For the isovector $\Xi_c\bar D_2^*$ states with $J^P=3/2^+$ and $5/2^+$, the attractive interaction arises from the $\sigma$ exchange, while the $\rho$ and $\omega$ exchanges give the repulsion potential. In Table \ref{XicD2}, we collect the obtained bound state solutions for the $\Xi_c\bar D_2^*$ system by considering the $S$-$D$ wave mixing case and the coupled channel case.

\renewcommand\tabcolsep{0.15cm}
\renewcommand{\arraystretch}{1.50}
\begin{table}[!htbp]
\centering
\caption{The cutoff values dependence of the bound state solutions for the $\Xi_c\bar D_2^*$ system by considering the $S$-$D$ wave mixing case and the coupled channel case. Here, the dominant contribution channel is shown in bold font, while the units of the cutoff $\Lambda$, binding energy $E$, and root-mean-square radius $r_{{\rm RMS}}$ are $\rm{GeV}$, $\rm {MeV}$, and $\rm {fm}$, respectively.}\label{XicD2}
\begin{tabular}{c|cccc}\toprule[1.0pt]\toprule[1.0pt]
\multicolumn{5}{c}{$S$-$D$ wave mixing case}\\\midrule[1.0pt]
$I(J^P)$&$\Lambda$  &$E$ &$r_{\rm RMS}$ &P(${}^4\mathbb{S}_{\frac{3}{2}}/{}^4\mathbb{D}_{\frac{3}{2}}/{}^6\mathbb{D}_{\frac{3}{2}}$)\\\hline
\multirow{3}{*}{$0(\frac{3}{2}^+)$}&1.32&$-0.32$ &4.63&\textbf{100.00}/$o(0)$/$o(0)$     \\
&1.49&$-4.89$ &1.54&\textbf{100.00}/$o(0)$/$o(0)$      \\
&1.65&$-12.86$ &1.03&\textbf{100.00}/$o(0)$/$o(0)$     \\\midrule[1.0pt]
$I(J^P)$&$\Lambda$ &$E$ &$r_{\rm RMS}$ &P(${}^6\mathbb{S}_{\frac{5}{2}}/{}^4\mathbb{D}_{\frac{5}{2}}/{}^6\mathbb{D}_{\frac{5}{2}}$)\\\hline
\multirow{3}{*}{$0(\frac{5}{2}^+)$}&1.32&$-0.32$ &4.63&\textbf{100.00}/$o(0)$/$o(0)$\\
&1.49&$-4.89$ &1.54&\textbf{100.00}/$o(0)$/$o(0)$\\
&1.65&$-12.86$ &1.03&\textbf{100.00}/$o(0)$/$o(0)$\\\midrule[1.0pt]
\multicolumn{5}{c}{Coupled channel case}\\\midrule[1.0pt]
$I(J^P)$&$\Lambda$ &$E$ &$r_{\rm RMS}$ &P($\Xi_c\bar D_2^*/\Xi_c^{\prime}\bar D_1/\Xi_c^{\prime}\bar D_2^*/\Xi_c^{*}\bar D_1/\Xi_c^{*}\bar D_2^*$)\\\hline
\multirow{3}{*}{$0(\frac{3}{2}^+)$}&  1.06&$-0.29$&4.69&\textbf{98.02}/0.28/0.57/0.06/1.07\\
&1.10&$-3.68$&1.63&\textbf{91.56}/0.90/1.02/0.47/6.04\\
&1.13&$-10.56$&0.98&\textbf{81.06}/1.03/0.15/1.79/15.98\\\midrule[1.0pt]
\multirow{3}{*}{$1(\frac{3}{2}^+)$}& 1.86&$-1.34$& 2.14&\textbf{78.06}/17.92/0.28/2.17/1.56\\
& 1.87&$-3.92$&1.19&\textbf{68.33}/25.96/0.38/3.19/2.14\\
& 1.88&$-7.30$&0.85&\textbf{62.02}/31.19/0.44/3.88/2.47\\\midrule[1.0pt]
$I(J^P)$&$\Lambda$ &$E$ &$r_{\rm RMS}$ &P($\Xi_c\bar D_2^*/\Xi_c^{\prime}\bar D_2^*/\Xi_c^{*}\bar D_1/\Xi_c^{*}\bar D_2^*$)\\\hline
\multirow{3}{*}{$0(\frac{5}{2}^+)$}&  1.05&$-0.40$&4.25&\textbf{97.77}/0.46/0.28/1.49\\
&1.09&$-4.53$&1.48&\textbf{91.10}/1.73/1.11/6.06\\
&1.12&$-11.14$&0.98&\textbf{84.77}/2.83/1.90/10.50\\\midrule[1.0pt]
\multirow{3}{*}{$1(\frac{5}{2}^+)$}&1.58&$-0.15$&5.09&\textbf{94.31}/1.64/0.66/3.39\\
&1.60&$-5.30$&1.10&\textbf{76.87}/7.17/2.59/13.37\\
&1.61&$-9.76$&0.80&\textbf{71.47}/9.08/3.14/16.31\\
\bottomrule[1.0pt]\bottomrule[1.0pt]
\end{tabular}
\end{table}

In the context of the $S$-$D$ wave mixing analysis, the bound state solutions for the isoscalar $\Xi_c\bar D_2^*$ states with $J^P=3/2^+$ and $5/2^+$ appear when the cutoff values $\Lambda$ are tuned larger than $1.32$ GeV, which is the reasonable cutoff value. Moreover, the probabilities for the $D$-wave channels are zero, since there does not exist the contribution of the tensor force mixing the $S$-wave and $D$-wave components in the OBE effective potentials for the $\Xi_c\bar D_2^*$ system. Based on our obtained numerical results, the isoscalar $\Xi_c\bar D_2^*$ states with $J^P=3/2^+$ and $5/2^+$ can be recommended as the most promising hidden-charm molecular pentaquark candidates with strangeness. Similar to the case of the isoscalar $\Xi_c\bar D_1$ bound states with $J^P=1/2^+$ and $3/2^+$, the numerical results shown in Table \ref{XicD2} indicate that the isoscalar $\Xi_c\bar D_2^*$ bound states with $J^P=3/2^+$ and $5/2^+$ also exist the phenomenon of the mass degeneration when we take the same cutoff value as input in the $S$-$D$ wave mixing case.  For the isovector $\Xi_c\bar D_2^*$ states with $J^P=3/2^+$ and $5/2^+$, we have not found the bound state solutions when the cutoff values lie between 0.8 and 2.5 GeV. In addition, the $\Xi_c\bar D_1$ and $\Xi_c\bar D_2^*$ systems have the same interactions, but the binding energies of the isoscalar $\Xi_c\bar D_2^*$ states with $J^P=3/2^+$ and $5/2^+$ are larger than those of the isoscalar $\Xi_c\bar D_1$ states with $J^P=1/2^+$ and $3/2^+$ if we adopt the same cutoff value, since the hadrons with heavier masses are more easily form the bound states due to the relatively small kinetic terms.

Furthermore, we take into account the role of the coupled channel effect for the $\Xi_c\bar D_2^*$ system. As indicated in Table \ref{XicD2}, the bound state solutions for the isoscalar $\Xi_c\bar D_2^*$ states with $J^P=3/2^+$ and $5/2^+$ can be obtained with the cutoff values $\Lambda$ above 1.06 GeV and 1.05 GeV, respectively, where the dominant channel is the $\Xi_c\bar D_2^*$ with the probability over 80\%. Different from the single channel and $S$-$D$ wave mixing cases, the isoscalar $\Xi_c\bar D_2^*$ states with $J^P=3/2^+$ and $5/2^+$ have different bound state properties when taking the same cutoff value after including the influence of the coupled channel effect. Such a case is particularly interesting, and it is a good place to test the role of the coupled channel effect for studying the hadron-hadron interactions. Besides, our study indicates that the contribution from the coupled channel effect is crucial for the formation of the isovector $\Xi_c\bar D_2^*$ bound states with $J^P=3/2^+$ and $5/2^+$, and their bound state solutions can be found when the cutoff values are fixed to be larger than 1.86 GeV and 1.58 GeV, respectively. For both bound states, the $\Xi_c\bar D_2^*$ component is dominant and decreases as the cutoff parameter increases, while the contributions of other coupled channels are also important in generating both bound states.

As we can see, the isoscalar $\Xi_c\bar D_2^*$ states with $J^P=3/2^+$ and $5/2^+$ are expected to be the most promising hidden-charm molecular pentaquark candidates with strangeness, while the isovector $\Xi_c\bar D_2^*$ states with $J^P=3/2^+$ and $5/2^+$ may be the possible candidates of the hidden-charm molecular pentaquarks with strangeness.

\subsection{$\Xi_c^{\prime}\bar D_1$ system}

For the $\Xi_c^{\prime}\bar D_1$ system, the $\pi$ and $\eta$ exchanges also contribute to the total effective potential, except for the $\sigma$, $\rho$, and $\omega$ exchange interactions. In addition, the relevant channels for the $\Xi_c^{\prime}\bar D_1$ system with the same total angular momentum $J$ and parity $P$ but the different spins $S$ and orbital angular momenta $L$ can mix each other due to the existence of the tensor force operator in the OBE effective potential, which leads to the contribution of the $S$-$D$ wave mixing effect. These features are obviously different from the $\Xi_c\bar D_1$ and $\Xi_c\bar D_2^*$ systems. In Table \ref{XicprimeD1}, we give the obtained bound state properties for the $\Xi_c^{\prime}\bar D_1$ system by considering the single channel case, the $S$-$D$ wave mixing case, and the coupled channel case.

\renewcommand\tabcolsep{0.22cm}
\renewcommand{\arraystretch}{1.50}
\begin{table}[!htbp]
\centering
\caption{The cutoff values dependence of the bound state solutions for the $\Xi_c^{\prime}\bar D_1$ system by considering the single channel case, the $S$-$D$ wave mixing case, and the coupled channel case. Here, the dominant contribution channel is shown in bold font, while the units of the cutoff $\Lambda$, binding energy $E$, and root-mean-square radius $r_{{\rm RMS}}$ are $\rm{GeV}$, $\rm {MeV}$, and $\rm {fm}$, respectively.}\label{XicprimeD1}
\begin{tabular}{c|cccc}\toprule[1.0pt]\toprule[1.0pt]
\multicolumn{5}{c}{Single channel case}\\\midrule[1.0pt]
$I(J^P)$&$\Lambda$  &$E$ &$r_{\rm RMS}$ \\\hline
\multirow{3}{*}{$0(\frac{1}{2}^+)$}&0.94&$-0.34$ &4.40\\
&1.00&$-4.81$ &1.47\\
&1.05&$-12.80$ &0.98\\\midrule[1.0pt]
\multirow{3}{*}{$0(\frac{3}{2}^+)$}&1.92&$-0.29$ &4.88\\
&2.21&$-1.77$ &2.49\\
&2.50&$-4.15$ &1.74\\\midrule[1.0pt]
\multicolumn{5}{c}{$S$-$D$ wave mixing case}\\\midrule[1.0pt]
$I(J^P)$&$\Lambda$  &$E$ &$r_{\rm RMS}$ &P(${}^2\mathbb{S}_{\frac{1}{2}}/{}^4\mathbb{D}_{\frac{1}{2}}$)\\\hline
\multirow{3}{*}{$0(\frac{1}{2}^+)$}&0.93&$-0.35$ &4.39&\textbf{99.69}/0.31     \\
&0.99&$-4.55$ &1.52&\textbf{99.51}/0.49     \\
&1.04&$-12.07$ &1.01&\textbf{99.51}/0.49     \\\midrule[1.0pt]
$I(J^P)$&$\Lambda$ &$E$ &$r_{\rm RMS}$ &P(${}^4\mathbb{S}_{\frac{3}{2}}/{}^2\mathbb{D}_{\frac{3}{2}}/{}^4\mathbb{D}_{\frac{3}{2}}$)\\\hline
\multirow{3}{*}{$0(\frac{3}{2}^+)$}&1.53&$-0.30$ &4.87&\textbf{98.78}/0.20/1.02\\
&1.95&$-4.95$ &1.65&\textbf{96.78}/0.50/2.72\\
&2.37&$-12.64$ &1.15&\textbf{95.81}/0.65/3.54\\\midrule[1.0pt]
\multicolumn{5}{c}{Coupled channel case}\\\midrule[1.0pt]
$I(J^P)$&$\Lambda$  &$E$ &$r_{\rm RMS}$ &P($\Xi_c^{\prime}\bar D_1/\Xi_c^{*}\bar D_1/\Xi_c^{*}\bar D_2^*$)\\\hline
\multirow{3}{*}{$0(\frac{1}{2}^+)$}&0.91&$-0.21$&4.94&\textbf{99.40}/0.48/0.12     \\
                           &0.96&$-4.30$&1.52&\textbf{96.38}/2.98/0.65 \\
                           &1.00&$-12.00$&0.97&\textbf{91.96}/6.73/1.31 \\\midrule[1.0pt]
$I(J^P)$&$\Lambda$ &$E$ &$r_{\rm RMS}$ &P($\Xi_c^{\prime}\bar D_1/\Xi_c^{\prime}\bar D_2^*/\Xi_c^{*}\bar D_1/\Xi_c^{*}\bar D_2^*$)\\\hline
\multirow{3}{*}{$0(\frac{3}{2}^+)$}& 1.07&$-3.13$ &1.01&23.01/\textbf{54.40}/9.45/13.14\\
&1.08&$-7.68$&  0.68&13.39/\textbf{59.97}/10.42/16.22\\
&1.09&$-12.86$&0.59& 9.51/\textbf{61.29}/10.86/18.34\\\midrule[1.0pt]
\multirow{3}{*}{$1(\frac{3}{2}^+)$}& 1.99&$-0.14$&5.08&\textbf{97.58}/0.09/2.09/0.25\\
& 2.02&$-4.97$&1.13&\textbf{91.25}/0.33/7.53/0.89\\
& 2.04&$-10.55$& 0.77&\textbf{88.95}/0.43/9.49/1.13\\
\bottomrule[1.0pt]\bottomrule[1.0pt]
\end{tabular}
\end{table}

For the isoscalar $\Xi_c^{\prime}\bar D_1$ state with $J^P=1/2^+$, the solutions of the bound state can be found when we set the cutoff value to be 0.96 GeV in the single channel case, and the binding energies become large with the increase of the cutoff values. If considering the $S$-$D$ wave mixing effect with channels mixing among the $|{}^2\mathbb{S}_{{1}/{2}}\rangle$ and $|{}^4\mathbb{D}_{{1}/{2}}\rangle$, we can obtain the bound state solutions for the isoscalar $\Xi_c^{\prime}\bar D_1$ state with $J^P=1/2^+$ when the cutoff value $\Lambda$ is fixed larger than 0.93 GeV, where the $|{}^2\mathbb{S}_{{1}/{2}}\rangle$ channel has the dominant contribution with the probability over 99\%. In other words, the role of the $S$-$D$ wave mixing effect is tiny for the formation of this bound state. After including the coupled channel effect from the $\Xi_c^{\prime}\bar D_1$, $\Xi_c^{*}\bar D_1$, and $\Xi_c^{*}\bar D_2^*$ channels, the bound state solutions can be found when we choose the cutoff value around 0.91 GeV, where the $\Xi_c^{\prime}\bar D_1$ channel contribution is dominant with the probability greater than 90\% and the remaining channels have small probabilities. Since the isoscalar $\Xi_c^{\prime}\bar D_1$ bound state with $J^P=1/2^+$ has the small binding energy and the large size with the reasonable cutoff value around 1.0 GeV, the isoscalar $\Xi_c^{\prime}\bar D_1$ state with $J^P=1/2^+$ can be assigned as the most promising hidden-charm molecular pentaquark candidate with strangeness.

\begin{figure}[htbp]
\includegraphics[scale=0.25]{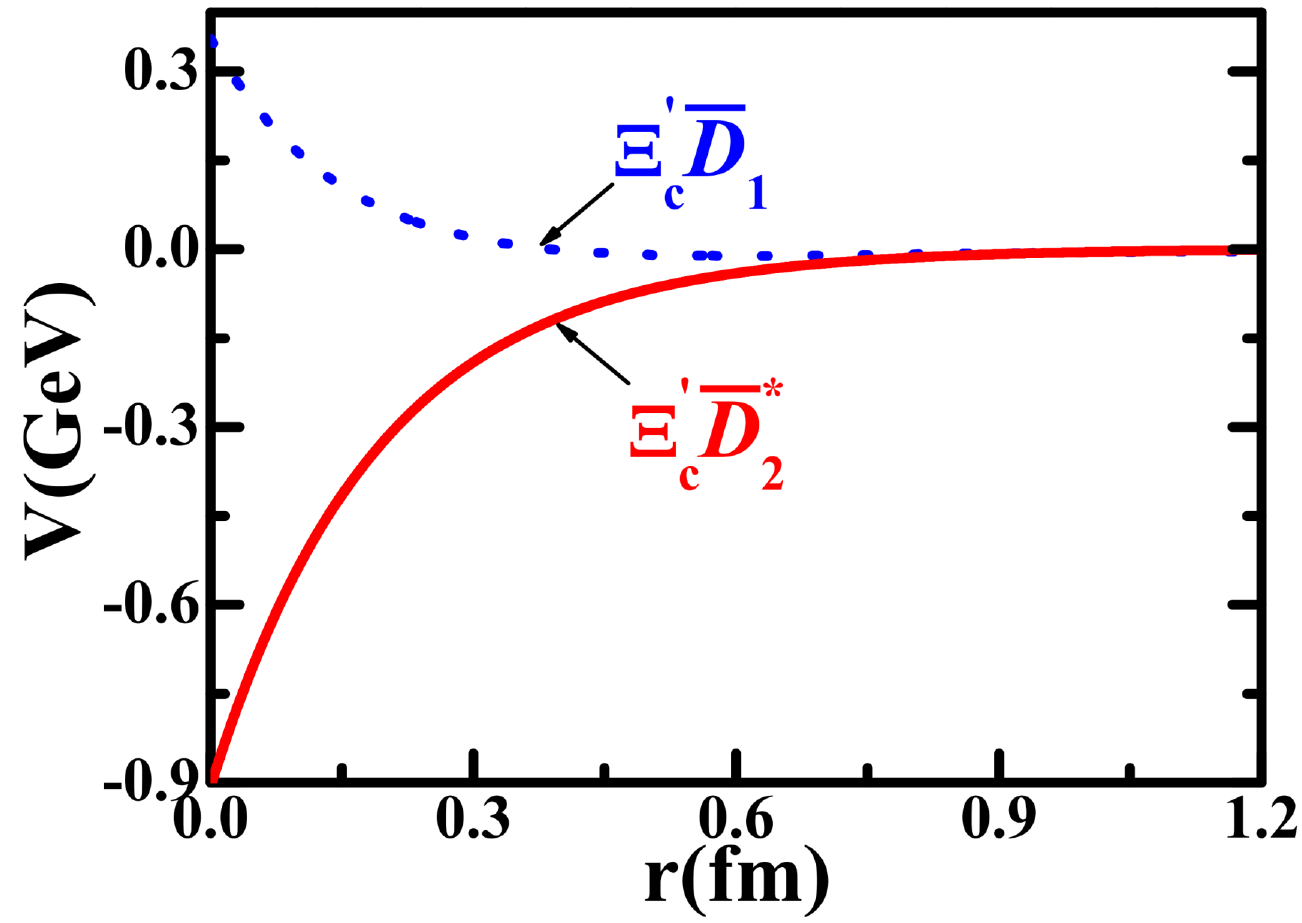}
\caption{The OBE effective potentials for the $\Xi_c^{\prime}\bar D_1$ and $\Xi_c^{\prime}\bar D_2^*$ states with $I(J^P)=0(3/2^+)$, where the cutoff value $\Lambda$ is taken as $1.08$ GeV for an illustration.}\label{potentialshape2}
\end{figure}

For the isoscalar $\Xi_c^{\prime}\bar D_1$ state with $J^P=3/2^+$, the existence of the bound state solutions requires that the cutoff value should be at least larger than 1.92 GeV in the single channel case. After adding the contributions of the $D$-wave channels, there exists the bound state solutions with the cutoff value around 1.53 GeV for the isoscalar $\Xi_c^{\prime}\bar D_1$ state with $J^P=3/2^+$, where the contribution of the $S$-wave channel is over 95\%. By comparing the obtained bound state solutions for the isoscalar $\Xi_c^{\prime}\bar D_1$ state with $J^P=3/2^+$ in the single channel and $S$-$D$ wave mixing cases, the cutoff value in the $S$-$D$ wave mixing analysis is smaller than that in the single channel analysis when obtaining the same binding energy, which means that the $S$-$D$ wave mixing effect plays the important role in generating the isoscalar $\Xi_c^{\prime}\bar D_1$ bound state with $J^P=3/2^+$. Furthermore, the isoscalar $\Xi_c^{\prime}\bar D_1$ bound state with $J^P=3/2^+$ has the small binding energy and the suitable size under the reasonable cutoff value after considering the $S$-$D$ wave mixing effect. Thus, the isoscalar $\Xi_c^{\prime}\bar D_1$ state with $J^P=3/2^+$ may be the promising candidate of the hidden-charm molecular pentaquark with strangeness. After that, we also discuss the bound state properties for the isoscalar $\Xi_c^{\prime}\bar D_1$ state with $J^P=3/2^+$ by considering the coupled channel effect, but this coupled system is dominated by the $\Xi_c^{\prime}\bar D_2^*$ channel, which results a little small size for this bound state \cite{Chen:2017xat}. This can be attributed to the effective interaction of the isoscalar $\Xi_c^{\prime}\bar D_2^*$ state with $J^P=3/2^+$ is far stronger than that of the isoscalar $\Xi_c^{\prime}\bar D_1$ state with $J^P=3/2^+$ when we adopt the same cutoff value (see Fig. \ref{potentialshape2} for more details), and the thresholds of the $\Xi_c^{\prime}\bar D_1$ and $\Xi_c^{\prime}\bar D_2^*$ channels are very close with the difference is 39 MeV. As proposed in Ref. \cite{Wang:2022mxy}, the cutoff values for the involved coupled channels may be different in reality, which may result in the coupled channel effect only playing the role of decorating the bound state properties for the pure state. As discussed above, when existing the related experimental information, the bound state properties of the isoscalar $\Xi_c^{\prime}\bar D_1$ state with $J^P=3/2^+$ deserve further studies by including the coupled channel effect and adopting different cutoff values for the involved coupled channels in future, and this approach has been used to discuss the double peak structures of the $P_{\psi s}^{\Lambda}(4459)$ under the $\Xi_c\bar D^{*}$ molecule picture in Ref. \cite{Wang:2022mxy}.

For the isovector $\Xi_c^{\prime}\bar D_1$ state with $J^P=1/2^+$, the interaction is not strong enough to form the bound state even though we tune the cutoff values as high as 2.5 GeV and consider the coupled channel effect. Thus, our obtained numerical results disfavor the existence of the hidden-charm molecular pentaquark candidate with strangeness for the isovector $\Xi_c^{\prime}\bar D_1$ state with $J^P=1/2^+$. For the isovector $\Xi_c^{\prime}\bar D_1$ state with $J^P=3/2^+$, there is no bound state solutions in the single channel and the $S$-$D$ wave mixing cases by scanning the cutoff values from $0.8$ to $2.5\,{\rm GeV}$. When further adding the role of the coupled channel effect from the $\Xi_c^{\prime}\bar D_1$, $\Xi_c^{\prime}\bar D_2^*$, $\Xi_c^{*}\bar D_1$, and $\Xi_c^{*}\bar D_2^*$ channels, there exists the bound state solutions with the cutoff value $\Lambda$ slightly below 2.0 GeV, where the probability of the $\Xi_c^{\prime}\bar D_1$ channel is more than 88\%. However, such cutoff parameter is a little away from the typical value around 1.0 GeV, which indicates that the isovector $\Xi_c^{\prime}\bar D_1$ state with $J^P=3/2^+$ can be treated as the potential candidate of the hidden-charm molecular pentaquark with strangeness, rather than the most promising candidate.

\begin{figure}[htbp]
\centering
\includegraphics[scale=0.36]{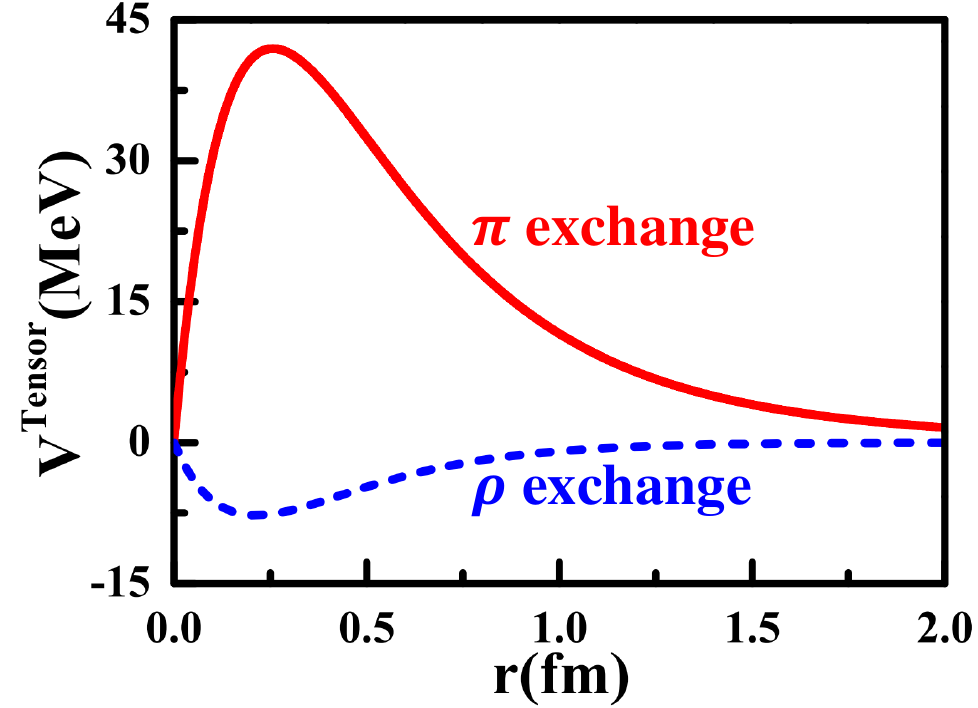}
\caption{The tensor interactions of the $\pi$ and $\rho$ exchange potentials for the isoscalar $\Xi_c^{\prime}\bar D_1$ state with $J^P=1/2^+$, where the cutoff $\Lambda$ is fixed as the typical value $1.0$ GeV.}\label{pirho}
\end{figure}

In the following, we discuss the tensor interactions of the $\pi$ and $\rho$ exchange potentials for the isoscalar $\Xi_c^{\prime}\bar D_1$ state with $J^P=1/2^+$. In Fig. \ref{pirho}, we present the tensor interactions of the $\pi$ and $\rho$ exchange potentials for the isoscalar $\Xi_c^{\prime}\bar D_1$ state with $J^P=1/2^+$. As shown in Fig. \ref{pirho}, the tensor interaction is optimized by the balance of the $\pi$ and $\rho$ exchange potentials for the isoscalar $\Xi_c^{\prime}\bar D_1$ state with $J^P=1/2^+$, which is similar to the case of the $NN$ interaction.

\subsection{$\Xi_c^{\prime}\bar D_2^*$ system}

For the $S$-wave $\Xi_c^{\prime}\bar D_2^*$ system, the total effective potential arises from the $\sigma$, $\pi$, $\eta$, $\rho$, and $\omega$ exchanges within the OBE model, while the allowed quantum numbers contain $I(J^P)=0({3}/{2}^+)$, $0({5}/{2}^+)$, $1({3}/{2}^+)$, and $1({5}/{2}^+)$. In Table \ref{XicprimeD2}, we collect the obtained bound state solutions for the $\Xi_c^{\prime}\bar D_2^*$ system by considering the single channel case, the $S$-$D$ wave mixing case, and the coupled channel case.

\renewcommand\tabcolsep{0.32cm}
\renewcommand{\arraystretch}{1.50}
\begin{table}[!htbp]
\centering
\caption{The cutoff values dependence of the bound state solutions for the $\Xi_c^{\prime}\bar D_2^*$ system by considering the single channel case, the $S$-$D$ wave mixing case, and the coupled channel case. Here, the dominant contribution channel is shown in bold font, while the units of the cutoff $\Lambda$, binding energy $E$, and root-mean-square radius $r_{{\rm RMS}}$ are $\rm{GeV}$, $\rm {MeV}$, and $\rm {fm}$, respectively.}\label{XicprimeD2}
\begin{tabular}{c|cccc}\toprule[1.0pt]\toprule[1.0pt]
\multicolumn{5}{c}{Single channel case}\\\midrule[1.0pt]
$I(J^P)$&$\Lambda$  &$E$ &$r_{\rm RMS}$ \\\hline
\multirow{3}{*}{$0(\frac{3}{2}^+)$}&0.96&$-0.29$ &4.59\\
&1.02&$-4.44$ &1.52\\
&1.08&$-13.78$ &0.95\\\midrule[1.0pt]
\multirow{3}{*}{$0(\frac{5}{2}^+)$}&1.99&$-0.28$ &4.92\\
&2.26&$-1.57$ &2.63\\
&2.50&$-3.35$ &1.91\\\midrule[1.0pt]
\multirow{3}{*}{$1(\frac{5}{2}^+)$}&2.38&$-0.34$ &4.04\\
&2.41&$-5.02$ &1.12\\
&2.43&$-10.00$ &0.78\\\midrule[1.0pt]
\multicolumn{5}{c}{$S$-$D$ wave mixing case}\\\midrule[1.0pt]
$I(J^P)$&$\Lambda$  &$E$ &$r_{\rm RMS}$ &P(${}^4\mathbb{S}_{\frac{3}{2}}/{}^4\mathbb{D}_{\frac{3}{2}}/{}^6\mathbb{D}_{\frac{3}{2}}$)\\\hline
\multirow{3}{*}{$0(\frac{3}{2}^+)$}&0.94&$-0.27$ &4.73&\textbf{99.49}/0.07/0.43     \\
&1.00&$-3.88$ &1.64&\textbf{99.11}/0.13/0.76      \\
&1.06&$-12.18$ &1.02&\textbf{99.11}/0.13/0.76     \\\midrule[1.0pt]
$I(J^P)$&$\Lambda$ &$E$ &$r_{\rm RMS}$ &P(${}^6\mathbb{S}_{\frac{5}{2}}/{}^4\mathbb{D}_{\frac{5}{2}}/{}^6\mathbb{D}_{\frac{5}{2}}$)\\\hline
\multirow{3}{*}{$0(\frac{5}{2}^+)$}&1.56&$-0.31$ &4.83&\textbf{98.63}/0.41/0.96\\
&1.99&$-4.91$ &1.67&\textbf{96.34}/1.07/2.59\\
&2.42&$-12.89$ &1.15&\textbf{95.19}/1.39/3.43\\\midrule[1.0pt]
\multirow{3}{*}{$1(\frac{5}{2}^+)$}&2.33&$-0.25$ &4.44&\textbf{99.79}/0.06/0.15\\
&2.36&$-4.52$ &1.19&\textbf{99.43}/1.15/0.42\\
&2.38&$-9.20$ &0.84&\textbf{99.29}/0.19/0.52\\\midrule[1.0pt]
\multicolumn{5}{c}{Coupled channel case}\\\midrule[1.0pt]
$I(J^P)$&$\Lambda$ &$E$ &$r_{\rm RMS}$ &P($\Xi_c^{\prime}\bar D_2^*/\Xi_c^{*}\bar D_1/\Xi_c^{*}\bar D_2^*$)\\\hline
\multirow{3}{*}{$0(\frac{3}{2}^+)$}& 0.93& $-0.51$&3.81&\textbf{98.44}/0.33/1.23\\
&0.97&$-4.27$&1.50& \textbf{93.92}/1.39/4.69 \\
&1.01&$-12.44$& 0.94&\textbf{86.77}/3.09/10.14\\\midrule[1.0pt]
\multirow{3}{*}{$0(\frac{5}{2}^+)$}& 1.26&$-0.36$&4.33&\textbf{92.19}/1.76/6.05\\
&1.30&$-4.88$& 1.29& \textbf{67.14}/7.07/25.79 \\
&1.33&$-11.67$&0.83&\textbf{52.74}/9.56/37.70\\\midrule[1.0pt]
\multirow{3}{*}{$1(\frac{5}{2}^+)$}&1.88&$-0.35$&4.02&\textbf{96.29}/1.71/2.00\\
&1.90&$-3.11$&1.43&\textbf{91.23}/4.36/4.41\\
&1.92&$-7.68$&0.91&\textbf{88.01}/6.27/5.73 \\
\bottomrule[1.0pt]\bottomrule[1.0pt]
\end{tabular}
\end{table}

In the single channel case, the OBE effective potentials are sufficient to form the $\Xi_c^{\prime}\bar D_2^*$ bound states with $I(J^P)=0({3}/{2}^+)$, $0({5}/{2}^+)$, and $1({5}/{2}^+)$ when the cutoff values are taken to be around 0.96, 1.99, and 2.38 GeV, respectively. However, we fail to find the bound state solutions for the $\Xi_c^{\prime}\bar D_2^*$ state with $I(J^P)=1({3}/{2}^+)$ when the cutoff values $\Lambda$ are scanned from 0.8 to 2.5 GeV. In the following, we continue to discuss the bound state properties for the $\Xi_c^{\prime}\bar D_2^*$ system by considering the $S$-$D$ wave mixing effect and the coupled channel effect.

For the $\Xi_c^{\prime}\bar D_2^*$ state with $I(J^P)=0({3}/{2}^+)$, we can consider the $S$-$D$ wave mixing effect from the $|{}^4\mathbb{S}_{{3}/{2}}\rangle$, $|{}^4\mathbb{D}_{{3}/{2}}\rangle$, and $|{}^6\mathbb{D}_{{3}/{2}}\rangle$ channels, and there exists the  bound state solutions when the cutoff value $\Lambda$ should be at least 0.94 GeV, where the $|{}^4\mathbb{S}_{{3}/{2}}\rangle$ is the dominant channel with the probability over 99\%. However, when the $S$-$D$ wave mixing effect is included, the conclusion of the absence of the bound state solutions does not change for the $\Xi_c^{\prime}\bar D_2^*$ state with $I(J^P)=1({3}/{2}^+)$ if we fix the cutoff values $\Lambda$ smaller than 2.5 GeV. After adding the contribution of the $S$-$D$ wave mixing effect among the $|{}^6\mathbb{S}_{{5}/{2}}\rangle$, $|{}^4\mathbb{D}_{{5}/{2}}\rangle$, and $|{}^6\mathbb{D}_{{5}/{2}}\rangle$ channels, the $\Xi_c^{\prime}\bar D_2^*$ states with $I(J^P)=0({5}/{2}^+)$ and $1({5}/{2}^+)$ have the bound state solutions when the cutoff values $\Lambda$ are larger than 1.56 GeV and 2.33 GeV, respectively. Compared to the obtained bound state solutions in the single channel case, the $S$-$D$ wave mixing effect plays the important role for forming the $\Xi_c^{\prime}\bar D_2^*$ bound state with $I(J^P)=0({5}/{2}^+)$. Nevertheless, the bound state properties of the $\Xi_c^{\prime}\bar D_2^*$ states with $I(J^P)=0({3}/{2}^+)$ and $1({5}/{2}^+)$ change slightly after considering the role of the $S$-$D$ wave mixing effect, and the total probability of the $D$-wave channels is less than 1\%, which provides the negligible contributions.

Meanwhile, we consider the influence of the coupled channel effect from the $\Xi_c^{\prime}\bar D_2^*$, $\Xi_c^{*}\bar D_1$, and $\Xi_c^{*}\bar D_2^*$ channels for the $\Xi_c^{\prime}\bar D_2^*$ system. For the $\Xi_c^{\prime}\bar D_2^*$ state with $I(J^P)=0({3}/{2}^+)$, the bound state solutions can be found when the cutoff value is fixed to be larger than 0.93 GeV, where the $\Xi_c^{\prime}\bar D_2^*$ channel provides the dominant contribution with the probability over 86\%. Moreover, the coupled channel effect plays the important role for forming the $\Xi_c^{\prime}\bar D_2^*$ bound states with $I(J^P)=0({5}/{2}^+)$ and $1({5}/{2}^+)$, and their bound state solutions appear when the cutoff values are larger than 1.26 GeV and 1.88 GeV, respectively. Correspondingly, the $\Xi_c^{\prime}\bar D_2^*$ is the dominant channel, and the contributions of other coupled channels increase with the cutoff values. For the $\Xi_c^{\prime}\bar D_2^*$ state with $I(J^P)=1({3}/{2}^+)$, we also cannot find the bound state solutions corresponding to the cutoff values $\Lambda=0.8\sim 2.5\,{\rm GeV}$ even if including the coupled channel effect.

According to our quantitative analysis, the $\Xi_c^{\prime}\bar D_2^*$ states with $I(J^P)=0({3}/{2}^+)$ and $0({5}/{2}^+)$ can be recommended as the most promising hidden-charm molecular pentaquark candidates with strangeness, the $\Xi_c^{\prime}\bar D_2^*$ state with $I(J^P)=1({5}/{2}^+)$ may be the possible candidate of the hidden-charm molecular pentaquark with strangeness, and the $\Xi_c^{\prime}\bar D_2^*$ state with $I(J^P)=1({3}/{2}^+)$ is not considered as the hidden-charm molecular pentaquark candidate with strangeness.

\subsection{$\Xi_c^{*}\bar D_1$ system}

For the $S$-wave $\Xi_c^{*}\bar D_1$ system, the $\sigma$, $\pi$, $\eta$, $\rho$, and $\omega$ exchanges contribute to the total effective potential, while the allowed quantum numbers are $I(J^P)=0({1}/{2}^+)$, $0({3}/{2}^+)$, $0({5}/{2}^+)$, $1({1}/{2}^+)$, $1({3}/{2}^+)$, and $1({5}/{2}^+)$. Here, we perform the comprehensive and systematic analysis of the bound state properties for the $\Xi_c^{*}\bar D_1$ system by conducting the single channel analysis, the $S$-$D$ wave mixing analysis, and the coupled channel analysis.

In Table \ref{XicstarD1SSD}, we present the obtained bound state properties for the $\Xi_c^{*}\bar D_1$ system by considering the single channel case and the $S$-$D$ wave mixing case. From the numerical results listed in Table \ref{XicstarD1SSD}, the $\Xi_c^{*}\bar D_1$ states with $I(J^P)=0({1}/{2}^+)$ and $0({3}/{2}^+)$ exist the bound state solutions when we choose the cutoff values about 0.88 GeV and 1.08 GeV in the single channel analysis, respectively. As the cutoff values are increased, both bound states bind deeper and deeper. When further adding the contributions from the $D$-wave channels, the bound state solutions also appear for the $\Xi_c^{*}\bar D_1$ states with $I(J^P)=0({1}/{2}^+)$ and $0({3}/{2}^+)$ with the cutoff values around 0.86 GeV and 1.04 GeV, respectively, where the $S$-wave percentage is more than 98\% and plays the important role to generate both bound states. Compared to the obtained numerical results in the single channel case, the bound state properties for the $\Xi_c^{*}\bar D_1$ states with $I(J^P)=0({1}/{2}^+)$ and $0({3}/{2}^+)$ do not change too much after including the $S$-$D$ wave mixing effect. Moreover, the $\Xi_c^{*}\bar D_1$ states with $I(J^P)=0({5}/{2}^+)$ and $1({5}/{2}^+)$ can form the bound states when the cutoff values $\Lambda$ are set to be around 2.06 GeV in the context of the single channel analysis. After including the $S$-$D$ wave mixing effect among the $|{}^6\mathbb{S}_{{5}/{2}}\rangle$, $|{}^2\mathbb{D}_{{5}/{2}}\rangle$, $|{}^4\mathbb{D}_{{5}/{2}}\rangle$, and $|{}^6\mathbb{D}_{{5}/{2}}\rangle$ channels, the bound state properties for the $\Xi_c^{*}\bar D_1$ states with $I(J^P)=0({5}/{2}^+)$ and $1({5}/{2}^+)$ will change, and we can obtain their bound state solutions when the cutoff values $\Lambda$ are lowered down 1.56 GeV and 2.04 GeV, respectively. The contributions of the $D$-wave channels are quite small, and the probability of the dominant $S$-wave channel is over 99\% for the $\Xi_c^{*}\bar D_1$ bound state with $I(J^P)=1({5}/{2}^+)$. Comparing the obtained numerical results, the $S$-$D$ wave mixing effect is salient in generating the $\Xi_c^{*}\bar D_1$ bound state with $I(J^P)=0({5}/{2}^+)$, and the contribution of the $|{}^6\mathbb{D}_{{5}/{2}}\rangle$ channel is important, except for the $|{}^6\mathbb{S}_{{5}/{2}}\rangle$ channel. Unfortunately, there does not exist the bound state solutions for the $\Xi_c^{*}\bar D_1$ states with $I(J^P)=1({1}/{2}^+)$ and $1({3}/{2}^+)$ when the cutoff values are chosen between 0.8 to 2.5 GeV and the $S$-$D$ wave mixing effect is included.

\renewcommand\tabcolsep{0.30cm}
\renewcommand{\arraystretch}{1.50}
\begin{table}[!htbp]
\centering
\caption{The cutoff values dependence of the bound state solutions for the $\Xi_c^{*}\bar D_1$ system by considering the single channel case and the $S$-$D$ wave mixing case. Here, the dominant contribution channel is shown in bold font, while the units of the cutoff $\Lambda$, binding energy $E$, and root-mean-square radius $r_{{\rm RMS}}$ are $\rm{GeV}$, $\rm {MeV}$, and $\rm {fm}$, respectively.}\label{XicstarD1SSD}
\begin{tabular}{c|cccc}\toprule[1.0pt]\toprule[1.0pt]
\multicolumn{5}{c}{Single channel case}\\\midrule[1.0pt]
$I(J^P)$&$\Lambda$  &$E$ &$r_{\rm RMS}$ \\\hline
\multirow{3}{*}{$0(\frac{1}{2}^+)$}&0.88&$-0.28$ &4.61\\
&0.94&$-4.78$ &1.46\\
&0.99&$-13.33$ &0.96\\\midrule[1.0pt]
\multirow{3}{*}{$0(\frac{3}{2}^+)$}&1.08&$-0.37$ &4.34\\
&1.15&$-4.22$ &1.57\\
&1.22&$-12.31$ &1.00\\\midrule[1.0pt]
\multirow{3}{*}{$0(\frac{5}{2}^+)$}&2.06&$-0.28$ &4.95\\
&2.28&$-1.26$ &2.91\\
&2.50&$-2.81$ &2.07\\\midrule[1.0pt]
\multirow{3}{*}{$1(\frac{5}{2}^+)$}&2.06&$-0.41$ &3.81\\
&2.09&$-4.66$ &1.19\\
&2.11&$-9.22$ &0.84\\\midrule[1.0pt]
\multicolumn{5}{c}{$S$-$D$ wave mixing case}\\\midrule[1.0pt]
$I(J^P)$&$\Lambda$  &$E$ &$r_{\rm RMS}$ &P(${}^2\mathbb{S}_{\frac{1}{2}}/{}^4\mathbb{D}_{\frac{1}{2}}/{}^6\mathbb{D}_{\frac{1}{2}}$)\\\hline
\multirow{3}{*}{$0(\frac{1}{2}^+)$}&0.86&$-0.26$ &4.75&\textbf{99.46}/0.32/0.21     \\
&0.92&$-4.06$ &1.60&\textbf{99.07}/0.57/0.36     \\
&0.98&$-13.47$ &0.98&\textbf{99.09}/0.56/0.35     \\\midrule[1.0pt]
$I(J^P)$&$\Lambda$ &$E$ &$r_{\rm RMS}$ &P(${}^4\mathbb{S}_{\frac{3}{2}}/{}^2\mathbb{D}_{\frac{3}{2}}/{}^4\mathbb{D}_{\frac{3}{2}}/{}^6\mathbb{D}_{\frac{3}{2}}$)\\\hline
\multirow{3}{*}{$0(\frac{3}{2}^+)$}&1.04&$-0.36$ &4.41&\textbf{99.16}/0.23/0.55/0.06\\
&1.12&$-4.52$ &1.56&\textbf{98.48}/0.42/0.99/0.11\\
&1.19&$-12.23$ &1.03&\textbf{98.38}/0.45/1.05/0.12\\\midrule[1.0pt]
$I(J^P)$&$\Lambda$ &$E$ &$r_{\rm RMS}$ &P(${}^6\mathbb{S}_{\frac{5}{2}}/{}^2\mathbb{D}_{\frac{5}{2}}/{}^4\mathbb{D}_{\frac{5}{2}}/{}^6\mathbb{D}_{\frac{5}{2}}$)\\\hline
\multirow{3}{*}{$0(\frac{5}{2}^+)$}&1.56&$-0.29$ &4.95&\textbf{98.38}/0.93/0.05/1.48\\
&1.98&$-4.70$ &1.72&\textbf{95.53}/0.23/0.13/4.12\\
&2.40&$-12.77$ &1.18&\textbf{94.03}/0.29/0.17/5.51\\\midrule[1.0pt]
\multirow{3}{*}{$1(\frac{5}{2}^+)$}&2.04&$-0.89$ &2.75&\textbf{99.80}/0.01/0.01/0.18\\
&2.06&$-3.75$ &1.33&\textbf{99.67}/0.02/0.01/0.30\\
&2.08&$-8.00$ &0.91&\textbf{99.58}/0.03/0.01/0.38\\
\bottomrule[1.0pt]\bottomrule[1.0pt]
\end{tabular}
\end{table}

After that, we also study the bound state properties for the $\Xi_c^{*}\bar D_1$ system by adding the role of the coupled channel effect from the $\Xi_c^{*}\bar D_1$ and $\Xi_c^{*}\bar D_2^*$ channels, and the relevant numerical results are collected in Table \ref{XicstarD1coupled}. The $\Xi_c^{*}\bar D_1$ states with $I(J^P)=0({1}/{2}^+)$ and $0({3}/{2}^+)$ exist the small binding energies and the suitable sizes with the cutoff values around 0.90 GeV and 1.05 GeV, respectively, where the $\Xi_c^{*}\bar D_1$ channel has the dominant contribution. In addition, the $\Xi_c^{*}\bar D_2^*$ channel is important for the formation of the $\Xi_c^{*}\bar D_1$ bound state with $I(J^P)=0({3}/{2}^+)$, and whose contribution is over 30\% when the corresponding binding energy increases to be $-12\,{\rm MeV}$. For the $\Xi_c^{*}\bar D_1$ state with $I(J^P)=0({5}/{2}^+)$, we can obtain the bound state solutions when the cutoff value is taken to be larger than 1.42 GeV, which is the mixture state formed by the $\Xi_c^{*}\bar D_1$ and $\Xi_c^{*}\bar D_2^*$ channels. For the $\Xi_c^{*}\bar D_1$ state with $I(J^P)=1({5}/{2}^+)$, there exists the bound state solutions when the cutoff value is tuned larger than 1.86 GeV, where the probability of the $\Xi_c^{*}\bar D_1$ channel is over 93\%. After considering the contribution of the coupled channel effect, the bound state solutions of the $\Xi_c^{*}\bar D_1$ states with $I(J^P)=1({1}/{2}^+)$ and $1({3}/{2}^+)$ still disappear when the cutoff values change from 0.8 to 2.5 GeV.

\renewcommand\tabcolsep{0.40cm}
\renewcommand{\arraystretch}{1.50}
\begin{table}[!htbp]
\centering
\caption{The cutoff values dependence of the bound state solutions for the $\Xi_c^{*}\bar D_1$ system when the coupled channel effect is considered. Here, the dominant contribution channel is shown in bold font, while the units of the cutoff $\Lambda$, binding energy $E$, and root-mean-square radius $r_{{\rm RMS}}$ are $\rm{GeV}$, $\rm {MeV}$, and $\rm {fm}$, respectively.}\label{XicstarD1coupled}
\begin{tabular}{c|cccc}\toprule[1.0pt]\toprule[1.0pt]
$I(J^P)$&$\Lambda$  &$E$ &$r_{\rm RMS}$ &P($\Xi_c^{*}\bar D_1/\Xi_c^{*}\bar D_2^*$)\\\hline
\multirow{3}{*}{$0(\frac{1}{2}^+)$}&0.88&$-0.51$&3.80&\textbf{99.53}/0.47\\
                           &0.93&$-4.75$&1.45&\textbf{97.63}/2.37       \\
                           &0.97&$-11.89$&0.98&\textbf{94.39}/5.61\\\midrule[1.0pt]
\multirow{3}{*}{$0(\frac{3}{2}^+)$}&1.02&$-0.52$&3.77&\textbf{95.89}/4.11\\
&1.06&$-4.82$&1.36&\textbf{82.40}/17.60\\
&1.09&$-11.39$&0.91&\textbf{69.78}/30.22\\\midrule[1.0pt]
\multirow{3}{*}{$0(\frac{5}{2}^+)$}& 1.42&$-0.23$&4.88& \textbf{91.01}/8.99\\
&1.46&$-2.89$&1.59&\textbf{62.21}/37.79\\
&1.50&$-8.48$&0.90&41.36/\textbf{58.64}\\\midrule[1.0pt]
\multirow{3}{*}{$1(\frac{5}{2}^+)$}& 1.86&$-0.37$& 3.98&\textbf{97.90}/2.10\\
&1.89&$-4.68$&1.18&\textbf{94.51}/5.49\\
&1.91&$-9.41$&0.84&\textbf{93.17}/6.83\\
\bottomrule[1.0pt]\bottomrule[1.0pt]
\end{tabular}
\end{table}

To summarize, our obtained numerical results indicate that the $\Xi_c^{*}\bar D_1$ states with $I(J^P)=0({1}/{2}^+)$, $0({3}/{2}^+)$, and $0({5}/{2}^+)$ are favored to be the most promising hidden-charm molecular pentaquark candidates with strangeness since they have the small binding energies and the suitable sizes under the reasonable cutoff values, the $\Xi_c^{*}\bar D_1$ state with $I(J^P)=1({5}/{2}^+)$ may be viewed as the possible candidate of the hidden-charm molecular pentaquark with strangeness, while the $\Xi_c^{*}\bar D_1$ states with $I(J^P)=1({1}/{2}^+)$ and $1({3}/{2}^+)$ as the hidden-charm molecular pentaquark candidates with strangeness can be excluded.

\subsection{$\Xi_c^{*}\bar D_2^*$ system}

For the $S$-wave $\Xi_c^{*}\bar D_2^*$ system, there exists the $\sigma$, $\pi$, $\eta$, $\rho$, and $\omega$ exchange interactions within the OBE model, and the allowed quantum numbers are more abundant, which include $I(J^P)=0({1}/{2}^+)$, $0({3}/{2}^+)$, $0({5}/{2}^+)$, $0({7}/{2}^+)$, $1({1}/{2}^+)$, $1({3}/{2}^+)$, $1({5}/{2}^+)$, and $1({7}/{2}^+)$. In Table \ref{XicstarD2}, the obtained bound state solutions for the $\Xi_c^{*}\bar D_2^*$ system by considering the single channel case and the $S$-$D$ wave mixing case are presented.

\renewcommand\tabcolsep{0.22cm}
\renewcommand{\arraystretch}{1.50}
\begin{table}[!htbp]
\centering
\caption{The cutoff values dependence of the bound state solutions for the $\Xi_c^{*}\bar D_2^*$ system by considering the single channel case and the $S$-$D$ wave mixing case. Here, the dominant contribution channel is shown in bold font, while the units of the cutoff $\Lambda$, binding energy $E$, and root-mean-square radius $r_{{\rm RMS}}$ are $\rm{GeV}$, $\rm {MeV}$, and $\rm {fm}$, respectively.}\label{XicstarD2}
\begin{tabular}{c|cccc}\toprule[1.0pt]\toprule[1.0pt]
\multicolumn{5}{c}{Single channel case}\\\midrule[1.0pt]
$I(J^P)$&$\Lambda$  &$E$ &$r_{\rm RMS}$ \\\hline
\multirow{3}{*}{$0(\frac{1}{2}^+)$}&0.86&$-0.36$ &4.25\\
&0.92&$-5.08$ &1.42\\
&0.97&$-13.97$ &0.94\\\midrule[1.0pt]
\multirow{3}{*}{$0(\frac{3}{2}^+)$}&0.96&$-0.42$ &4.11\\
&1.02&$-4.88$ &1.45\\
&1.07&$-12.57$ &0.98\\\midrule[1.0pt]
\multirow{3}{*}{$0(\frac{5}{2}^+)$}&1.23&$-0.22$ &5.00\\
&1.35&$-4.23$ &1.59\\
&1.47&$-12.72$ &1.00\\\midrule[1.0pt]
\multirow{3}{*}{$0(\frac{7}{2}^+)$}&2.08&$-0.27$ &4.99\\
&2.29&$-1.22$ &2.95\\
&2.50&$-2.76$ &2.10\\\midrule[1.0pt]
\multirow{3}{*}{$1(\frac{7}{2}^+)$}&1.84&$-0.61$ &3.27\\
&1.87&$-4.73$ &1.20\\
&1.89&$-9.07$ &0.87\\\midrule[1.0pt]
\multicolumn{5}{c}{$S$-$D$ wave mixing case}\\\midrule[1.0pt]
$I(J^P)$&$\Lambda$  &$E$ &$r_{\rm RMS}$ &P(${}^2\mathbb{S}_{\frac{1}{2}}/{}^4\mathbb{D}_{\frac{1}{2}}/{}^6\mathbb{D}_{\frac{1}{2}}$)\\\hline
\multirow{3}{*}{$0(\frac{1}{2}^+)$}&0.84&$-0.46$ &3.98&\textbf{99.23}/0.39/0.38     \\
&0.90&$-4.63$ &1.52&\textbf{98.83}/0.60/0.57     \\
&0.95&$-12.42$ &1.01&\textbf{98.84}/0.60/0.56     \\\midrule[1.0pt]
$I(J^P)$&$\Lambda$ &$E$ &$r_{\rm RMS}$ &P(${}^4\mathbb{S}_{\frac{3}{2}}/{}^2\mathbb{D}_{\frac{3}{2}}/{}^4\mathbb{D}_{\frac{3}{2}}/{}^6\mathbb{D}_{\frac{3}{2}}/{}^8\mathbb{D}_{\frac{3}{2}}$)\\\hline
\multirow{3}{*}{$0(\frac{3}{2}^+)$}&0.93&$-0.39$ &4.26&\textbf{99.16}/0.22/0.39/0.08/0.15\\
&0.99&$-4.10$ &1.61&\textbf{98.62}/0.37/0.64/0.13/0.24\\
&1.05&$-12.33$ &1.02&\textbf{98.58}/0.39/0.66/0.13/0.24\\\midrule[1.0pt]
$I(J^P)$&$\Lambda$ &$E$ &$r_{\rm RMS}$ &P(${}^6\mathbb{S}_{\frac{5}{2}}/{}^2\mathbb{D}_{\frac{5}{2}}/{}^4\mathbb{D}_{\frac{5}{2}}/{}^6\mathbb{D}_{\frac{5}{2}}/{}^8\mathbb{D}_{\frac{5}{2}}$)\\\hline
\multirow{3}{*}{$0(\frac{5}{2}^+)$}&1.15&$-0.32$ &4.61&\textbf{98.86}/0.16/0.06/0.89/0.03\\
&1.27&$-4.44$ &1.61&\textbf{97.65}/0.34/0.13/1.83/0.05\\
&1.39&$-13.04$ &1.04&\textbf{97.31}/0.40/0.15/2.08/0.06\\\midrule[1.0pt]
$I(J^P)$&$\Lambda$ &$E$ &$r_{\rm RMS}$ &P(${}^8\mathbb{S}_{\frac{7}{2}}/{}^4\mathbb{D}_{\frac{7}{2}}/{}^6\mathbb{D}_{\frac{7}{2}}/{}^8\mathbb{D}_{\frac{7}{2}}$)\\\hline
\multirow{3}{*}{$0(\frac{7}{2}^+)$}&1.57&$-0.31$ &4.84&\textbf{98.10}/0.08/0.02/1.79\\
&1.96&$-4.74$ &1.73&\textbf{94.84}/0.19/0.06/4.92\\
&2.38&$-12.79$ &1.19&\textbf{93.11}/0.23/0.08/6.59\\\midrule[1.0pt]
\multirow{3}{*}{$1(\frac{7}{2}^+)$}&1.83&$-0.80$ &2.92&\textbf{99.89}/0.01/$o(0)$/0.10\\
&1.86&$-5.17$ &1.15&\textbf{99.81}/0.01/$o(0)$/0.18\\
&1.89&$-9.68$ &0.85&\textbf{99.76}/0.01/$o(0)$/0.23\\
\bottomrule[1.0pt]\bottomrule[1.0pt]
\end{tabular}
\end{table}

In the single channel analysis, the $\Xi_c^{*}\bar D_2^*$ states with $I(J^P)=0({1}/{2}^+)$, $0({3}/{2}^+)$, and $0({5}/{2}^+)$ can be bound together to form the bound states when we set the cutoff values to be around 0.86 GeV, 0.96 GeV, and 1.23 GeV, respectively, which are the reasonable cutoff values. For the $\Xi_c^{*}\bar D_2^*$ states with $I(J^P)=0({7}/{2}^+)$ and $1({7}/{2}^+)$, there exist the bound state solutions when the cutoff values are taken to be around 2.08 and 1.84 GeV, respectively. For the $\Xi_c^{*}\bar D_2^*$ states with $I(J^P)=1({1}/{2}^+)$, $1({3}/{2}^+)$, and $1({5}/{2}^+)$, the bound state solutions cannot be found if we fix the cutoff values between 0.8 to 2.5 GeV.

Now we further take into account the role of the $S$-$D$ wave mixing effect for the $\Xi_c^{*}\bar D_2^*$ system. For the $\Xi_c^{*}\bar D_2^*$ states with $I(J^P)=0({1}/{2}^+)$, $0({3}/{2}^+)$, $0({5}/{2}^+)$, and $1({7}/{2}^+)$, the $S$-$D$ wave mixing effect plays the positive but minor role for the formation of these bound states, where the contribution of the dominant $S$-wave channel is over 97\%. However, the role of the tensor force from the $S$-$D$ wave mixing effect plays the important role in generating the $\Xi_c^{*}\bar D_2^*$ bound state with $I(J^P)=0({7}/{2}^+)$, and the corresponding bound state solutions can be obtained when we tune the cutoff value to be around 1.57 GeV. In addition, we still cannot find the bound state solutions for the $\Xi_c^{*}\bar D_2^*$ states with $I(J^P)=1({1}/{2}^+)$, $1({3}/{2}^+)$, and $1({5}/{2}^+)$ when the cutoff values are chosen between 0.8 to 2.5 GeV and the role of the $S$-$D$ wave mixing effect is introduced.

In short summary, the $\Xi_c^{*}\bar D_2^*$ states with $I(J^P)=0({1}/{2}^+)$, $0({3}/{2}^+)$, $0({5}/{2}^+)$, and $0({7}/{2}^+)$ can be  considered as the prime hidden-charm molecular pentaquark candidates with strangeness, and the $\Xi_c^{*}\bar D_2^*$ state with $I(J^P)=1({7}/{2}^+)$ may be the potential candidate of the hidden-charm molecular pentaquark with strangeness. In addition, our quantitative analysis does not support the $\Xi_c^{*}\bar D_2^*$ states with $I(J^P)=1({1}/{2}^+)$, $1({3}/{2}^+)$, and $1({5}/{2}^+)$ as the hidden-charm molecular pentaquark candidates with strangeness.

\section{Discussions and conclusions}\label{sec4}

Since 2003, numerous exotic hadron states have been reported by various experiments, sparking the interest in exploring these exotic hadrons and establishing them as a  research frontier within the hadron physics. Notably, in 2019, LHCb announced the discoveries of the $P_{\psi}^{N}(4312)$, $P_{\psi}^{N}(4440)$, and $P_{\psi}^{N}(4457)$ states, providing robust experimental evidence supporting the existence of the hidden-charm baryon-meson molecular pentaquark states. This progress has fueled our enthusiasm for constructing the family of the hidden-charm molecular pentaquarks. Inspired by the discoveries of the $P_{\psi s}^\Lambda(4459)$ and $P_{\psi s}^\Lambda(4338)$ as the potential $\Xi_c\bar D^{(*)}$ molecules, we have undertaken an investigation of the interactions between the charmed baryons $\Xi_c^{(\prime,*)}$ and the anticharmed mesons $\bar D_1/\bar D_2^*$. This work aims to explore a novel class of molecular $P_{\psi s}^{\Lambda/\Sigma}$ pentaquark candidates, which are composed of the charmed baryons $\Xi_c^{(\prime,*)}$ and the anticharmed mesons $\bar D_1/\bar D_2^*$ and possess masses ranging from approximately $4.87$ to $5.10$ GeV. Through our study, we anticipate predicting the existence of these intriguing pentaquark states.

\begin{figure}[htbp]
\includegraphics[scale=0.43]{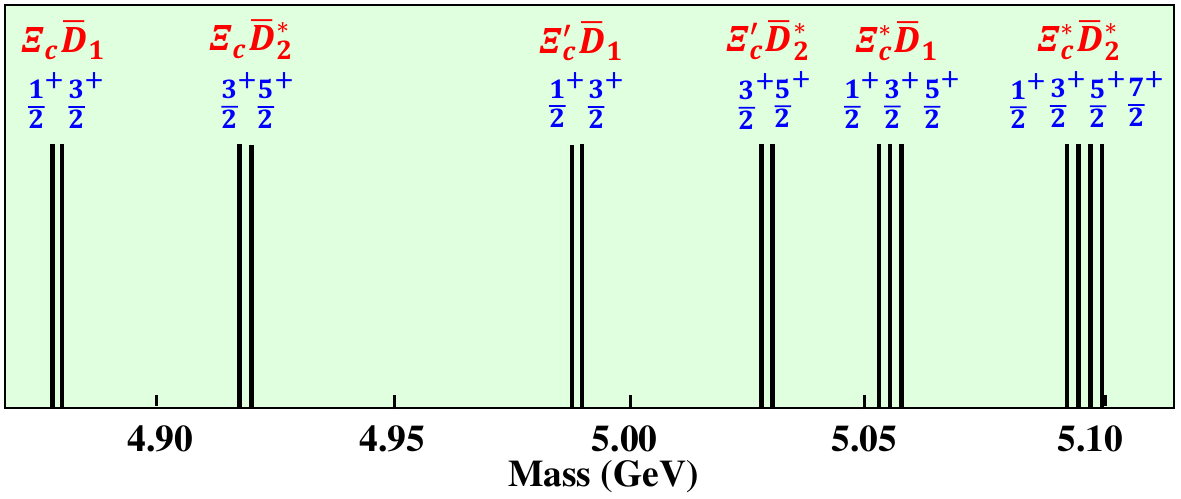}
\caption{The characteristic spectrum of the most promising molecular $P_{\psi s}^{\Lambda}$ pentaquark candidates for the $\Xi_c^{(\prime,*)}\bar D_1/\Xi_c^{(\prime,*)}\bar D_2^*$ systems.}\label{characteristicspectrum}
\end{figure}

In our concrete calculations, we have determined the effective potentials of the $\Xi_c^{(\prime,*)}\bar D_1/\Xi_c^{(\prime,*)}\bar D_2^*$ systems using the OBE model. These potentials incorporate the contributions from the exchange of the $\sigma$, $\pi$, $\eta$, $\rho$, and $\omega$ particles. Furthermore, we have taken into account both the $S$-$D$ wave mixing effect and the coupled channel effect. By solving the coupled channel Schr$\ddot{\rm o}$dinger equation, we have obtained the bound state properties of the discussed systems. Based on these obtained results, we propose that the following states can be considered as the most promising molecular $P_{\psi s}^{\Lambda}$ pentaquark candidates: the $\Xi_c\bar D_1$ states with $I(J^P)=0({1}/{2}^+,\,{3}/{2}^+)$, the $\Xi_c\bar D_2^*$ states with $I(J^P)=0({3}/{2}^+,\,{5}/{2}^+)$, the $\Xi_c^{\prime}\bar D_1$ states with $I(J^P)=0({1}/{2}^+,\,{3}/{2}^+)$, the $\Xi_c^{\prime}\bar D_2^*$ states with $I(J^P)=0({3}/{2}^+,\,{5}/{2}^+)$, the $\Xi_c^{*}\bar D_1$ states with $I(J^P)=0({1}/{2}^+,\,{3}/{2}^+,\,{5}/{2}^+)$, and the $\Xi_c^{*}\bar D_2^*$ states with $I(J^P)=0({1}/{2}^+,\,{3}/{2}^+,\,{5}/{2}^+,\,{7}/{2}^+)$. These findings align with the conclusions drawn in Ref. \cite{Dong:2021juy} and are depicted in Fig. \ref{characteristicspectrum}. Meanwhile, the $\Xi_c\bar D_1$ states with $I(J^P)=1({1}/{2}^+,\,{3}/{2}^+)$, the $\Xi_c\bar D_2^*$ states with $I(J^P)=1({3}/{2}^+,\,{5}/{2}^+)$, the $\Xi_c^{\prime}\bar D_1$ state with $I(J^P)=1({3}/{2}^+)$, the $\Xi_c^{\prime}\bar D_2^*$ state with $I(J^P)=1({5}/{2}^+)$, the $\Xi_c^{*}\bar D_1$ state with $I(J^P)=1({5}/{2}^+)$, and the $\Xi_c^{*}\bar D_2^*$ state with $I(J^P)=1({7}/{2}^+)$ may serve as the potential molecular $P_{\psi s}^{\Sigma}$ pentaquark candidates. However, the remaining states can be excluded as the hidden-charm molecular pentaquark candidates with strangeness. It is noteworthy that the $S$-$D$ mixing effect and the coupled channel effect are crucial for the formation of several hidden-charm molecular pentaquark candidates with strangeness. Notably, the spectroscopic behavior of the isoscalar $\Xi_c\bar D_1$ and $\Xi_c\bar D_2^*$ systems resembles that of the isoscalar $\Xi_c\bar D^*$ system \cite{Wang:2022mxy}, which can split into two distinct states due to the influence of the coupled channel effect.

It is very intriguing and important to search for these predicted hidden-charm molecular pentaquark candidates with strangeness experimentally, which can be detected in their allowed two-body strong decay channels. The two-body strong decay final states of our predicted most promising molecular $P_{\psi s}^\Lambda$ pentaquark candidates contain the baryon $\Lambda$ plus the charmonium state, the baryon $\Lambda_c$ plus the meson $\bar D_s$, the baryon $\Xi_c$ plus the meson $\bar D$, and so on. Furthermore, the two-body strong decay channels of our predicted potential molecular $P_{\psi s}^\Sigma$ pentaquark candidates include the baryon $\Sigma$ plus the charmonium state, the baryon $\Sigma_c$ plus the meson $\bar D_s$, the baryon $\Xi_c$ plus the meson $\bar D$, and so on. Here, the baryons and mesons in these final states stand for either the ground states or the excited states. These possible two-body strong decay information can give the crucial information to detect our predicted  hidden-charm molecular pentaquark candidates with strangeness in future experiments, specifically focusing on the two-body hidden-charm strong decay channels.

With the higher statistical data accumulation at the LHCb's Run II and Run III status \cite{Bediaga:2018lhg}, LHCb has the potential to detect these predicted hidden-charm molecular pentaquarks with strangeness by the $\Xi_b$ baryon weak decay in the near future\footnote{The $B$ meson weak decay is also the suitable process to produce the molecular $P_{\psi s}^{\Lambda/\Sigma}$ pentaquarks, and the maximum mass of the molecular $P_{\psi s}^{\Lambda/\Sigma}$ pentaquarks should be 4.34 GeV by the $B$ meson weak decay production. However, the masses of our predicted molecular $P_{\psi s}^{\Lambda/\Sigma}$ pentaquarks are around $4.87 \sim 5.10~{\rm GeV}$.}, which is the same as the production process of the $P_{\psi s}^\Lambda(4459)$ \cite{LHCb:2020jpq}. In addition, we hope that the joint effort from the theorists can give more abundant and reliable suggestions for future experimental searches for the hidden-charm molecular pentaquark states. Obviously, more and more hidden-charm molecular pentaquark candidates can be reported at the forthcoming experiments, and more opportunities and challenges are waiting for both theorists and experimentalists in the community of the hadron physics.

\section*{ACKNOWLEDGMENTS}

This work is supported by the China National Funds for Distinguished Young Scientists under Grant No. 11825503, National Key Research and Development Program of China under Contract No. 2020YFA0406400, the 111 Project under Grant No. B20063, the fundamental Research Funds for the Central Universities, the project for top-notch innovative talents of Gansu province, and the National Natural Science Foundation of China under Grant Nos. 12247155 and 12247101. F.L.W. is also supported by the China Postdoctoral Science Foundation under Grant No. 2022M721440.

\appendix

\section{The related interaction vertices}\label{app00}

\begin{figure}[htbp]
\centering
\includegraphics[scale=0.43]{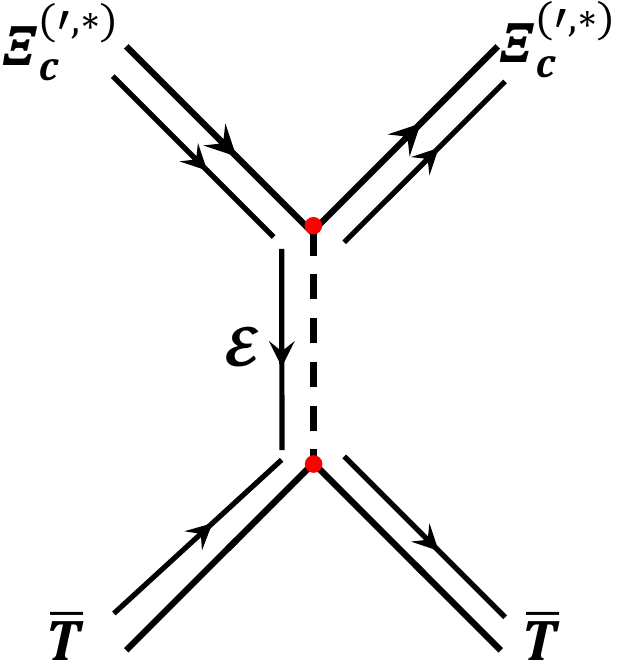}
\caption{The corresponding Feynman diagram for the $\Xi_c^{(\prime,*)}\bar T \to \Xi_c^{(\prime,*)}\bar T$ scattering process. Here, $T$ represents either $D_1$ or $D_2^*$.}\label{Feynmandiagram}
\end{figure}

For the $\Xi_c^{(\prime,*)}\bar T \to \Xi_c^{(\prime,*)}\bar T$ scattering process,  we provide the corresponding Feynman diagram in Fig. \ref{Feynmandiagram}.

In our concrete calculations, we can extract the interaction vertices for the $\Xi_c^{(\prime,*)}\bar T \to \Xi_c^{(\prime,*)}\bar T$ scattering process from the constructed effective Lagrangians. The explicit expressions for the related interaction vertex functions are given below
\begin{eqnarray*}
\Xi_c \Xi_c \sigma&:&2l_Bm_{\Xi_c}\chi^{\dagger}_3\chi_1,\\
\Xi_c^{\prime} \Xi_c^{\prime} \sigma&:&-2l_Sm_{\Xi_c^{\prime}}\chi^{\dagger}_3\chi_1,\\
\Xi_c^{*} \Xi_c^{*} \sigma&:&-2 l_Sm_{\Xi_c^{*}}\mathcal{A}{\bm\epsilon^{\dagger m_2^\prime}_{3}}\cdot{\bm\epsilon_{1}^{m_2}}\chi^{\dagger m_1^\prime}_3\chi_1^{m_1},\\
\Xi_c^{*} \Xi_c^{\prime} \sigma&:&\frac{2 l_S}{\sqrt{3}}\sqrt{m_{\Xi_c^{*}}m_{\Xi_c^{\prime}}}\mathcal{B}\chi^{\dagger m_1}_3{\bm \sigma}\cdot{\bm\epsilon^{\dagger m_2}_{3}}\chi_1,\\
\Xi_c^{\prime} \Xi_c^{\prime} \mathbb{P}&:&-\frac{2ig_1}{f_{\pi}}m_{\Xi_c^{\prime}}{\bm \sigma}\cdot{\bm q}\chi^{\dagger}_3\chi_1,\\
\Xi_c^{*} \Xi_c^{*} \mathbb{P}&:&-\frac{3g_1}{f_{\pi}}m_{\Xi_c^{*}}\mathcal{A}\left({\bm\epsilon^{\dagger m_2^\prime}_{3}}\times{\bm\epsilon_{1}^{m_2}}\right)\cdot{\bm q}\chi^{\dagger m_1^\prime}_3\chi_1^{m_1},\\
\Xi_c^{\prime} \Xi_c \mathbb{P}&:&-\frac{2ig_4}{\sqrt{3}f_{\pi}}\sqrt{m_{\Xi_c^{\prime}}m_{\Xi_c}}\chi^{\dagger}_3{\bm \sigma}\cdot{\bm q}\chi_1,\\
\Xi_c^{*} \Xi_c  \mathbb{P}&:&\frac{2ig_4}{f_{\pi}}\sqrt{m_{\Xi_c^{*}}m_{\Xi_c}}\mathcal{B}{\bm\epsilon^{\dagger m_2}_{3}}\cdot{\bm q}\chi^{\dagger m_1}_3\chi_1,\\
\Xi_c^{*} \Xi_c^{\prime}  \mathbb{P}&:&\frac{\sqrt{3}g_1}{f_{\pi}}\sqrt{m_{\Xi_c^{*}}m_{\Xi_c^{\prime}}}\mathcal{B}\chi^{\dagger m_1}_3\left({\bm\epsilon^{\dagger m_2}_{3}}\times{\bm \sigma}\right)\cdot{\bm q}\chi_1,\\
\Xi_c \Xi_c \mathbb{V}&:&\sqrt{2}\beta_B g_Vm_{\Xi_c}\chi^{\dagger}_3\chi_1,\\
\Xi_c^{\prime} \Xi_c^{\prime} \mathbb{V}&:&-\sqrt{2}\beta_S g_Vm_{\Xi_c^{\prime}}\chi^{\dagger}_3\chi_1+\frac{2\sqrt{2}i\lambda_S g_V}{3}m_{\Xi_c^{\prime}}\chi^{\dagger}_3\left({\bm \sigma}\times{\bm q}\right)_{\mu}\chi_1,\\
\Xi_c^{*} \Xi_c^{*} \mathbb{V}&:&-\sqrt{2}\beta_S g_Vm_{\Xi_c^{*}}\mathcal{A}{\bm\epsilon^{\dagger m_2^\prime}_{3}}\cdot{\bm\epsilon_{1}^{m_2}}\chi^{\dagger m_1^\prime}_3\chi_1^{m_1}\nonumber\\
&&+\sqrt{2}\lambda_S g_Vm_{\Xi_c^{*}}\mathcal{A}\nonumber\\
&&\times\left[\left({\bm\epsilon^{\dagger m_2^\prime}_{3}}\cdot{\bm q}\right)\epsilon_{1\mu}^{m_2}-{\epsilon^{\dagger m_2^\prime}_{3\mu}}\left(\bm \epsilon_{1}^{m_2}\cdot{\bm q}\right)\right]\chi^{\dagger m_1^\prime}_3\chi_1^{m_1},\\
\Xi_c^{\prime} \Xi_c \mathbb{V}&:&\frac{4i\lambda_I g_V}{\sqrt{6}}\sqrt{m_{\Xi_c^{\prime}}m_{\Xi_c}}\varepsilon^{0\nu\lambda\mu}\chi^{\dagger}_3\sigma_{\nu}q_{\lambda}\chi_1,\\
\Xi_c^{*} \Xi_c \mathbb{V}&:&-2\sqrt{2}i\lambda_I g_V\sqrt{m_{\Xi_c^{*}}m_{\Xi_c}}\mathcal{B}\varepsilon^{0\nu\lambda\mu}\epsilon_{3\nu}^{\dagger m_2}q_{\lambda}\chi^{\dagger m_1}_3\chi_1,\\
\Xi_c^{*} \Xi_c^{\prime} \mathbb{V}&:&\frac{2\beta_S g_V}{\sqrt{6}}\sqrt{m_{\Xi_c^{*}}m_{\Xi_c^{\prime}}}\mathcal{B}\chi^{\dagger m_1}_3{\bm\epsilon^{\dagger m_2}_{3}}\cdot{\bm \sigma}\chi_1\nonumber\\
&&-\frac{2\lambda_S g_V}{\sqrt{6}}\sqrt{m_{\Xi_c^{*}}m_{\Xi_c^{\prime}}}\mathcal{B}\chi^{\dagger m_1}_3\nonumber\\
&&\times\left[\left({\bm\epsilon^{\dagger m_2}_{3}}\cdot{\bm q}\right)\sigma_{1\mu}-{\epsilon^{\dagger m_2}_{3\mu}}\left(\bm \sigma\cdot{\bm q}\right)\right]\chi_1,\\
\bar D_1 \bar D_1 \sigma&:&2g''_{\sigma}m_{D_1}{\bm\epsilon^{\dagger}_4}\cdot{\bm\epsilon_2},\\
\bar D_2^* \bar D_2^* \sigma&:&2g''_{\sigma}m_{D_2^*}\mathcal{C}\left({\bm\epsilon^{\dagger}_{4m}}\cdot{\bm\epsilon_{2a}}\right)\left({\bm\epsilon^{\dagger}_{4n}}\cdot {\bm\epsilon_{2b}}\right),\\
\bar D_1 \bar D_1 \mathbb{P}&:&\frac{-5k}{3f_{\pi}}m_{D_1}\left({\bm\epsilon^{\dagger}_4}\times{\bm\epsilon_2}\right)\cdot {\bm q},\\
\bar D_2^* \bar D_2^* \mathbb{P}&:&\frac{-2k}{f_{\pi}}m_{D_2^*}\mathcal{C}\left({\bm\epsilon^{\dagger}_{4m}}\cdot{\bm\epsilon_{2a}}\right)\left[\left({\bm\epsilon^{\dagger}_{4n}}\times {\bm\epsilon_{2b}}\right)\cdot{\bm q}\right],\\
\bar D_2^* \bar D_1 \mathbb{P}&:&-i\sqrt{\frac{2}{3}}\frac{k}{f_{\pi}}\sqrt{m_{D_2^*}m_{D_1}}\mathcal{D}\left({\bm\epsilon^{\dagger}_{4m}}\cdot{\bm\epsilon_{2}}\right)\left({\bm\epsilon^{\dagger}_{4n}}\cdot{\bm q}\right),\\
\bar D_1 \bar D_1 \mathbb{V}&:&-\sqrt{2}\beta^{\prime\prime} g_Vm_{D_1}{\bm\epsilon^{\dagger}_4}\cdot{\bm\epsilon_2}\nonumber\\
&&-\frac{5\sqrt{2}\lambda^{\prime\prime} g_V}{3}m_{D_1}\left[\left({\bm\epsilon^{\dagger}_{4}}\cdot{\bm q}\right)\epsilon^{\nu}_{2}-\epsilon^{\nu\dagger}_{4}\left({\bm\epsilon_{2}}\cdot{\bm q}\right)\right],\\
\bar D_2^* \bar D_2^* \mathbb{V}&:&-\sqrt{2}\beta^{\prime\prime} g_Vm_{D_2^*}\mathcal{C}\left({\bm\epsilon^{\dagger}_{4m}}\cdot{\bm\epsilon_{2a}}\right)\left({\bm\epsilon^{\dagger}_{4n}}\cdot {\bm\epsilon_{2b}}\right)\nonumber\\
&&-2\sqrt{2}\lambda^{\prime\prime} g_Vm_{D_2^*}\mathcal{C}\left({\bm\epsilon^{\dagger}_{4m}}\cdot{\bm\epsilon_{2a}}\right)\nonumber\\
&&\times\left[\left({\bm\epsilon^{\dagger}_{4n}}\cdot{\bm q}\right)\epsilon^{\nu}_{2b}-\epsilon^{\nu\dagger}_{4n}\left({\bm\epsilon_{2b}}\cdot{\bm q}\right)\right],\\
\bar D_2^* \bar D_1 \mathbb{V}&:&\frac{2i\lambda^{\prime\prime} g_V}{\sqrt{3}}\sqrt{m_{D_2^*}m_{D_1}}\mathcal{D}\left[3\varepsilon^{\mu 0\nu\lambda}\left({\bm\epsilon^{\dagger}_{4m}}\cdot{\bm\epsilon_{2}}\right)\epsilon_{4n\lambda}^{\dagger}q_{\mu}\right.\nonumber\\
&&\left.+2\varepsilon^{\mu \lambda 0\nu}\left({\bm\epsilon^{\dagger}_{4m}}\cdot{\bm q}\right)\epsilon_{4n\mu}^{\dagger}{\epsilon_{2\lambda}}-2\left({\bm\epsilon^{\dagger}_{4m}}\times{\bm\epsilon_{2}}\right)\cdot{\bm q}\epsilon_{4n}^{\nu\dagger}\right].
\end{eqnarray*}
In the above interaction vertex functions, we take the notations $\mathcal{A}=\sum_{m_1,m_2,m_1^\prime,m_2^\prime}C^{\frac{3}{2},m_1+m_2}_{\frac{1}{2}m_1,1m_2}C^{\frac{3}{2},m_1^\prime+m_2^\prime}_{\frac{1}{2}m_1^\prime,1m_2^\prime}$, $\mathcal{B}=\sum_{m_1,m_2}C^{\frac{3}{2},m_1+m_2}_{\frac{1}{2}m_1,1m_2}$, $\mathcal{C}=\sum_{m,n,a,b}C^{2,m+n}_{1m,1n}C^{2,a+b}_{1a,1b}$, and $\mathcal{D}=\sum_{m,n}C^{2,m+n}_{1m,1n}$.

\section{The OBE effective potentials for the $\Xi_c^{(\prime,*)}\bar D_1/\Xi_c^{(\prime,*)}\bar D_2^*$ systems}\label{app01}

In this appendix, we collect the obtained OBE effective potentials for the $\Xi_c^{(\prime,*)}\bar D_1/\Xi_c^{(\prime,*)}\bar D_2^*$ systems when considering the single channel analysis and the $S$-$D$ wave mixing analysis. Before listing the OBE effective potentials for the $\Xi_c^{(\prime,*)}\bar D_1/\Xi_c^{(\prime,*)}\bar D_2^*$ systems, several operators adopted in the present work are defined as
\begin{eqnarray*}
\mathcal{O}_{1}&=&{\bm\epsilon^{\dagger}_4}\cdot{\bm\epsilon_2}\chi^{\dagger}_3\chi_1,\nonumber\\
\mathcal{O}_{2}&=&\chi^{\dagger}_3 \left[{\bm\sigma}\cdot\left(i{\bm\epsilon^{\dagger}_4}\times {\bm\epsilon_2}\right)\right]\chi_1,\nonumber\\
\mathcal{O}_{3}&=&\chi^{\dagger}_3T({\bm\sigma},i{\bm\epsilon^{\dagger}_4}
\times{\bm\epsilon_2})\chi_1,\nonumber\\
\mathcal{O}_{4}&=&\mathcal{A}\left({\bm\epsilon^{\dagger}_{4m}}
\cdot{\bm\epsilon_{2a}}\right)\left({\bm\epsilon^{\dagger}_{4n}}\cdot {\bm\epsilon_{2b}}\right)\chi^{\dagger}_3\chi_1,\nonumber\\
\mathcal{O}_{5}&=&\mathcal{A}\left({\bm\epsilon^{\dagger}_{4m}}
\cdot{\bm\epsilon_{2a}}\right)\chi^{\dagger}_3[{\bm\sigma}\cdot(i{\bm\epsilon^{\dagger}_{4n}}
\times {\bm\epsilon_{2b}})]\chi_1,\nonumber\\
\mathcal{O}_{6}&=&\mathcal{A}\left({\bm\epsilon^{\dagger}_{4m}}
\cdot{\bm\epsilon_{2a}}\right)\chi^{\dagger}_3T({\bm\sigma},i{\bm \epsilon^{\dagger}_{4n}}\times{\bm\epsilon_{2b}})\chi_1,\nonumber\\
\mathcal{O}_{7}&=&\mathcal{B}\left({\bm\epsilon^{\dagger m_2^\prime}_{3}}\cdot{\bm\epsilon_{1}}^{m_2}\right)\left({\bm\epsilon^{\dagger}_{4}}\cdot {\bm\epsilon_{2}}\right)\chi^{\dagger m_1^\prime}_3\chi_1^{m_1},\nonumber\\
\mathcal{O}_{8}&=&\mathcal{B}\left[\left({\bm\epsilon^{\dagger m_2^\prime}_{3}}\times{\bm\epsilon_{1}}^{m_2}\right)\cdot\left({\bm\epsilon^{\dagger}_{4}}\times {\bm\epsilon_{2}}\right)\right]\chi^{\dagger m_1^\prime}_3\chi_1^{m_1},\nonumber\\
\mathcal{O}_{9}&=&\mathcal{B}T({\bm\epsilon^{\dagger m_2^\prime}_{3}}\times{\bm\epsilon_{1}}^{m_2},{\bm\epsilon^{\dagger}_{4}}\times {\bm\epsilon_{2}})\chi^{\dagger m_1^\prime}_3\chi_1^{m_1},\nonumber\\
\mathcal{O}_{10}&=&\mathcal{A}\mathcal{B}
\left({\bm\epsilon^{\dagger m_2^\prime}_{3}}\cdot{\bm\epsilon_{1}}^{m_2}\right)\left({\bm\epsilon^{\dagger}_{4m}}
\cdot{\bm\epsilon_{2a}}\right)\left({\bm\epsilon^{\dagger}_{4n}}\cdot {\bm\epsilon_{2b}}\right)\chi^{\dagger m_1^\prime}_3\chi_1^{m_1},\nonumber\\
\mathcal{O}_{11}&=&\mathcal{A}\mathcal{B}
\left({\bm\epsilon^{\dagger}_{4m}}
\cdot{\bm\epsilon_{2a}}\right)\left[\left({\bm\epsilon^{\dagger m_2^\prime}_{3}}\times{\bm\epsilon_{1}}^{m_2}\right)\cdot\left({\bm\epsilon^{\dagger}_{4n}}\times {\bm\epsilon_{2b}}\right)\right]\chi^{\dagger m_1^\prime}_3\chi_1^{m_1},\nonumber\\
\mathcal{O}_{12}&=&\mathcal{A}\mathcal{B}
\left({\bm\epsilon^{\dagger}_{4m}}\cdot{\bm\epsilon_{2a}}\right)T({\bm\epsilon^{\dagger m_2^\prime}_{3}}\times{\bm\epsilon_{1}}^{m_2},{\bm\epsilon^{\dagger}_{4n}}\times {\bm\epsilon_{2b}})\chi^{\dagger m_1^\prime}_3\chi_1^{m_1}.
\end{eqnarray*}
In the above defined operators, we use the notations $\mathcal{A}=\sum_{m,n,a,b}C^{2,m+n}_{1m,1n}C^{2,a+b}_{1a,1b}$ and $\mathcal{B}=\sum_{m_1,m_2,m_1^\prime,m_2^\prime}C^{\frac{3}{2},m_1+m_2}_{\frac{1}{2}m_1,1m_2}C^{\frac{3}{2},m_1^\prime+m_2^\prime}_{\frac{1}{2}m_1^\prime,1m_2^\prime}$. In addition, the tensor force operator $T({\bm x},{\bm y})$ is $T({\bm x},{\bm y})= 3\left(\hat{\bm r} \cdot {\bm x}\right)\left(\hat{\bm r} \cdot {\bm y}\right)-{\bm x} \cdot {\bm y}$. In the concrete calculations, these defined operators $\mathcal{O}_i$ should be sandwiched by the corresponding spin-orbital wave functions $|{}^{2S+1}L_{J}\rangle$ of the initial and final states listed in Eq. (\ref{spinorbitalwavefunctions}), where the obtained operator matrix elements are summarized in Table~\ref{matrix}.
\renewcommand\tabcolsep{0.00cm}
\renewcommand{\arraystretch}{1.50}
\begin{table*}[!htbp]
  \caption{The obtained operator matrix elements in the OBE effective potentials when considering the $S$-$D$ wave mixing effect.}\label{matrix}
  \begin{tabular}{c|cccc}\toprule[1pt]\toprule[1pt]
   {{{$J^P$}}}
                    & $1/2^+$     & $3/2^+$     & $5/2^+$  & $7/2^+$ \\\midrule[1pt]
 $\mathcal{O}_1$
            &Diag(1,1) &Diag(1,1,1) &&\\
 $\mathcal{O}_2$
            &Diag(2,$-1$) &Diag($-1$,2,$-1$) &&\\
 $\mathcal{O}_3$
           &$\left(\begin{array}{cc} 0 & \sqrt{2} \\ \sqrt{2} & 2\end{array}\right)$&$\left(\begin{array}{ccc} 0 & -1& -2 \\ -1 & 0& 1 \\ -2 & 1& 0 \end{array}\right)$&\\
 $\mathcal{O}_4$
            & &Diag(1,1,1) &Diag(1,1,1)&\\
 $\mathcal{O}_5$
            & &Diag($\frac{3}{2}$,$\frac{3}{2}$,$-1$) &Diag($-1$,$\frac{3}{2}$,$-1$)&\\
 $\mathcal{O}_6$
           &
           &$\left(\begin{array}{ccc} 0 & \frac{3}{5}& \frac{3\sqrt{21}}{10} \\ \frac{3}{5} & 0& \frac{3\sqrt{21}}{14} \\ \frac{3\sqrt{21}}{10} & \frac{3\sqrt{21}}{14}& \frac{4}{7} \end{array}\right)$
           &$\left(\begin{array}{ccc} 0 & -\frac{3\sqrt{14}}{10}& -\frac{2\sqrt{14}}{5} \\ -\frac{3\sqrt{14}}{10} & \frac{3}{7}& \frac{3}{7} \\ -\frac{2\sqrt{14}}{5} & \frac{3}{7}& -\frac{4}{7} \end{array}\right)$&\\
 $\mathcal{O}_7$
            &Diag(1,1,1) &Diag(1,1,1,1) &Diag(1,1,1,1)&\\
 $\mathcal{O}_8$
            &Diag($\frac{5}{3}$,$\frac{2}{3}$,$-1$) &Diag($\frac{2}{3}$,$\frac{5}{3}$,$\frac{2}{3}$,$-1$) &Diag($-1$,$\frac{5}{3}$,$\frac{2}{3}$,$-1$)& \\
 $\mathcal{O}_9$
           &$\left(\begin{array}{ccc} 0 & -\frac{7}{3\sqrt{5}}& \frac{2}{\sqrt{5}} \\ -\frac{7}{3\sqrt{5}} & \frac{16}{15}& -\frac{1}{5} \\ \frac{2}{\sqrt{5}} &-\frac{1}{5}& \frac{8}{5} \end{array}\right)$
           &$\left(\begin{array}{cccc} 0 & \frac{7}{3\sqrt{10}}& -\frac{16}{15}& -\frac{\sqrt{7}}{5\sqrt{2}}\\ \frac{7}{3\sqrt{10}} & 0& -\frac{7}{3\sqrt{10}} & -\frac{2}{\sqrt{35}} \\ -\frac{16}{15} & -\frac{7}{3\sqrt{10}}& 0& -\frac{1}{\sqrt{14}} \\-\frac{\sqrt{7}}{5\sqrt{2}}&-\frac{2}{\sqrt{35}} &-\frac{1}{\sqrt{14}}&\frac{4}{7}\end{array}\right)$
           &$\left(\begin{array}{cccc} 0 & \frac{2}{\sqrt{15}}& \frac{\sqrt{7}}{5\sqrt{3}}& -\frac{2\sqrt{14}}{5}\\ \frac{2}{\sqrt{15}} & 0& \frac{\sqrt{7}}{3\sqrt{5}} & -\frac{4\sqrt{2}}{\sqrt{105}} \\ \frac{\sqrt{7}}{5\sqrt{3}} & \frac{\sqrt{7}}{3\sqrt{5}}& -\frac{16}{21}& -\frac{\sqrt{2}}{7\sqrt{3}} \\-\frac{2\sqrt{14}}{5}&-\frac{4\sqrt{2}}{\sqrt{105}} &-\frac{\sqrt{2}}{7\sqrt{3}}&-\frac{4}{7}\end{array}\right)$&\\
 $\mathcal{O}_{10}$
            &Diag(1,1,1)  &Diag(1,1,1,1,1)  &Diag(1,1,1,1,1) &Diag(1,1,1,1)  \\
 $\mathcal{O}_{11}$
            &Diag($\frac{3}{2}$,1,$\frac{1}{6}$)&Diag(1,$\frac{3}{2}$,1,$\frac{1}{6}$,$-1$)
            &Diag($\frac{1}{6}$,$\frac{3}{2}$,1,$\frac{1}{6}$,$-1$) &Diag($-1$,1,$\frac{1}{6}$,$-1$)  \\
  $\mathcal{O}_{12}$
   &$\left(\begin{array}{ccc} 0 & \frac{9}{10}& \frac{\sqrt{21}}{5} \\ \frac{9}{10} & \frac{4}{5}& \frac{3\sqrt{21}}{70} \\ \frac{\sqrt{21}}{5} & \frac{3\sqrt{21}}{70}& \frac{116}{105} \end{array}\right)$
           &$\left(\begin{array}{ccccc}0 & -\frac{9\sqrt{2}}{20} & -\frac{4}{5}& \frac{3\sqrt{6}}{20} & \frac{2\sqrt{2}}{5}\\ -\frac{9\sqrt{2}}{20} & 0& \frac{9\sqrt{2}}{20} & -\frac{\sqrt{3}}{5}& 0\\ -\frac{4}{5} & \frac{9\sqrt{2}}{20}& 0& \frac{3\sqrt{6}}{28}& -\frac{4\sqrt{2}}{35}\\ \frac{3\sqrt{6}}{20} & -\frac{\sqrt{3}}{5}&\frac{3\sqrt{6}}{28}& \frac{58}{147} & \frac{18\sqrt{3}}{245}\\ \frac{2\sqrt{2}}{5} & 0& -\frac{4\sqrt{2}}{35}& \frac{18\sqrt{3}}{245}& \frac{60}{49}\end{array}\right)$
           &$\left(\begin{array}{ccccc}0 & \frac{\sqrt{7}}{5} & -\frac{3}{10}& -\frac{29\sqrt{2}}{15\sqrt{7}} & \frac{\sqrt{6}}{5\sqrt{7}}\\ \frac{\sqrt{7}}{5} & 0& \frac{-9}{10\sqrt{7}} & -\frac{2\sqrt{2}}{5}& 0\\ -\frac{3}{10} & -\frac{9}{10\sqrt{7}}& -\frac{4}{7}& \frac{3}{7\sqrt{14}}& -\frac{12\sqrt{6}}{35\sqrt{7}}\\ -\frac{29\sqrt{2}}{15\sqrt{7}} & -\frac{2\sqrt{2}}{5}&\frac{3}{7\sqrt{14}}& -\frac{58}{147} & \frac{17\sqrt{3}}{245}\\ \frac{\sqrt{6}}{5\sqrt{7}} & 0& -\frac{12\sqrt{6}}{35\sqrt{7}}& \frac{17\sqrt{3}}{245}& \frac{10}{49}\end{array}\right)$
           &$\left(\begin{array}{cccc} 0 & \frac{2}{5}& -\frac{3}{5\sqrt{14}}& -\frac{2\sqrt{21}}{7}\\ \frac{2}{5} & \frac{8}{35}& -\frac{\sqrt{27}}{35\sqrt{14}} & -\frac{4\sqrt{21}}{49} \\ -\frac{3}{5\sqrt{14}} & -\frac{27}{35\sqrt{14}}& -\frac{493}{735}& \frac{\sqrt{3}}{49\sqrt{2}} \\-\frac{2\sqrt{21}}{7}&-\frac{4\sqrt{21}}{49} &\frac{\sqrt{3}}{49\sqrt{2}}&-\frac{32}{49}\end{array}\right)$\\
           \bottomrule[1pt]\bottomrule[1pt]
  \end{tabular}
\end{table*}

For simplicity, we define the following relations in the obtained OBE effective potentials
\begin{eqnarray}
\mathcal{H}(I)Y_{\mathbb{P}}&=&\mathcal{H}_1(I)Y_\pi+\mathcal{H}_2(I)Y_\eta,\\
\mathcal{G}(I)Y_{\mathbb{V}}&=&\mathcal{G}_1(I)Y_\rho+\mathcal{G}_2(I)Y_\omega.
\end{eqnarray}
Here, the isospin factors $\mathcal{H}(I)$ and $\mathcal{G}(I)$ are introduced for the $\Xi_c^{(\prime,*)}\bar D_1/\Xi_c^{(\prime,*)}\bar D_2^*$ systems, and $I$ stands for the corresponding isospin quantum number. For the $\Xi_c\bar D_1/\Xi_c\bar D_2^*$ systems, we can obtain
\begin{eqnarray}
\mathcal{G}_1(I=0)&=&-3/2,~~~~\mathcal{G}_1(I=1)=1/2,\nonumber\\
\mathcal{G}_2(I=0)&=&1/2,~~~~~~~\mathcal{G}_2(I=1)=1/2.
\end{eqnarray}
For the $\Xi_c^{\prime(*)}\bar D_1/\Xi_c^{\prime(*)}\bar D_2^*$ systems, we can get
\begin{eqnarray}
\mathcal{H}_1(I=0)&=&-3/4,~~~~~~\mathcal{H}_1(I=1)=1/4,\nonumber\\
\mathcal{H}_2(I=0)&=&-1/12,~~~~\mathcal{H}_2(I=1)=-1/12,\nonumber\\
\mathcal{G}_1(I=0)&=&-3/4,~~~~~~~\mathcal{G}_1(I=1)=1/4,\nonumber\\
\mathcal{G}_2(I=0)&=&1/4,~~~~~~~~~~\mathcal{G}_2(I=1)=1/4.
\end{eqnarray}
In the above defined relations, the Yukawa potential considering the monopole-type form factor $Y_{\mathcal{E}}$ can be written as
\begin{eqnarray}
Y_{\mathcal{E}}=\frac{e^{-m_{\mathcal{E}}r}-e^{-\Lambda r}}{4\pi r}-\frac{\Lambda^2-m_{\mathcal{E}}^2}{8 \pi\Lambda}e^{-\Lambda r},
\end{eqnarray}
where $\Lambda$ is the cutoff parameter in the monopole-type form factor, and $m_{\mathcal{E}}$ is the mass of the exchanged light meson $\mathcal{E}$.

The OBE effective potentials in the coordinate space for the $\Xi_c^{(\prime,*)}\bar D_1/\Xi_c^{(\prime,*)}\bar D_2^*$ systems are given by
\begin{eqnarray}
\mathcal{V}^{\Xi_{c}\bar D_{1}\rightarrow\Xi_{c}\bar D_{1}}&=&-2l_{B}g_{\sigma}^{\prime\prime}\mathcal{O}_1Y_\sigma-\frac{1}{2}\beta_B \beta^{\prime\prime} g_{V}^2\mathcal{O}_1\mathcal{G}(I)Y_{\mathbb{V}},\\
\mathcal{V}^{\Xi_{c}\bar D_{2}^{*}\rightarrow\Xi_{c}\bar D_{2}^{*}}&=&-2l_{B}g_{\sigma}^{\prime\prime}\mathcal{O}_4Y_\sigma-\frac{1}{2}\beta_B \beta^{\prime\prime} g_{V}^2\mathcal{O}_4\mathcal{G}(I)Y_{\mathbb{V}},\\
\mathcal{V}^{\Xi_{c}^{\prime}\bar D_{1}\rightarrow\Xi_{c}^{\prime}\bar D_{1}}&=&l_{S}g_{\sigma}^{\prime\prime}\mathcal{O}_1Y_\sigma+\frac{1}{2}\beta_S \beta^{\prime\prime} g_{V}^2\mathcal{O}_1\mathcal{G}(I)Y_{\mathbb{V}}\nonumber\\
&&+\frac{5}{18}\frac{g_1 k}{f_\pi^2}\left[\mathcal{O}_2\nabla^2+\mathcal{O}_3r\frac{\partial}{\partial r}\frac{1}{r}\frac{\partial}{\partial r}\right]\mathcal{H}(I)Y_{\mathbb{P}}\nonumber\\
&&-\frac{5}{27}\lambda_S \lambda^{\prime\prime}g_V^2\left[2\mathcal{O}_2\nabla^2-\mathcal{O}_3 r\frac{\partial}{\partial r}\frac{1}{r}\frac{\partial}{\partial r}\right]\mathcal{G}(I)Y_{\mathbb{V}},\nonumber\\\\
\mathcal{V}^{\Xi_{c}^{\prime}\bar D_{2}^{*}\rightarrow\Xi_{c}^{\prime}\bar D_{2}^{*}}&=&l_{S}g_{\sigma}^{\prime\prime}\mathcal{O}_4Y_\sigma+\frac{1}{2}\beta_S \beta^{\prime\prime} g_{V}^2\mathcal{O}_4\mathcal{G}(I)Y_{\mathbb{V}}\nonumber\\
&&+\frac{1}{3}\frac{g_1 k}{f_\pi^2}\left[\mathcal{O}_5\nabla^2+\mathcal{O}_6r\frac{\partial}{\partial r}\frac{1}{r}\frac{\partial}{\partial r}\right]\mathcal{H}(I)Y_{\mathbb{P}}\nonumber\\
&&-\frac{2}{9}\lambda_S \lambda^{\prime\prime}g_V^2\left[2\mathcal{O}_5\nabla^2-\mathcal{O}_6 r\frac{\partial}{\partial r}\frac{1}{r}\frac{\partial}{\partial r}\right]\mathcal{G}(I)Y_{\mathbb{V}},\nonumber\\\\
\mathcal{V}^{\Xi_{c}^{*}\bar D_{1}\rightarrow\Xi_{c}^{*}\bar D_{1}}&=&l_{S}g_{\sigma}^{\prime\prime}\mathcal{O}_7Y_\sigma+\frac{1}{2}\beta_S \beta^{\prime\prime} g_{V}^2\mathcal{O}_7\mathcal{G}(I)Y_{\mathbb{V}}\nonumber\\
&&+\frac{5}{12}\frac{g_1 k}{f_\pi^2}\left[\mathcal{O}_8\nabla^2+\mathcal{O}_9r\frac{\partial}{\partial r}\frac{1}{r}\frac{\partial}{\partial r}\right]\mathcal{H}(I)Y_{\mathbb{P}}\nonumber\\
&&-\frac{5}{18}\lambda_S \lambda^{\prime\prime}g_V^2\left[2\mathcal{O}_8\nabla^2-\mathcal{O}_9 r\frac{\partial}{\partial r}\frac{1}{r}\frac{\partial}{\partial r}\right]\mathcal{G}(I)Y_{\mathbb{V}},\nonumber\\\\
\mathcal{V}^{\Xi_{c}^{*}\bar D_{2}^{*}\rightarrow\Xi_{c}^{*}\bar D_{2}^{*}}&=&l_{S}g_{\sigma}^{\prime\prime}\mathcal{O}_{10}Y_\sigma+\frac{1}{2}\beta_S \beta^{\prime\prime} g_{V}^2\mathcal{O}_{10}\mathcal{G}(I)Y_{\mathbb{V}}\nonumber\\
&&+\frac{1}{2}\frac{g_1 k}{f_\pi^2}\left[\mathcal{O}_{11}\nabla^2+\mathcal{O}_{12}r\frac{\partial}{\partial r}\frac{1}{r}\frac{\partial}{\partial r}\right]\mathcal{H}(I)Y_{\mathbb{P}}\nonumber\\
&&-\frac{1}{3}\lambda_S \lambda^{\prime\prime}g_V^2\left[2\mathcal{O}_{11}\nabla^2-\mathcal{O}_{12}r\frac{\partial}{\partial r}\frac{1}{r}\frac{\partial}{\partial r}\right]\mathcal{G}(I)Y_{\mathbb{V}}.\nonumber\\
\end{eqnarray}
In the above OBE effective potentials, the superscript is used to mark the corresponding scattering process.

\end{document}